\documentclass[prd,preprint,tightenlines,11pt,amsmath,amssymb]{revtex4}

\usepackage{graphicx}       
\usepackage{dcolumn}        
\usepackage{bm}             
\usepackage{psfrag}

\newcommand{\beq}{\begin{equation}}
\newcommand{\eeq}{\end{equation}}
\newcommand{\mass}{\mu}
\newcommand{\Gret}{G_{\text{ret}}}
\newcommand{\Gadv}{G_{\text{adv}}}
 
\newcommand{\GS}{G_{S}}
\newcommand{\tauret}{\tau_{\text{ret}}}
\newcommand{\tauadv}{\tau_{\text{adv}}}
\newcommand{\Phires}{\Phi_{\mathcal{R}}}
\newcommand{\Phiresloc}{\Phi_{\hat{\mathcal{R}}}}
\newcommand{\Psires}{\Psi_{\mathcal{R}}}
\newcommand{\Phirestil}{\tilde{\Phi}_{\mathcal{R}}}
\newcommand{\PhiRtil}{\tilde{\Phi}_{R}}
\newcommand{\Phipunc}{\Phi_{\mathcal{P}}}
\newcommand{\Phipuncloc}{\Phi_{\hat{\mathcal{P}}}}
\newcommand{\Psirestil}{\tilde{\Psi}_{\mathcal{R}}}
\newcommand{\Psirettil}{\tilde{\Psi}_{\text{ret}}}
\newcommand{\Phiret}{\Phi_{\text{ret}}}
\newcommand{\Psiret}{\Psi_{\text{ret}}}
\newcommand{\Phiadv}{\Phi_{\text{adv}}}
\newcommand{\Fself}{F^{\text{self}}}
\newcommand{\Fselfdn}{F_{\text{self}}}
\newcommand{\Phicons}{\Phi^{\text{cons}}}
\newcommand{\Phidiss}{\Phi^{\text{diss}}}
\newcommand{\Fcons}{F^{\text{cons}}}
\newcommand{\Fdiss}{F^{\text{diss}}}
\newcommand{\Ftil}{F}
\newcommand{\Seff}{S_{\text{eff}}}
\newcommand{\Zeff}{Z_{\text{eff}}^m}
\newcommand{\p}{\rho}
\newcommand{\dr}{\delta r}
\newcommand{\dth}{\delta \theta}
\newcommand{\dph}{\delta \varphi}
\newcommand{\Prr}{P_{rr}}
\newcommand{\Pth}{P_{\theta \theta}}
\newcommand{\Pph}{P_{\varphi \varphi}}
\newcommand{\Qrr}{Q_{rr}}
\newcommand{\Qth}{Q_{\theta \theta}}
\newcommand{\Qph}{Q_{\varphi \varphi}}

\newcommand{\DelPhi}{\Delta \Phirestil}
\newcommand{\DelF}{\Delta F_r}

\newcommand{\Rrr}{U_{rr}}
\newcommand{\Rqq}{U_{\theta\theta}}
\newcommand{\Rpp}{U_{\varphi\varphi}}
\newcommand{\Rrq}{U_{r\theta}}
\newcommand{\Rrp}{U_{r\varphi}}
\newcommand{\Rqp}{U_{\theta \varphi}}

\newcommand{\Trr}{V_{rr}}
\newcommand{\Tqq}{V_{\theta\theta}}
\newcommand{\Tpp}{V_{\varphi\varphi}}
\newcommand{\Trq}{V_{r\theta}}
\newcommand{\Trp}{V_{r\varphi}}
\newcommand{\Tqp}{V_{\theta \varphi}}

\newcommand{\Xrr}{X_{rr}}
\newcommand{\Xqq}{X_{\theta\theta}}
\newcommand{\Xpp}{X_{\varphi\varphi}}
\newcommand{\Xrq}{X_{r\theta}}
\newcommand{\Xrp}{X_{r\varphi}}
\newcommand{\Xqp}{X_{\theta \varphi}}

\newcommand{\Yrr}{Y_{rr}}
\newcommand{\Yqq}{Y_{\theta\theta}}
\newcommand{\Ypp}{Y_{\varphi\varphi}}
\newcommand{\Yrq}{Y_{r\theta}}
\newcommand{\Yrp}{Y_{r\varphi}}
\newcommand{\Yqp}{Y_{\theta \varphi}}

\newcommand{\Psim}{\Psi^m}
\newcommand{\tmax}{t_\text{max}}

\newcommand{\intl}{\{\text{num.}\}}
\newcommand{\Psinum}{\Psi^\m_{\mathcal{R}}}

\newcommand{\mmin}{m_{\text{min}}}
\newcommand{\mmax}{m_{\text{max}}}

\newcommand{\m}{m}
\newcommand{\twr}{\Gamma_r}
\newcommand{\twrs}{\Gamma_{r*}}
\newcommand{\twth}{\Gamma_{\theta}}
\mathchardef\mhyphen="2D

\providecommand{\e}[1]{\ensuremath{\times 10^{-#1}}}

\newcommand{\ar}{\alpha_{\text{res}}}
\newcommand{\nr}{n_{\text{res}}}
\newcommand{\ratio}{\chi}
\newcommand{\ratiolog}{\chi_{\text{log}}}
\newcommand{\etal}{\emph{et al.}}

\newcommand{\ellK}{\text{ellipK}}
\newcommand{\ellE}{\text{ellipE}}

\begin{document}

\preprint{}

 \title{Self force via m-mode regularization and 2+1D evolution: \\ 
 Foundations and a scalar-field implementation on Schwarzschild}

\author{Sam R. Dolan}
 \email{s.dolan@soton.ac.uk}
 \affiliation{%
 School of Mathematics, University of Southampton, Southampton SO17 1BJ, United Kingdom. \\
}%
\author{Leor Barack}
 \email{l.barack@soton.ac.uk}
 \affiliation{%
 School of Mathematics, University of Southampton, Southampton SO17 1BJ, United Kingdom. \\
}%

\date{\today} 

\begin{abstract}

To model the radiative evolution of extreme mass-ratio binary inspirals (a key target of the LISA mission), the community needs efficient methods for computation of the gravitational self-force (SF) on the Kerr spacetime. Here we further develop a practical `$m$-mode regularization' scheme for SF calculations, and give details of a first implementation. The key steps in the method are (i) removal of a singular part of the perturbation field with a suitable `puncture' to leave a sufficiently regular residual within a finite worldtube surrounding the particle's worldline, (ii) decomposition in azimuthal ($m$-)modes, (iii) numerical evolution of the $m$-modes in 2+1D with a finite difference scheme, and (iv) reconstruction of the SF from the mode sum. The method relies on a judicious choice of puncture, based on the Detweiler--Whiting decomposition. We give a working definition for the `order' of the puncture, and show how it determines the convergence rate of the $m$-mode sum. The dissipative piece of the SF displays an exponentially convergent mode sum, while the $m$-mode sum for the conservative piece converges with a power law. In the latter case the individual modal contributions fall off at large $m$ as $m^{-n}$ for even $n$ and as $m^{-n+1}$ for odd $n$, where $n$ is the puncture order. We describe an $m$-mode implementation with a 4th-order puncture to compute the scalar-field SF along circular geodesics on Schwarzschild. In a forthcoming companion paper we extend the calculation to the Kerr spacetime.

\end{abstract}

\pacs{}
\maketitle

%
%
%
%

\section{Introduction}

The closely-related notions of ``self force'' \cite{Poisson:2004, Barack:2009}, ``radiation reaction'' \cite{Dirac:1938}, and ``radiation damping'' \cite{Havas:1957} have a long and interesting history in physics. A motivating example arises in classical electromagnetism when one considers a point particle undergoing acceleration. An accelerated charge produces electromagnetic radiation; hence the particle loses energy and must therefore experience a braking force. The braking force may be interpreted as arising from the interaction of the particle with its own radiative field. This interpretation is not straightforward mathematically, since the electromagnetic field is formally infinite at the particle. Dirac \cite{Dirac:1938} showed how to remove a divergent (time-symmetric) component of the field, to isolate the finite (non-symmetric) part responsible for radiation reaction.
 
The self-force idea now finds a modern application in the study of Extreme Mass Ratio Inspirals (EMRIs), which are a key target of the LISA mission \cite{LISA, Barack:Cutler:2004, Amaro-Seone:etal:2007}. An EMRI is a special case of the gravitational two-body problem, in which a compact body (e.g., a stellar-mass black hole or neutron star) of mass $\mass$ is gravitationally bound to a massive black hole of mass $M$, such that the mass ratio $\mass / M$ is very small. If $\mass / M$ is vanishingly small, the smaller body follows a geodesic in the spacetime of the larger body \cite{Gralla:Wald:2008}. For a small but non-zero $\mass$, the smaller body perturbs the spacetime geometry at $\mathcal{O}(\mass / M)$, and the resulting back-reaction force is referred to as the first-order ``gravitational self-force''. 

It was appreciated long ago \cite{Havas:1957, DeWitt:Brehme:1960, Hobbs:1968} in the electromagnetic context that self-force calculations on curved spacetimes are more challenging than in flat space. In principle, on a curved spacetime, the self force (henceforth SF) depends on the entire past history of the motion. Hence the challenge is not merely to derive formal expressions for the SF, but also to implement practical schemes for their evaluation (see  \cite{Poisson:2004, Barack:2009} for reviews). 

Following the foundational work of Dirac \cite{Dirac:1938} and DeWitt and Brehme \cite{DeWitt:Brehme:1960} on the electromagnetic SF, a key step forward came in 1997 with the derivation of equations for the (first-order) gravitational SF \cite{Mino:Sasaki:Tanaka:1997, Quinn:Wald:1997}, now known as the ``MiSaTaQuWa'' equations. Alternative derivations and extensions have appeared over the subsequent years, for example in the works of Detweiler and Whiting \cite{Detweiler:Whiting:2003}, Gralla and Wald \cite{Gralla:Wald:2008}, Harte \cite{Harte:2008,Harte:2010}, Gralla \etal~\cite{Gralla:Harte:Wald:2009}, and Pound \cite{Pound:2010}.  Equations for the scalar-field SF \cite{Quinn:2000, Poisson:2004} and the electromagnetic SF \cite{DeWitt:Brehme:1960, Hobbs:1968, Gralla:Harte:Wald:2009, Harte:2009} have also been obtained.

It is a non-trivial task to compute the gravitational SF from the MiSaTaQuWa equations, and first they must be cast into a form amenable to practical computation. One standard method is the so-called ``$l$-mode regularization scheme'', outlined in \cite{Barack:Ori:2000,Barack:Mino:2002}. This scheme has been applied in a range of studies of SF on the Schwarzschild spacetime, for example, the scalar SF for radial infall \cite{Barack:Burko:2000}, circular orbits \cite{Burko:2000, Diaz-Rivera:2004, Haas:Poisson:2006, Canizares:Sopuerta:2009} and eccentric orbits \cite{Haas:2007, Canizares:Sopuerta:Jaramillo:2010}; electromagnetic SF for eccentric orbits \cite{Haas:2008}; and the gravitational SF for radial infall \cite{Barack:Lousto:2002}, circular \cite{Barack:Sago:2007, Detweiler:2008, Sago:Barack:Detweiler:2008, Sago:2009, Keidl:Shah:Friedman:2010} and eccentric orbits \cite{Barack:Sago:2010}. Other approaches under development include \cite{Lousto:Nakano:2008, Casals:Dolan:Ottewill:Wardell:2009, Field:Hesthaven:Lau:2009, Vega:Detweiler:2008, Vega:Diener:Tichy:Detweiler:2009, Hopper:Evans:2010}.
 
There are promising signs that the SF programme is approaching maturity. For instance, gravitational SF results are now being compared against results of post-Newtonian theory \cite{Blanchet:Detweiler:LeTiec:Whiting:2010a, Blanchet:Detweiler:LeTiec:Whiting:2010b, Damour:2009, Barack:Damour:Sago:2010, Favata:2010}, and used to calibrate functions in the Effective One-Body (EOB) theory in the strong-field regime \cite{Damour:2009, Barack:Damour:Sago:2010}. The shift in the innermost stable circular orbit (ISCO) due to the conservative part of the gravitational SF was recently computed \cite{Barack:Sago:2009}. SF results are also being used to inform data analysis strategies for the mock LISA data challenge \cite{Huerta:Gair:2009}; an improved understanding of strong-field SF-related phenomena such as resonances \cite{Hinderer:Flanagan:2008, Flanagan:Hinderer:2010} will be undoubtedly prove central to this effort. There now arises the possibility that SF results will soon be meaningfully compared against Numerical Relativity simulations, which are pushing into the intermediate mass-ratio regime \cite{Gonzalez:Sperhake:Brugmann:2008, Lousto:Nakano:Zlochower:Campanelli:2010a, Lousto:Nakano:Zlochower:Campanelli:2010b, Lousto:Zlochower:2010}. 

Whilst most studies thus far have assumed the central black hole to be non-rotating (Schwarzschild-type), it is reasonable to expect that in astrophysically-relevant EMRIs it will be rotating (Kerr-type). At the time of writing, few calculations have been attempted for the more technically-demanding Kerr case. An exception is the recent work in Ref.~\cite{Warburton:Barack:2010}, in which the $l$-mode regularization scheme is applied in the frequency domain to compute the scalar SF for equatorial circular orbits on Kerr. Unfortunately, the lack of separability of the gravitational field equations means the $l$-mode scheme cannot be applied in a straightforward way to gravitational SF calculations on Kerr. This motivates the development of alternative methods. 

SF calculations may be performed in either the frequency or time domains. In the frequency domain, the SF is reconstructed from a sum over frequency modes, with a spectrum of frequencies which are integer multiples of the fundamental orbital frequencies. The frequency-domain approach works well for circular orbits and low-to-moderately eccentric orbits ($e \lesssim 0.7$ \cite{Warburton:Barack:2011}). It can give highly accurate results, because a complete decomposition (into frequency and angular modes) leaves one with an ordinary differential equation for the radial functions, which may be solved to high precision. 

The frequency-domain approach has limitations, however. It is not suitable if the field equations cannot be separated (as in the important case of metric perturbations on Kerr in Lorenz gauge), or if the orbit has high eccentricity \cite{Barton:Lazar:Kennefick:2008}, is highly generic (on Kerr), or is unbound. Importantly, frequency domain methods seem much less well suited to the challenge of evolving an orbit perturbed by a SF in a self-consistent manner. This has motivated the development of a range of time-domain approaches.

The `$m$-mode regularization method' introduced in \cite{Barack:Golbourn:2007, Barack:Golbourn:Sago:2007} provides a general framework for time-domain SF calculations in axisymmetric spacetimes (such as Kerr), which may be applied in the scalar, electromagnetic and gravitational cases. Let us briefly examine the similarities and differences between the $l$- and $m$-mode approaches, which are both based on a decomposition in angular modes. Whereas the $l$-mode decomposition in spherical harmonics results in $1+1$D modal equations, the $m$-mode decomposition in azimuthal modes leads to $2+1$D equations. Here lies a key difference: field modes in $2+1$D are divergent on the worldline, whereas $1+1$D modes are continuous (albeit not differentiable) there. This obviously poses a challenge to numerical schemes, and motivates `regularization' of the $m$ modes with an analytically-determined `puncture field' which is used to remove the divergence. In Ref.~\cite{Barack:Golbourn:2007} it was shown that, with a simple (`1st-order'; see below) puncture field, the $2+1$D field modes could be evolved numerically. The original idea behind \cite{Barack:Golbourn:2007} was to use the $m$-modes to obtain (via integration over $\theta$) the $l$-modes which are needed as input to the standard $l$-mode regularization scheme. However, it was later shown in \cite{Barack:Golbourn:Sago:2007} that, with an improved (`2nd-order') puncture field, the SF could be recovered directly from a sum of the gradients of the $m$ modes themselves. The current work describes the first implementation of this idea: we use the $m$-mode regularization scheme to compute the scalar-field SF in Schwarzschild. In a companion paper we will describe a scalar-field implementation on the Kerr spacetime.

The aim of this work is to lay the necessary foundations for, and to demonstrate the computational feasibility of, accurate $m$-mode $2+1$D time-domain SF calculations using high-order punctures. The punctures employed in \cite{Barack:Golbourn:2007, Barack:Golbourn:Sago:2007} are motivated by the Detweiler-Whiting split \cite{Detweiler:Whiting:2003} of the retarded field into `radiative' (R) and `singular' (S) parts. We give a method for constructing puncture fields from finite-order local expansions of the S field (see also Ref.~\cite{Wardell:2010}), and demonstrate that the order of the expansion directly affects the convergence rate of the $m$-mode reconstruction of the SF. We carefully describe the features of the $m$-mode scheme that will be common to all future implementations, such as the puncture formulation, the puncture order, the $m$-mode convergence rate for dissipative and conservative parts, the worldtube formulation, initial and boundary conditions, and the modelling and mitigation of various sources of numerical error. In this work the Schwarzschild spacetime serves as a testing ground in which to explore the features that do not depend on the precise form of the field equations. In the companion paper, we describe an implementation that explores the issues specific to Kerr, such as stability of finite-differencing schemes. 


Other time-domain approaches are under active development. Vega \etal~\cite{Vega:Diener:Tichy:Detweiler:2009} have outlined a framework for time-domain SF calculations in 3+1D. They have demonstrated that, with a little modification, codes originally written for applications in Numerical Relativity can be used for SF calculations. Furthermore, they have computed the scalar SF for circular orbits on Schwarzschild to within $\sim 1\%$ accuracy. Pushing the accuracy towards the one-part-per-million benchmark achieved by Thornburg \cite{Thornburg:2009} in a 1+1D time-domain code with adaptive mesh refinement represents a considerable challenge, and an ongoing community effort is underway. We believe the $m$-mode approach represents a competitive alternative to the 3+1D scheme, since it exploits the axisymmetry to achieve substantial gains in computational efficiency. Of course, the price to pay for decomposition is that the SF must then be reconstructed from a sum over modes; but, as we aim to show here, the convergence of the mode sum is now well understood. 



The paper is arranged as follows. In Sec.~\ref{sec:exposition} we outline the theoretical basis of our approach. Here we cover the Detweiler-Whiting split into `radiative' and `singular' fields (\ref{subsec:DW}) which motivates the `puncture' scheme (\ref{subsec:puncture}), the definition for `puncture order' (\ref{subsec:puncture-order}) and its effect on mode sum convergence properties (\ref{subsec:mmode}), and the worldtube formulation (\ref{subsec:worldtube}). In Sec.~\ref{sec:implementation} we give details of the first 2+1D implementation for a SF calculation, for circular orbits in Schwarzschild. We describe the calculation of the puncture function and the effective source at orders $2$, $3$, and $4$ (\ref{subsec:puncture-scheme}), the code architecture, the finite difference scheme and numerical stability (\ref{subsec:simulation-details}), and the method for reconstructing the SF from data extracted from multiple 2+1D `runs' (\ref{subsec:data-extraction}--\ref{subsec:error-sources}). In Sec.~\ref{sec:results} we present a sample of numerical results. After exploring the various sources of error and strategies for error mitigation (\ref{subsec:results-modes}--\ref{subsec:modesums}), and numerically testing various predictions of  Sec.~\ref{sec:exposition}, we present in Sec.~\ref{subsec:modesumresults} the $m$-mode SF results, which we compare against the frequency domain results of Ref.~\cite{Diaz-Rivera:2004}. We conclude in Sec.~\ref{sec:conclusions} by reviewing progress and outlining themes for future work. The Appendices make explicit some of the more elaborate technical parts of our calculations. Throughout we adopt the metric signature $( -1, 1, 1, 1 )$ and set $G=c=1$.

\section{Theoretical Exposition\label{sec:exposition}}
Consider a test particle with scalar charge $q$ moving along a worldline $\gamma$ in the vacuum exterior of a black hole. Neglecting SF effects, we assume that the worldline $\gamma$ is a geodesic on the background, parameterised by $z(\tau) \equiv z^\mu(\tau)$ where $\tau$ is the proper time, with a tangent vector $u^\mu = dz^\mu / d\tau$. The charge acts as a source for a scalar field $\Phi(x)$, which satisfies the minimally-coupled Klein-Gordon equation,
\beq
\Box \Phi \equiv \nabla_\mu \nabla^\mu \Phi  
= S(x) ,  \label{wave-eq}
\eeq
where $\nabla_\mu$ denotes the covariant derivative with respect to the background metric $g_{\mu \nu}$. The source term is given by
\beq
S(x) \equiv - 4 \pi \rho(x) \equiv -4 \pi q \int_{-\infty}^{\infty} \left[-g(x)\right]^{-1/2} \, \delta^4\left( x  - z(\tau') \right) d \tau',   \label{Sdef}
\eeq
where $\rho$ is the charge density, $g$ is the metric determinant, and $\delta^4$ is the four-dimensional Dirac delta distribution. The retarded solution to this wave equation may be expressed as
\beq
\Phi_{\text{ret}}(x) \equiv \int \Gret (x, x') \rho(x') d^4 x'  =  q \int_{-\infty}^{\infty} \Gret (x,z(\tau')) \left(-g(z)\right)^{-1/2} d \tau'  ,  \label{Phi-ret}
\eeq
where $\Gret (x,x')$ is the retarded Green function defined by 
\beq
\Box_x \Gret (x,x')  = -4 \pi \delta^4 \left( x-x' \right),   \label{Geq}
\eeq
with appropriate retarded boundary conditions. 
The retarded Green function has the Hadamard form \cite{Hadamard:1923}
\beq
\Gret(x,x') = \Theta_-(x,x') \left[ U(x,x') \delta( \sigma ) + V(x, x') \Theta( - \sigma ) \right] ,  \label{Gret-hadamard}
\eeq
where $U$ and $V$ are regular symmetric biscalars, and $\sigma \equiv \sigma(x,x')$ is Synge's world function \cite{Synge:1960}, defined to be half the squared geodesic distance between spacetime points $x$ and $x'$, with $\sigma$ being negative (positive) when $x$ and $x'$ are connected by a timelike (spacelike) geodesic, and zero in the null case. Here $\Theta(- \sigma)$ is the Heaviside step function, and $\Theta_-(x,x')$ is unity if $x'$ is in the causal past of $x$ and zero otherwise. Note that here we have adopted the sign convention for $V(x)$ of \cite{Poisson:2004}, which is opposite to, e.g., \cite{Detweiler:Whiting:2003, Ottewill:Wardell}.

Note that the Hadamard form, Eq.~(\ref{Gret-hadamard}), is only valid when $x'$ may be connected to $x$ by a unique non-spacelike geodesic. More precisely, Eq.~(\ref{Gret-hadamard}) is valid if and only if $x$ and $x'$ lie within a \emph{convex normal neighbourhood} \cite{Friedlander:1975}. Particularly, in the presence of a black hole, this condition is not valid for points $x'$ on the worldline in the `distant past' of $x$ (see, e.g.,  \cite{Casals:Dolan:Ottewill:Wardell:2009} for a discussion) and so Eq.~(\ref{Gret-hadamard}) is of restricted utility.

The particle obeys the equation of motion 
\beq
u_\nu \nabla^\nu ( \mass u^\mu ) = \Fselfdn^\mu ,
\eeq
where $\Fselfdn^\mu$ is the scalar SF. The component of $\Fselfdn^\mu$ orthogonal to $u^\mu$ gives rise to a self-acceleration; the component tangential gives rise to a change of mass (in the case of circular motion it turns out there is no tangential component and thus no change of mass). From a naive application of a Lagrangian principle \cite{Wiseman:2000}, one would expect the scalar SF to be obtained as the gradient of the retarded scalar field, $q \nabla_\mu \Phi_{\text{ret}}$. However, this expression of course becomes meaningless when evaluated on the worldline, where $\Phiret$ and $\nabla_\mu \Phiret$ diverge. A more careful and considered analysis is needed. It has been shown \cite{Quinn:2000, Poisson:2004} that the scalar SF along a geodesic on the vacuum exterior of a Kerr black hole is given by the integral
\beq
\Fself_\mu(\tau) = q^2 \lim_{\epsilon \rightarrow 0^+} \int_{-\infty}^{\tau - \epsilon} \left. \nabla_\mu \Gret \left( x, z(\tau^\prime) \right) \right|_{x=z(\tau')} d \tau^\prime .  \label{Ftail}
\eeq
It is difficult to evaluate this expression, because it involves an integral over the entire past history of the particle's motion. Next we consider an alternative expression more amenable to practical computation. 

\subsection{Detweiler-Whiting R--S decomposition\label{subsec:DW}}
In an influential work on classical electromagnetism in flat spacetime, Dirac \cite{Dirac:1938} showed that the physical radiation reaction force could be obtained if one removes a certain singular and time-symmetric component  from the physical (retarded) vector potential, to leave behind a `radiative' field. Detweiler and Whiting \cite{Detweiler:Whiting:2003} elegantly extended Dirac's argument to curved spacetimes (for the scalar, electromagnetic and gravitational cases). In particular, they have shown that the scalar SF equation (\ref{Ftail}) is alternatively obtained by taking the derivative on the worldline of a certain radiative/regular (R) field,
\beq
\Fself_\mu(\tau) = q \lim_{x \rightarrow z(\tau)} \nabla_\mu \Phi_R(x),   \label{sf-defn}
\eeq 
where $\Phi_{R}(x)$ is an homogeneous solution of Eq.~(\ref{wave-eq}) (i.e. with $S=0$) given by
\beq
\Phi_R(x) = \Phi_{\text{ret}}(x) - \Phi_{S} (x) .
\eeq
Here the symmetric/singular (S) field is a particular solution to the sourced wave equation given by
\beq
\Phi_S(x) \equiv q \int_{\gamma} \GS(x, z(\tau)) d\tau, \label{Phi-S}
\eeq
where the symmetric Green function $\GS$ is defined through its Hadamard form
\beq
\GS(x,x') = \frac{1}{2} \left[ U(x,x') \delta(\sigma) - V(x,x') \Theta( \sigma)  \right].  \label{eq:Hadamard}
\eeq
Here $U$ and $V$ are the same biscalars that feature in Eq.~(\ref{Gret-hadamard}). By construction, $\GS$ is zero inside the future and past light-cones.  
Inserting Eq.~(\ref{eq:Hadamard}) into (\ref{Phi-S}) leads to
\beq
q^{-1} \Phi_S(x) = \left[ \frac{U(x, z(\tau))}{2 \dot{\sigma}} \right]_{\tauret(x)} + \left[ \frac{U(x, z(\tau))}{2 \dot{\sigma}} \right]_{\tauadv(x)} - \frac{1}{2} \int_{\tauret}^{\tauadv} V(x, z(\tau)) d \tau ,  \label{S-hadamard}
\eeq
where $z(\tauret)$ and $z(\tauadv)$ are the points on the worldline in the causal past and future of $x$ that are connected to $x$ by a null geodesic, and $\tauret$ and $\tauadv$ are the corresponding proper times along the worldline. Note that $\tauret(x)$ and $\tauadv(x)$ are non-smooth functions of the field point $x$ that are not differentiable on the worldline, and that strictly speaking (\ref{S-hadamard}) is only well-defined if $x$ and $z$ lie inside a convex normal neighbourhood.


\subsection{Dissipative and conservative parts of SF\label{subsec:disscons}}
The SF can be split into `dissipative' and `conservative' parts, as follows. First, let us introduce the `advanced' field $\Phiadv$, which is defined in an analogous way to $\Phiret$, i.e., via Eq.~(\ref{Phi-ret}) with a Green function $\Gadv (x,x')$ governed by Eq.~(\ref{Geq}) with appropriate advanced boundary conditions. Next we may define retarded and advanced radiative (R) fields via $\Phi_R^{\text{ret}} = \Phiret - \Phi_S$ and $\Phi_R^{\text{adv}} = \Phiadv - \Phi_S$. Note that the same singular (S) field is used in both cases, since it represents the `symmetric' part of the field. Then we may define the `conservative' and `dissipative' combinations
\begin{eqnarray}
\Phicons &\equiv&  \frac{1}{2}  \left( \Phi_R^{\text{ret}} + \Phi_R^{\text{adv}} \right)  = \frac{1}{2} \left( \Phi_{\text{ret}} +\Phi_{\text{adv}}  - 2 \Phi_S  \right) , \label{Phi-cons} \\
\Phidiss &\equiv&  \frac{1}{2}  \left( \Phi_R^{\text{ret}} - \Phi_R^{\text{adv}} \right)  =  \frac{1}{2} \left( \Phi_{\text{ret}} - \Phi_{\text{adv}}  \right) ,  \label{Phi-diss}
\end{eqnarray}
and the corresponding `conservative' and `dissipative' parts of the scalar SF follow from 
\beq
F^{\text{cons/diss}}_\mu(\tau) = q \lim_{x \rightarrow z(\tau)} \nabla_\mu \Phi^{\text{cons/diss}} .
\eeq
A key point here is that the dissipative component is found from the field difference $\Phi_{\text{ret}} - \Phi_{\text{adv}}$, which is known to be a smooth function \cite{Detweiler:Whiting:2003} even on the worldline. In other words, provided we can obtain the advanced field the dissipative component of the SF does not require regularization. On the other hand, to compute the conservative component one requires knowledge of the $S$ field, which is singular on the worldline. 

\subsection{Puncture scheme\label{subsec:puncture}}
Computing $\Phi_S$ is not straightforward. Unfortunately, a closed form expression for the biscalars $U$ and $V$ is not available, and generally one falls back on approximation methods. Fortunately, methods exist to expand the biscalars $U$ and $V$ as covariant series in $\sigma$, and ultimately as series in coordinate separations. 
The Hadamard-expansion method is now well advanced for several spacetimes of physical relevance, such as Schwarzschild and Kerr \cite{Anderson:Hu, Anderson:Flanagan:Ottewill:2005, Ottewill:Wardell, Wardell:thesis}.

Knowledge of the Hadamard expansion of $\Phi_S$ allows the introduction of a `puncture' ($\mathcal{P}$) field $\Phipunc$. To be of use, $\Phipunc$ must be defined `globally' (or at least everywhere within a region surrounding the worldline), and in the vicinity of the worldline it must have the same local behaviour as the S field of Detweiler and Whiting [Eq.~(\ref{S-hadamard})], up to a certain order (to be made precise below). We then define a `residual' ($\mathcal{R}$) field to be
\beq
\Phires(x) = \Phiret(x) - \Phipunc(x).  \label{Phires-def}
\eeq
The residual field $\Phires$ will have the same local expansion as $\Phi_R$, up to a certain order, and (if the order is sufficient) the scalar SF may be obtained from evaluating the derivatives of $\Phires$ on the worldline (see Sec.~\ref{subsec:sf-from-res}). This basic idea is also applicable in the electromagnetic and gravitational cases \cite{Barack:Golbourn:Sago:2007}; in this work we focus on the scalar-field case.

\subsection{Order of the puncture function\label{subsec:puncture-order}}
Let us now give a working definition for the `order' of the puncture function, with reference to local coordinate expansions. Our classification of order should in fact be independent of the choice of coordinates, provided one works with a sufficiently regular coordinate system. A similar definition of order from a covariant point of view is given in \cite{Wardell:2010}.

We begin by making a rather subtle distinction between a field $\Phipuncloc(x;z)$ which is defined locally in the vicinity of a particular point $z$ on the worldline, and a puncture function $\Phipunc(x;\gamma)$ which must be defined everywhere in the vicinity of the worldline $\gamma$. Likewise, we should distinguish between the global residual field, defined by Eq.~(\ref{Phires-def}), and a local residual field $\Phiresloc(x; z) = \Phiret(x) - \Phipuncloc(x; z)$ which is again defined with reference to a particular point $z$ on the worldline. In Sec.~\ref{subsec:globaldef} we give a practical scheme for promoting a `local' expansion $\Phipuncloc$ to a `global' puncture field $\Phipunc$ in the simple case of circular orbits.

Denote a field point by $x^\mu$, denote a worldline point by $z^\mu$, and define the coordinate differences $\delta x^{\mu} = x^\mu - z^\mu$. 
It is convenient when discussing the order of the puncture to introduce a scaling parameter $\lambda$ and new coordinates $\delta \bar{x}^\mu$, through
\beq
\delta x^\mu = \lambda \, \delta \bar{x}^\mu .
\eeq
Taking the limit $\lambda \rightarrow 0$ with fixed $\delta \bar{x}^\mu$ is equivalent to approaching the point $z^\mu$ from a specific direction.

A lowest-order puncture was given in \cite{Barack:Golbourn:2007}:
\beq
\Phipuncloc^{[1]}(\lambda \delta \bar{x}) = q /  \epsilon_{[1]} , \quad \quad \text{where} \quad \epsilon_{[1]} = \left| \lambda \right| \mathcal{S}_0^{1/2} ,  \label{PhiP1}
\eeq
and
\beq
\mathcal{S}_0 \equiv \left. \left( g_{\mu \nu} + u_\mu u_\nu \right) \right|_z \delta \bar{x}^\mu \delta \bar{x}^\nu .
\eeq
Subtracting (\ref{PhiP1}) from the retarded field leaves a residual field $\Phiresloc^{[1]}$ which is $C^{-1}$ in the sense that it is bounded but discontinuous as $\delta x \rightarrow 0$, viz.
\beq
\lim_{\lambda \rightarrow 0^+} \Phiresloc^{[1]}( \lambda \delta \bar{x} ) \neq \lim_{\lambda \rightarrow 0^-} \Phiresloc^{[1]}( \lambda \delta \bar{x} )
\eeq
in general. 

The next-order puncture was obtained in \cite{Barack:Golbourn:Sago:2007}, and may be written
\beq
\Phipuncloc^{[2]} = q /  \epsilon_{[2]},  \quad \quad \text{where} \quad \epsilon_{[2]} = \left|\lambda\right| \left( \mathcal{S}_0 + \lambda S_1 \right)^{1/2},
\eeq
and
\beq
\mathcal{S}_1 \equiv \left. \left( u_\lambda u_\sigma {\Gamma^\lambda}_{\mu \nu} + g_{\mu \nu, \sigma} / 2 \right) \right|_z \delta \bar{x}^\mu \delta \bar{x}^\nu \delta \bar{x}^\sigma ,
\eeq
where ${\Gamma^\lambda}_{\mu \nu}$ are Christoffel symbols for the background metric. 
An alternative puncture of the same order may be defined as
\beq
q^{-1} \Phipuncloc^{[2,\text{alt}]} = \frac{1}{|\lambda|} \frac{1}{\mathcal{S}_0^{1/2}} - \frac{\lambda}{|\lambda|} \frac{\mathcal{S}_1}{2 \mathcal{S}_0^{3/2}}  ,
\eeq
so that $\Phipuncloc^{[2,\text{alt}]} - \Phipuncloc^{[2]} = \mathcal{O} ( |\lambda| )$. 
Subtracting either puncture from $\Phiret$ leaves a $C^{0}$ residual, i.e., a function $\Phiresloc^{[2]}$ which is continuous but not differentiable at $\delta x = 0$.

Starting with Eq.~(\ref{S-hadamard}), the Detweiler-Whiting S field may be expanded as
\beq
\Phi_S(\delta x) = \frac{1}{|\lambda| \mathcal{S}_0^{1/2}} \left( 1 + \lambda \frac{ \mathcal{F}_1(\delta \bar{x})}{ \mathcal{S}_0 } + \lambda^2 \frac{ \mathcal{F}_2(\delta \bar{x})}{ \mathcal{S}_0^2 } + \ldots   \right) ,   
\eeq
where $\mathcal{F}_k(\delta \bar x)$ are polynomials in $\delta \bar x$ of order $3k$. 
We will call $\Phipuncloc^{[n]}$ an ``$n$th order puncture'' if 
\beq
\Phipuncloc^{[n]} - \Phi_S = \mathcal{ O } \left(|\lambda| \lambda^{n-2} \right) \quad \text{and} \quad \lim_{\lambda \rightarrow 0} \left( \Phipuncloc^{[n]} - \Phi_S \right) / \left(|\lambda| \lambda^{n-2} \right) \neq 0  .   \label{nth:order}
\eeq
It follows that
\beq
\Phiresloc^{[n]} = \Phi_R + \mathcal{O} \left( |\lambda| \lambda^{n-2} \right) .
\eeq
Since $\Phi_R$ is a smooth function, the residual field $\Phiresloc^{[n]}$ is $C^{n-2}$. Hence, for example, a 2nd-order residual $\Phiresloc^{[2]}$ is continuous but not differentiable, and a 3rd-order residual $\Phiresloc^{[3]}$ is both continuous and differentiable.


\subsection{Global definition for the puncture function\label{subsec:globaldef}}
The covariant expansion method developed by Ottewill and Wardell \cite{Ottewill:Wardell, Wardell:thesis} enables one to compute expressions for $n$th order `local' punctures expressed in terms of coordinate differences, i.e.~$\Phi_P^{[n]}(\delta x^\mu)$. In this paper, we discuss 2nd, 3rd and 4th-order punctures; in principle, higher orders are possible (see Ref.~\cite{Wardell:2010} for a discussion).

Given a field point $x$, we are free to choose the worldline point $z$ to lie anywhere on the worldline between $z(\tauret(x))$ and $z(\tauadv(x))$. In order to obtain a puncture function which is globally-defined (or at least defined within the vicinity of the worldline for all $t$), we must allow $z$ to become a function of $x$.

There is more than one way to relate $x$ to $z$. For instance, we could choose $z$ to be the point on the worldline that is connected to $x$ by a spacelike geodesic orthogonal to the worldline at the point of intersection. Although a natural definition, this makes $z$ a rather complicated function of $x$. A simpler approach (and that used in e.g.~\cite{Barack:Golbourn:2007}) is to set the coordinate time of $z$ to be equal to the coordinate time of $x$, $z^0 = x^0 = t$. This is a coordinate-dependent construction; henceforth we will work in the Boyer-Lindquist system $\{t,r,\theta,\varphi\}$. Then 
\beq
\delta t = 0, \quad \delta r = r - r_p(t), \quad \delta \theta = \theta - \theta_p(t), \quad \delta \varphi = \varphi - \varphi_p(t),  \label{coord-differences}
\eeq
where $r_p(t), \theta_p(t), \varphi_p(t)$ are coordinate functions describing the worldline, and a globally-defined puncture function for use in our scheme is
\beq
\Phipunc^{[n]}(x^\mu) \equiv \Phipuncloc^{[n]}(0, r - r_p(t), \theta - \theta_p(t), \varphi - \varphi_p(t) )  . \label{global-punc-def}
\eeq

A puncture function of the form (\ref{global-punc-def}) will generally be ill-behaved at spatial infinity, and possibly elsewhere; we deal with this problem in Sec.~\ref{subsec:worldtube}.
There is still arbitrariness in the definition of the puncture function $\Phipunc^{[n]}$, since the only requirement is that the puncture field has the correct expansion [see Eq.~(\ref{nth:order})] in the vicinity of the worldline. 
For example, within the $m$-mode scheme we are motivated to replace $\delta \varphi$ with a smooth periodic function $f(\delta \varphi)$. This would not change the order of the puncture if $f(\varphi) = \delta \varphi + \mathcal{O}(\varphi^{n+1})$.

\subsection{Self force from the residual field\label{subsec:sf-from-res}}
Let us define the residual field through Eq.~(\ref{Phires-def}) with a `global' puncture field (\ref{global-punc-def}), and now consider its gradient near the worldline, i.e.~the quantity
\beq
\nabla_\mu \Phires^{[n]} =  \nabla_\mu \Phi_R + \mathcal{O}\left( |\lambda| \lambda^{n-3}  \right) .
\eeq
It is clear that this quantity is only guaranteed to be well-defined on the worldline if $n \ge 3$, i.e. if we use a 3rd-order puncture or higher. 
It is also clear that in this case, the gradient evaluated on the worldline leads to the correct SF via Eq.~(\ref{sf-defn}), i.e.
\beq
\Fself_\mu(\tau)  = q \lim_{x \rightarrow z(\tau)} \nabla_\mu \Phires^{[n\ge3]}(x) .
\eeq
However, this is not the complete story. If we employ a 2nd-order puncture, the gradient is discontinuous at the worldline, $\mathcal{O}(|\lambda| / \lambda)$. In other words, it depends on the direction in which the worldline is approached. Nevertheless, it was shown in \cite{Barack:Golbourn:Sago:2007} that the SF constructed from a sum over azimuthal $m$ modes \emph{is} in fact well-defined and correct. A related fact is that the convergence rate of the $m$-mode sum depends in a particular way on the order of the puncture function, as we now begin to discuss.

\subsection{$m$-mode decomposition and mode sum convergence\label{subsec:mmode}}

We may take advantage of the azimuthal symmetry of the background (i.e.~the Kerr spacetime) to decompose into $m$-modes, i.e.
\begin{eqnarray}
Q(t, r, \theta, \varphi) &=& \sum_{m = -\infty}^{\infty} Q^\m(t,r,\theta) e^{i m \varphi}  ,  \label{mmode-sum} \\
Q^\m(t,r,\theta) &=&  \frac{1}{2 \pi} \int_{-\pi}^{\pi} Q(t, r, \theta, \varphi) e^{- i m \varphi} d \varphi ,  \label{mmode-inversion} 
\end{eqnarray}
where $Q$ may be any member of the set $\{ \Phiret, \Phires^{[n]}, \Phipunc^{[n]}, \Seff^{[n]} \}$ [here $\Seff^{[n]}$ is the effective source to be defined in Eq.~(\ref{Seff}) below]. For later convenience, we define the `total $m$-mode contribution' $\tilde{Q}^\m(t,r,\theta,\varphi)$ (or `modal contribution' for brevity) to be the real quantity given by 
\beq
\tilde{Q}^\m(t,r,\theta,\varphi) \equiv \left\{ \begin{array}{ll}  Q^\m(t,r,\theta) e^{i m \varphi}  +  Q^{-m}(t,r,\theta) e^{-i m \varphi}, \quad \quad & m > 0 \\ Q^m(t,r,\theta), \quad & m = 0 \end{array} \right. .  \label{mmode-contribution-eq}
\eeq
The value of the radiative field $\Phi_R(z)$ at a point $z$ on the worldline may be found from a sum over the modal contributions of the residual field, 
\beq
\Phi_R(z) = \sum_{m \ge 0} \Phirestil^{[n\ge1]m}(z) .   \label{Phi-mmode}
\eeq
The SF is reconstructed from the gradient of the residual field modes evaluated at the worldline point $z$,
\beq
\Fself_\mu(z) = \sum_{m \ge 0} F_\mu^m(z), \quad \text{where} \quad F_\mu^m(z) \equiv q \lim_{x\rightarrow z}  \nabla_\mu \Phirestil^{[n>1]m} . \label{F-mmode}
\eeq



\subsubsection{Exponential convergence of the dissipative SF\label{subsec:exponential-convergence}}
In Sec.~\ref{subsec:disscons} we described the split of the SF into dissipative and conservative pieces. The dissipative piece $\Fdiss_\mu$ is found from the gradient of $\Phidiss$, where $\Phidiss$ is formed from the difference between retarded and advanced fields [see Eq.~(\ref{Phi-diss})]. The difference $\Phidiss$ is a smooth ($C^\infty$) function, hence its $m$-mode contributions decay faster than $m^{-k}$ (where $k$ is any positive integer) in the limit $m \rightarrow \infty$. We will call this behaviour `exponential convergence'. However, in order to construct the gradient of $\Phidiss$ in practice we need a method for calculating the gradient of the advanced field explicitly. In the case of eccentric geodesic orbits in the equatorial plane, it is straightforward to obtain $\nabla_\mu \Phiadv$ on the worldline by making use of the convenient symmetry relation [see Eq.~(2.80) in \cite{Hinderer:Flanagan:2008}]
\beq
\nabla_\mu \Phiadv (r, u_r) = - \epsilon_\mu \nabla_\mu \Phiret(r, -u_r) ,  \label{HF-symmetry}
\eeq
where $\epsilon_\mu = (1,-1,-1,1)$ and there is no summation over $\mu$. In other words, by identifying a point on the orbit conjugate to the point of interest (i.e.~a point at the same radius with equal and opposite radial velocity $u_r$), we may obtain the gradient of the advanced field directly from the gradient of the retarded field (see, e.g., Sec.~IIE in \cite{Barack:Sago:2010}).  

Let us consider the case of {\it circular orbits} in the equatorial plane in a little more detail. In this case, since $u_r = 0$, it follows via (\ref{HF-symmetry}) that $\nabla_\mu \Phiadv = \pm \nabla_\mu \Phiret$, where the gradients are evaluated at the same point on the worldline. Here the plus sign corresponds to the $r$ component, and the minus sign to the $t$ and $\varphi$ components. It follows immediately via (\ref{Phi-cons}) and (\ref{Phi-diss}) that the conservative part of the SF is purely radial, whereas the dissipative part has components only in the $t$ and $\varphi$ directions, i.e.~$\Fself_\mu = \left( \Fdiss_t, \Fcons_r, 0, \Fdiss_\varphi \right)$. Now, let us consider the (dissipative) $t$ component of the SF; by the above discussion it follows that 
\beq
F^m_t =  q \lim_{x\rightarrow z}  \nabla_t \Phirestil^{[n>1]m} = q \lim_{x\rightarrow z} \nabla_t \tilde{\Phi}^{\text{diss}, m} ,
\label{eqFt}
\eeq
where $\tilde{\Phi}^{\text{diss}, m}$ is the $m$-mode contribution to $\Phi^{\text{diss}}$. Since the field $\Phi^{\text{diss}}$ is a smooth function, it follows that the quantity on the right-hand side converges exponentially fast with $m$. Hence the SF modes $F_t^m$ must also convergence exponentially fast with $m$. A similar argument follows immediately for the (dissipative) $\varphi$ component of SF, which in fact is related to the $t$ component via
\beq
\Fself_t  + \omega \Fself_\varphi = 0 ,  \label{Ft-eq}
\eeq
where $\omega$ is the frequency of the circular orbit. Note that the argument for exponential convergence is independent of puncture order, and so the modal contributions $F_t^\m$ and $F_\varphi^\m$ are also independent of the puncture order (we check this in the numerical implementation of Sec.~\ref{sec:results}, cf.~Fig.~\ref{fig:modes-Fphi}).

\subsubsection{Power-law convergence of the conservative SF}
Let us now investigate the $m$-mode convergence properties of the conservative part of the SF. 
A careful analysis of convergence for the 2nd-order puncture scheme was given in \cite{Barack:Golbourn:Sago:2007}. In this section, we eschew a formal analysis in favour of a heuristic analysis which illustrates the key features. It remains for us to demonstrate that these features are supported by the results of our specific implementation, which we do in Sec.~\ref{sec:results}.   

Let $H^{[n]}(\varphi)$ be a smooth function on $-\pi\leq\varphi\leq\pi$, and across $\varphi = -\pi, \pi$ (i.e.~all its derivatives match there), except at $\varphi=0$, where it admits the local expansion
\begin{equation}
H^{[n]}=\frac{1}{|\varphi|}\sum_{k=n}^{\infty}\frac{h_k}{(k-1)!}\, \varphi^k .  \label{Fdef}
\end{equation}
Here $h_k$ are constant coefficients and $h_n,h_{n+1}\ne 0$. The function $H^{[n]}(\varphi)$ is akin to the $n$th-order residual field $\Phires^{[n]}$. Note $H^{[n]}$ is continuously differentiable $n-2$ times everywhere, but its $(n-1)$th derivative has a jump discontinuity at $\varphi=0$. $H^{[n]}$ has a mode-sum reconstruction of the form
\beq
H^{[n]} = \sum_{m \ge 0}^\infty \tilde{H}^{[n]\m}   .
\eeq
The $m$-mode contribution $\tilde{H}^{[n]\m}$ [defined as in Eq.~(\ref{mmode-contribution-eq})] is shown in Appendix \ref{appendix:toymodel} to have the following asymptotic behaviour in the limit of large $m$:
\begin{eqnarray}
\tilde{H}^{[n]\m}
&\sim&
\frac{2 h_n}{\pi m^n}\times 
\left\{
\begin{array}{ll}
(-)^{n/2} \cos m\varphi, & \text{$n$ even}, \\
(-)^{\frac{n-1}{2}}\sin m\varphi,  & \text{$n$ odd},
\end{array} \right.
\nonumber\\
&&+
\frac{2 h_{n+1}}{\pi m^{n+1}}\times 
\left\{
\begin{array}{ll}
(-)^{n/2} \sin m\varphi, & \text{$n$ even}, \\
(-)^{\frac{n-1}{2}}\cos m\varphi,  & \text{$n$ odd},
\end{array} \right.
+ \mathcal{O}\left(m^{-(n+2)}\right).   \label{toymodel-mmodes}
\end{eqnarray}

In the large-$m$ limit, the $m$-mode contributions (\ref{toymodel-mmodes}) decay as $\sim 1/m^n$, in general. However, at the irregular point $\varphi = 0$ (i.e.~on the worldline) they decay as $\sim 1/m^{n+1}$ if $n$ is odd. In other words, for odd $n$ the mode-sum reconstruction is ``one order more convergent" than would be naively expected. 

The toy model illustrates a well-known feature of Fourier theory: the smoother a function, the more rapid the convergence of its Fourier series. It also demonstrates a less obvious feature: odd-order punctures (i.e.~$n = 1$, $3$ etc.) will generate a residual field whose $m$-mode contributions decay one order faster than expected, i.e.~$\sim 1 / m^{n+1}$.

Now let us consider the gradient of the residual field (giving the SF), which is in general one order less differentiable than the field itself. Odd-order punctures give modal contributions to $\nabla_\mu \Phires^{[n]}$ which decay as expected, i.e.~as $\sim 1/m^{n-1}$. Even-order punctures (i.e. 2nd, 4th, etc.) give modal contributions for $\nabla_\mu \Phires^{[n]}$ which decay one order faster than naively expected, i.e.~as $\sim 1 / m^{n}$.  

The expected convergence behaviour of key quantities in our problem is summarised in Table \ref{table-modesum-convergence}. Let us note in passing that to obtain a mode sum for the conservative part of the SF whose terms fall off as $m^{-4}$, we must use a 4th-order puncture; the 3rd-order puncture gives only $m^{-2}$ modal fall off, just like the 2nd-order puncture. In Sec.~\ref{subsec:modesums} we give numerical evidence in support of this assertion (cf.~Fig.~\ref{fig:modes-Fr}). 

\begin{table}
\begin{tabular}{l | c c c c c c}
\hline
\hline 
Puncture Order & $\Phires^{[n]}$ \quad & smoothness of $\Phires^{[n]}$ \quad & $\quad \Phirestil^{[n]m} \quad$ & $\quad \nabla_r \Phirestil^{[n]m} \quad$ & $\quad \Seff \quad$ \\
\hline
$n=1$  &  $\left| \lambda \right| / \lambda$ & $C^{-1}$  & $m^{-2}$ & |  &  $\left| \lambda \right| / \lambda^3$ \\
$n=2$  &  $\left| \lambda \right|$ & $C^0$  & $m^{-2}$ & $m^{-2}$ & $\left| \lambda \right|  / \lambda^2$  \\
$n=3$  &  $\left| \lambda \right| \lambda$ & $C^1$  & $m^{-4}$ & $m^{-2}$ & $\left| \lambda \right|  / \lambda$ \\
$n=4$ &  $\left| \lambda \right| \lambda^2$ & $C^2$  & $m^{-4}$ & $m^{-4}$ & $\left| \lambda \right|$ \\
\hline
\hline 
\end{tabular}
\caption{Differentiability and large-$m$ behaviour of key quantities. As discussed in the text, the convergence rate of the $m$-mode series is related to the smoothness of $\Phires^{[n]}$ on the worldline, which depends on the order of puncture used. The parity of the puncture order is important; convergence of the modal contributions improves in jumps of $m^{-2}$ every second time the order ($n$) is increased by one. The 4th and 5th columns show the large-$m$ power-law behaviour of the individual $m$ modes, for the residual field and the conservative part of the SF, respectively. The modes of the dissipative part of SF converge exponentially fast (not shown; see text). The final column shows the smoothness of the effective source $\Seff$ (to be defined in Sec.~\ref{subsec:worldtube}) near the worldline.}
\label{table-modesum-convergence}
\end{table}


\subsection{Effective source, worldtube formulation and modal equations\label{subsec:worldtube}}
\begin{figure}
 \begin{center}
  \includegraphics[width=15cm]{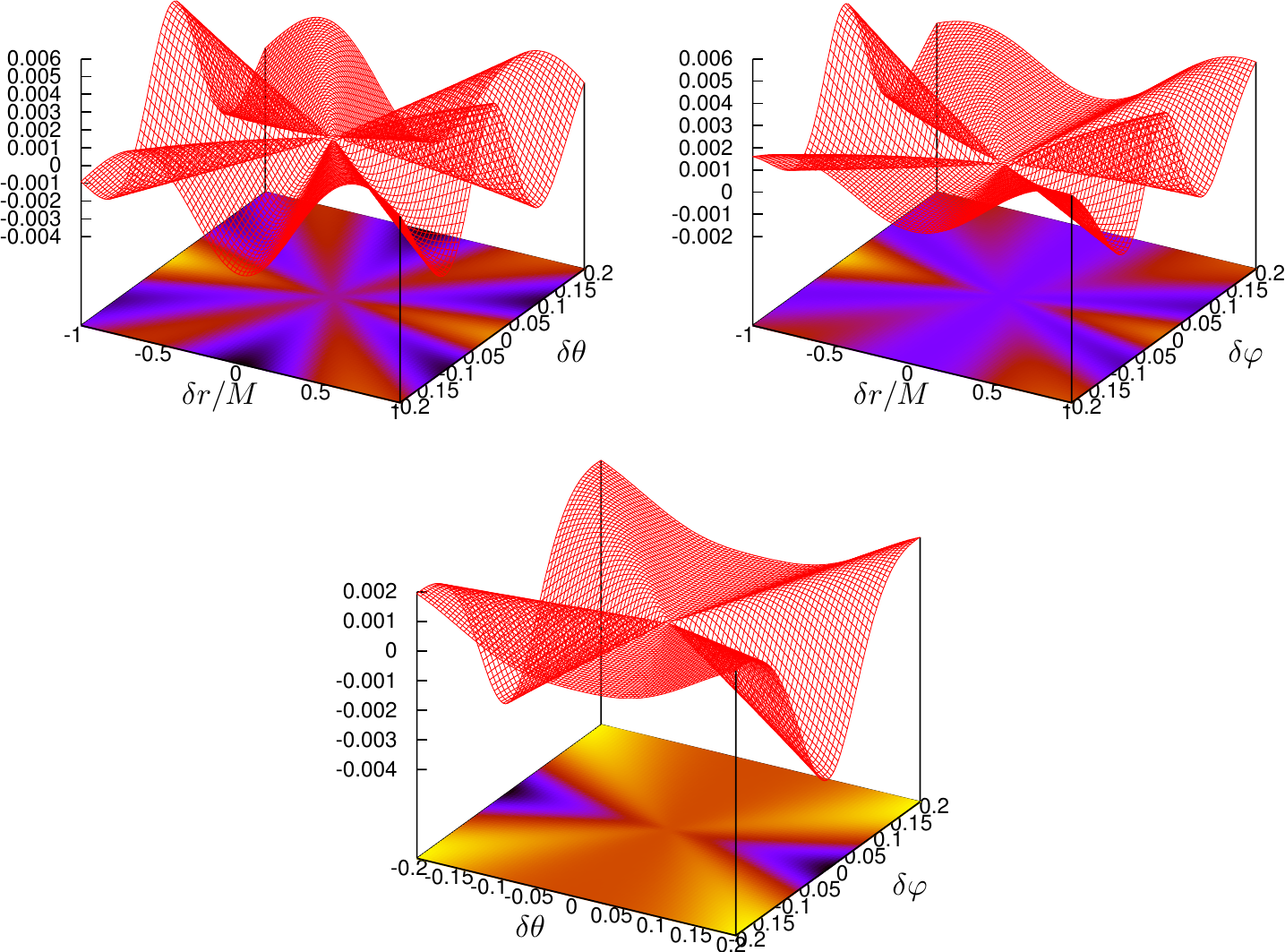}
 \end{center}
 \caption{Illustration of a typical effective source $\Seff^{[4]}$ derived via Eq.~(\ref{Seff}) from the 4th-order puncture field $\Phipunc^{[4]}$ [given in Eq.~(\ref{Phi-4thord})]. The plots show $M^3 q^{-1} \Seff^{[4]}$, for a particle on a circular orbit at $r_0=7M$ on Schwarzschild spacetime, as a function of the coordinate differences $\delta r$, $\delta \theta$, $\delta \varphi$, with the worldline being at $\dr = \dth = \dph = 0$. The source is shown on the spatial slices (clockwise from top left) $\delta \varphi = 0$, $\delta \theta = 0$ and $\delta r = 0$, with $t=\text{const}$. As expected, the 4th-order source is continuous but not differentiable at the worldline.}
 \label{fig:source3d}
\end{figure}

The residual field $\Phires^{[n]}$ is governed by an inhomogeneous wave equation,
\beq
\Box \Phires^{[n]} = \Seff^{[n]} , 
\eeq
where the effective source $\Seff^{[n]}$ is given by
\beq
\Seff^{[n]} (x) \equiv  S - \Box \Phipunc^{[n]} .  \label{Seff}
\eeq
The behaviour of $\Seff^{[n]}$ near $\gamma$ depends on the order ($n$) of the puncture field. The puncture field is such that the distributional component of the original source (i.e. the delta function) is eliminated. In the vicinity of the worldline, $\Seff^{[n]}$ (expressed in terms of coordinate differences $\delta x^\mu = \lambda \delta \bar{x}^\mu$) has a local expansion starting at $\mathcal{O}(|\lambda| \lambda^{n-4})$. In other words, $\Seff^{[n]}$ is divergent for $n = 1, 2$, discontinuous for $n=3$ and continuous ($C^0$) for $n=4$. An illustration of a typical 4th-order effective source $\Seff^{[4]}$ close to the worldline is shown in Fig.~\ref{fig:source3d}, for a particle on a circular geodesic orbit. 

The effective source $\Seff^{[n]}$ is generally divergent far from the worldline. To mitigate this unwanted behaviour, Vega \etal~\cite{Vega:Diener:Tichy:Detweiler:2009} multiply the 3+1D puncture field by a smooth windowing function, which has the effect of attenuating the effective source far from the worldline. Lousto and Nakano \cite{Lousto:Nakano:2008} have designed a puncture field with an effective source which is well-behaved at infinity, but is rather complicated to compute. In this work, we prefer to make use of a sharply-defined `worldtube', which we introduce in the 2+1D domain after decomposition in azimuthal modes.

The key idea was described in \cite{Barack:Golbourn:2007}. A worldtube $\mathcal{T}$ with boundary $\partial \mathcal{T}$ is constructed in the 2+1D domain, to enclose the worldline (note that a 2+1D tube may also be interpreted as a 3+1D which spans the full range of azimuthal angles). Outside the worldtube, we evolve the modes of the retarded field $\Phiret^m$, governed by the homogeneous wave equation. Inside the worldtube, we evolve the modes of the residual field $\Phires^m$, governed by the inhomogeneous wave equation sourced by the effective source modes $\Seff^{[n]m}$ [defined as in Eq.~(\ref{mmode-inversion})]. Across the boundary of the worldtube, one may convert between $\Phiret^m$ and $\Phires^{[n]m}$ using the $m$ modes of the puncture field, i.e.~$\Phiret^m = \Phires^{[n]m} + \Phipunc^{[n]m}$. To summarise,
\beq
\left\{
\begin{array}{l l}
\Box^{m} \Phires^m = S^m - \Box^m \Phipunc^m \equiv \Seff^m , & \text{ inside } \mathcal{T} ,\\
\Box^{m} \Phiret^m = 0 , &  \text{ outside } \mathcal{T}, \\
\Phires^m = \Phiret^m - \Phipunc^m , & \text{ across } \partial \mathcal{T} . \\
\end{array}
\right. 
\label{eqs-worldtube}
\eeq
Here $\Box^m$ is the d'Alembertian in the $2+1$D domain, obtained by making the replacement $\partial^k / \partial \phi^k \rightarrow (-i m)^k$ in the $3+1$D d'Alembertian $\Box$.

\begin{figure}
 \begin{center}
  \includegraphics[width=5cm]{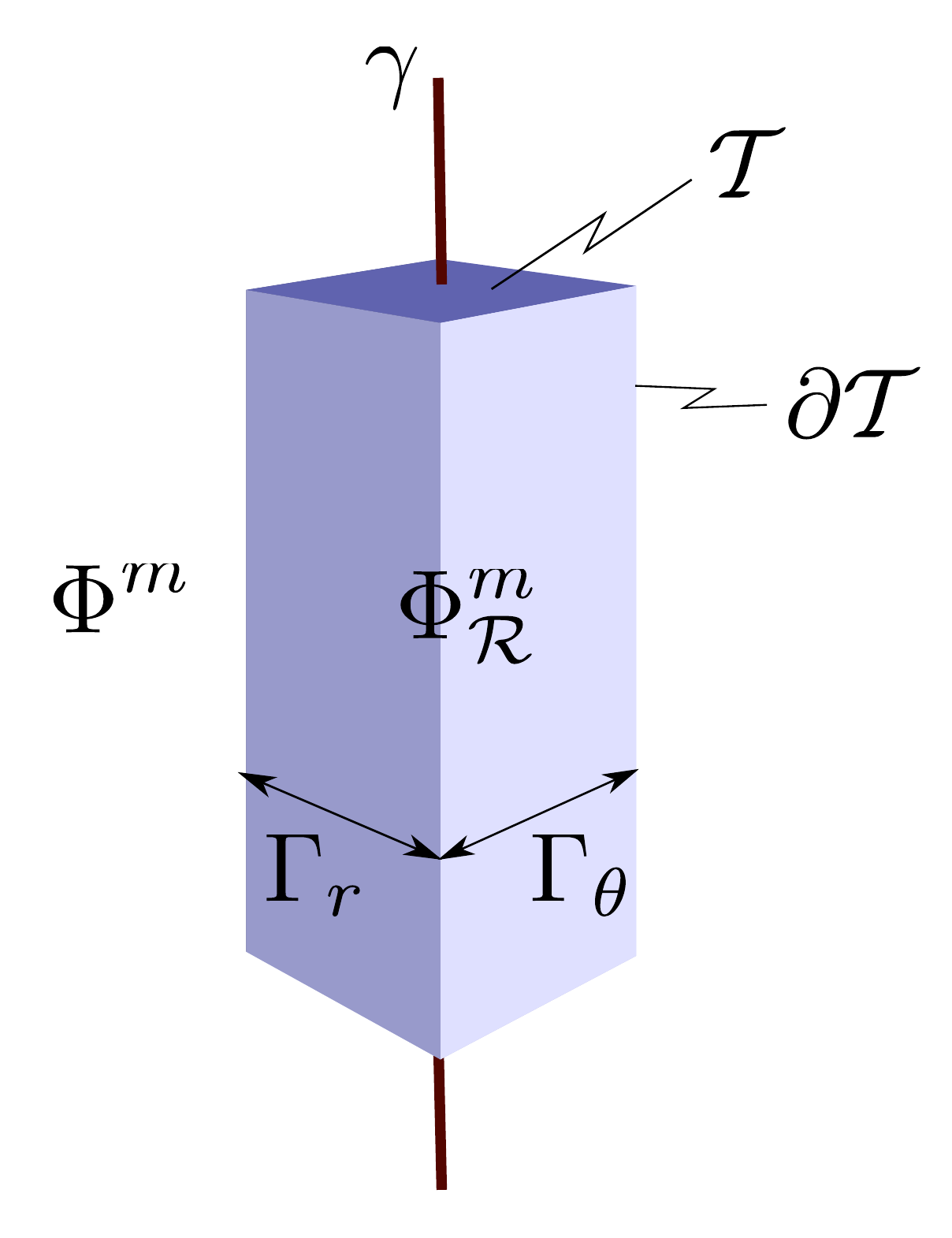}
 \end{center}
  \caption{
Visualization of the worldtube in 2+1D domain for a particle on a circular orbit. Here the $t$ axis runs vertically, and the worldtube $\mathcal{T}$ is shown as a boxed domain of fixed width $\{\Gamma_r, \Gamma_\theta\}$, centred on the worldline $\gamma$ at fixed $r=r_0$, $\theta = \pi/2$. Inside the tube $\mathcal{T}$ we evolve $\Phires^\m$, outside the tube we evolve $\Phiret^\m$, and across the tube boundary $\partial \mathcal{T}$ we convert between the two using $\Phiret^\m = \Phires^\m + \Phipunc^\m$.
}
  \label{fig:worldtube}
 \end{figure}

For circular orbits, it is simplest to construct a worldtube with fixed coordinate widths $\twr$ and $\twth$ in the  $r$ and $\theta$ directions. A worldtube of this form is illustrated in Fig.~\ref{fig:worldtube}.

\section{Implementation: Circular Orbits in Schwarzschild Spacetime\label{sec:implementation}}
Having laid out the general principles of the ``puncture method with $m$-mode regularization'' in Sec.~\ref{sec:exposition}, let us move on to describe a simple implementation for the special case of circular geodesic orbits in the Schwarzschild spacetime.

\subsection{Physical setup}
We consider the case of a pointlike test particle endowed with scalar charge $q$ moving along a circular geodesic orbit at radius $r = r_0$ on a Schwarzschild black hole background. 
We work in the standard Schwarzschild coordinate system $\{t, r, \theta, \varphi \}$. Let $z^\mu(\tau)$ denote the particle's worldline, parameterized by proper time $\tau$, and let $u^\mu = dz^\mu / d \tau$ denote the tangent vector. Without loss of generality, we assume that the orbit is in the equatorial plane, so that
\begin{eqnarray}
z^\mu(\tau) &=& \left[ t_p(\tau), r_0, \pi/2, \omega t_p(\tau) \right] ,  \\
u^\mu &=& \frac{\mathcal{E}}{f_0} \left(1, 0, 0, \omega \right) ,
\end{eqnarray}
where $t_p(\tau)$ is the coordinate time on the worldline, $\omega = ( M / r_0^3)^{1/2}$ is the angular frequency with respect to coordinate time $t$, and $\mathcal{E}$ is the specific energy given by
\beq
\mathcal{E} \equiv -u_{t} = f_0 \left( 1 - 3M/r_0 \right)^{-1/2} ,
\eeq
where $f = 1 - 2M/r$ and $f_0 = f(r_0)$. The retarded field $\Phiret$ is governed by Eq.~(\ref{wave-eq}), explicitly,
\beq
\Box \Phiret = (-g)^{-1/2} \left[ (-g)^{1/2} g^{\mu \nu} \Phi^{\text{ret}}_{, \nu} \right]_{,\mu} = S(x) ,   \label{eq:scalarfield}
\eeq
where $g^{\mu \nu} = \text{diag}[ -f^{-1}, f, 1/r^2, 1/(r^2 \sin^2 \theta)]$ is the contravariant Schwarzschild metric and $g = - r^4 \sin^2 \theta$ is the metric determinant.
The source term is defined by Eq.~(\ref{Sdef}), explicitly,
\beq
S = -\frac{4 \pi q}{r_0^2} \frac{f_0}{\mathcal{E}} \delta (r - r_0) \delta \left( \theta - \frac{\pi}{2} \right) \delta \left( \varphi - \omega t_p \right).
\eeq

\subsection{Puncture scheme\label{subsec:puncture-scheme}}
To briefly recap Sec.~\ref{subsec:puncture}--\ref{subsec:globaldef}, a puncture scheme involves the introduction of a puncture field $\Phipunc^{[n]}$, given analytically, whose singular structure is similar to that of $\Phiret$. The residual field $\Phires^{[n]}$ is found from the difference between the full field and the puncture, i.e., $\Phires^{[n]} = \Phiret - \Phipunc^{[n]}$, and the effective source $\Seff$ is found from the d'Alembertian of the puncture via Eq.~(\ref{Seff}). After $m$ mode decomposition, a worldtube (Fig.~\ref{fig:worldtube}) is constructed around the worldline whose dimensions $\{ \twr, \twth \}$ are kept as controllable numerical parameters, of order $M$. The field equations to be evolved are given in (\ref{eqs-worldtube}).

To proceed, we now require expressions for the $m$-mode decompositions of the puncture field and effective source, i.e., $\Phipunc^m$ and $\Seff^m$. In this section we give explicit expressions for $\Phipunc^{[n]}$ in the Schwarzschild spacetime for the 2nd, 3rd and 4th-order ($n=2,3,4$) schemes, and we describe how to obtain $\Phipunc^m$ and $\Seff^m$.

\subsubsection{1st-order puncture, $n=1$.}
A first-order puncture scheme was described in \cite{Barack:Golbourn:2007}. The first-order puncture field is simply
\beq
\Phipunc^{[1]} = q / \epsilon_{[1]}, \quad \quad \text{where} \quad \epsilon_{[1]} = \left( \Prr \dr^2 + \Pth \dth^2 + \Pph \dph^2 \right)^{1/2} ,
\eeq
with coordinate differences $\delta r = r-r_0, \delta \theta=\theta - \pi/2$, and $\delta \varphi=\varphi-\omega t$ as defined in (\ref{coord-differences}). Here the coefficients are
\beq
\Prr = f_0^{-1}, \quad \Pth = r_0^2, \quad \Pph = r_0^2 \, \frac{r_0-2M}{r_0-3M}.   \label{Pco-def}
\eeq
Unfortunately, as discussed in Sec.~\ref{subsec:sf-from-res}, a first-order scheme is not sufficient to extract the SF; let us therefore proceed immediately to consider higher-order schemes.

\subsubsection{2nd-order puncture, $n=2$.}
A 2nd-order puncture scheme was described in \cite{Barack:Golbourn:Sago:2007, Barack:2009}. A 2nd-order puncture is given by  
\beq 
\Phipunc^{[2]} = q / \epsilon_{[2]} , \label{punc-2nd-order}
\eeq 
where
\beq
\epsilon_{[2]}^2 = \epsilon_{[1]}^2 + \dr \left( \Qrr \dr^2 + \Qth \dth^2 + \Qph \dph^2 \right) , \label{punc-2nd-dphi}
\eeq
with components 
\beq
\Qrr = - \frac{M}{r_0^2 f_0^2}, \quad \Qth = r_0, \quad \Qph = r_0 \, \left( \frac{r_0 - M}{r_0 - 3M} \right).   \label{Qco-def}
\eeq
Within the $m$ mode scheme we are motivated to re-express our puncture in terms of analytic periodic functions of $\delta \varphi$. Smoothness across $\varphi = -\pi$, $\pi$ can be achieved by making the replacement
\beq
\delta \varphi^2 \rightarrow 2 \left( 1 - \cos \dph \right)   =  \dph^2 + \mathcal{O}(\dph^4) \label{phi-replacement-2nd}
\eeq
in Eq.~(\ref{punc-2nd-dphi}). This replacement does not affect the order of the 2nd-order puncture.
Note that the alternative replacement $\dph^2 \rightarrow \sin^2 \dph$ is unsuitable, because it leads to an additional zero in $\epsilon_{[2]}$ at $\dph=\pi$. 

The $m$-mode decomposition of the puncture field (\ref{punc-2nd-order}) and the effective source [obtained via (\ref{Seff})] is described in Appendix \ref{sec:appendix:second-order}. With the replacement (\ref{phi-replacement-2nd}), it turns out that the $m$-modes can be found explicitly in closed forms involving elliptic integrals. 

\subsubsection{3rd-order puncture, $n=3$.\label{subsec:punc-thirdorder}}
Expressions for the 3rd and 4th-order punctures can be obtained using the covariant expansion method of Wardell and collaborators \cite{Anderson:Flanagan:Ottewill:2005, Wardell:thesis, Ottewill:Wardell}. A 4th-order puncture is also employed in \cite{Vega:Diener:Tichy:Detweiler:2009}. A 3rd-order puncture field is given by \cite{Wardell:2010}
\beq 
\Phipunc^{[3]} = q \left( \frac{1}{ \epsilon_{[3]}} +  \frac{\alpha_{[3]}}{\epsilon_{[1]} \epsilon_{[3]}^2} \right)  ,\label{punc-3rd-order}
\eeq 
where
\beq
\alpha_{[3]} = 
\frac{M \left[ dr^2 + r_0^2 f_0(\dth^2 + \dph^2) \right] \left[ (2r_0-3M)r_0^{-1} \dr^2 - r_0^2 f_0 (\dth^2 + f_0 \dph^2 \right]}{6 r_0^2 f_0^2 (r_0-3M)} ,
\eeq
and
\beq
\epsilon_{[3]}^2 = \epsilon_{[2]}^2 +  \sum_{i=1}^{3} \sum_{j \ge i}^3 U_{ij} (\delta x_i)^2 (\delta x_j)^2 . 
\eeq
Here the indices $i$, $j$ run from 1 to 3, and we use the shorthand $\delta x_1 = \dr$, $\delta x_2 = \dth$ and $\delta x_3 = \dph$. The coefficients are
\begin{eqnarray}
\hspace{-0.5cm}&&\Rrr = \frac{(8r_0 - M)M}{12 r_0^4 f_0^3}  , \quad \Rqq = - \frac{r_0^2 f_0}{12} , \quad \Rpp = -\frac{r_0^2 f_0 (r_0 + M)}{12(r_0-3M)},   \\
\hspace{-0.5cm}&&\Rrq = - \frac{M}{6r_0 f_0}  , \quad \Rrp = \frac{M(5 r_0 - 11M)}{6r_0f_0(r_0-3M)} , \quad \Rqp = - \frac{r_0(3r_0-2M)(r_0-M)}{6(r_0 - 3M)}.
\end{eqnarray}

Again, we may replace the azimuthal variable to obtain a periodic function that is smooth across $\dph = -\pi$, $\pi$. We make the replacement
\beq
\delta \varphi^2  \rightarrow  \frac{5}{2} - \frac{8}{3} \cos \delta \varphi + \frac{1}{6} \cos 2 \delta \varphi  = \delta \varphi^2 + \mathcal{O} ( \delta \varphi^6 ) ,   \label{phi-repl-3rd4th}
\eeq
which preserves the order of the puncture.

Next we compute the $m$-mode decomposition of the puncture field via (\ref{mmode-inversion}), to obtain
\beq
\Phipunc^{[3] \m}(\dr,\dth) = \frac{e^{-i m \omega t}}{2\pi} \int_{-\pi}^{\pi} \Phipunc^{[3]}(\dr,\dth,\dph^\prime) \cos( m \dph^\prime ) d(\dph^\prime) \label{phipunc3m} .
\eeq
Note that here we have used $\varphi = \dph + \omega t$ to factor out the time-dependence, and we have also used the symmetry of the puncture under $\dph \rightarrow -\dph$ to eliminate the imaginary (sine) component of $e^{-im\dph}$, leaving the real (cosine) component. Analytic integrals are not readily available for computing (\ref{phipunc3m}) and so we resort to numerical methods. 
Thanks to the simple factorization of the $t$ dependence in (\ref{phipunc3m}), we do not need to perform new integrals for each value of $t$ in the simulation; nevertheless, we do need to compute the integrals numerically for every value of $\dr$ and $\dth$ within the worldtube. In future, it may be possible to find a way to obtain analytic representations for the $m$ modes, by casting the puncture into a more tractable form. It is likely this question will require further investigation when eccentric orbits are considered (in which case the factorization of the time dependence is not so straightforward). 

The effective source $\Seff$ is found from inserting the d'Alembertian of (\ref{punc-3rd-order}) into Eq.~(\ref{Seff}). We used a symbolic algebra package to help with this calculation. The $m$-modes $\Seff^m$ are obtained via numerical integration, in a similar manner to the above.

\subsubsection{4th-order puncture, $n=4$\label{subsec:4thordpunc}}
Finally, a 4th-order puncture field is given by \cite{Wardell:2010}
\beq
\Phipunc^{[4]} = q \left( \frac{1}{\epsilon_{[4]}} + \frac{\alpha_{[4]}}{\epsilon_{[2]}^{3}} + \frac{\beta_{[4]}}{\epsilon_{[1]}^{3}} \right) ,  \label{Phi-4thord}
\eeq
where
\beq
\epsilon^2_{[4]} = \epsilon^2_{[3]} + M^{-1} \dr 
 \sum_{i=1}^{3} \sum_{j \ge i}^3 V_{ij} (\delta x_i)^2 (\delta x_j)^2
\eeq
and
\begin{eqnarray}
\hspace{-1.0cm}&&\Trr = - \frac{(6r_0^2 - 2Mr_0 + M^2)M^2}{12 r_0^6 f_0^4} , \quad \Tqq = - \frac{(r_0-M)M}{12} , \quad \Tpp = - \frac{(r_0^2+4Mr_0-9M^2)M}{12(r_0-3M)},   \\
\hspace{-1.0cm}&&\Trq = \frac{M^2}{12 r_0^2 f_0^2} , \quad \Trp = \frac{(r_0^2 - 5Mr_0 + 8M^2)M^2}{12r_0^3 f_0^2 (r_0-3M)}, \quad \Tqp = - \frac{(3 r_0^2 + 2Mr_0 - 3M^2)M}{6(r_0 - 3M)} .
\end{eqnarray}
The remaining quantities are 
\beq
\alpha_{[4]} = \alpha_{[3]} + \frac{M^2 \dr}{6r_0^2 (r_0 - 3M)} 
 \sum_{i=1}^{3} \sum_{j \ge i}^3 X_{ij} (\delta x_i)^2 (\delta x_j)^2 ,
\eeq
with
\begin{eqnarray}
\hspace{-1.0cm}&&\Xrr = -\frac{2(2r_0 - 3M)}{r_0^3 f_0^3} , \quad \Xqq = -M^{-1} r_0^2(5r_0-3M), \quad \Xpp = -M^{-1}(5r_0-9M)r_0^2 ,   \\
\hspace{-1.0cm}&&\Xrq = -\frac{(2r_0^2 - 3Mr_0-3M^2)}{Mr_0f_0^2}, \quad \Xrp = -\frac{2r_0^2-7Mr_0+7M^2}{Mr_0f_0^2} , \quad \Xqp = -2M^{-1}r_0^2(5r_0 - 6M)   ,
\end{eqnarray}
and
\beq
\beta_{[4]} = \frac{M^2 \dr}{8r_0^2 (r_0-3M)} 
 \sum_{i=1}^{3} \sum_{j \ge i}^3 Y_{ij} (\delta x_i)^2 (\delta x_j)^2 ,
\eeq
with
\begin{eqnarray}
\hspace{-1.0cm}&&\Yrr = - \frac{2r_0 - 3M}{M r_0^2 f_0^2} , \quad \Yqq = M^{-1} r_0^2(3r_0 - 2M) , \quad \Ypp = \frac{r_0^3 f_0 (3r_0^2-24Mr_0+41M^2)}{M(r_0-3M)^2}   ,   \\
\hspace{-1.0cm}&&\Yrq = \frac{(r_0+M)}{Mf_0}, \quad \Yrp = \frac{(r_0^2-12Mr_0+21M^2)}{M(r_0-3M)f_0} , \quad \Yqp = \frac{2r_0^2(3r_0^2-16Mr_0+18M^2)}{M(r_0 -  3M)}   .
\end{eqnarray}
As in Sec.~\ref{subsec:punc-thirdorder}, we used the replacement (\ref{phi-repl-3rd4th}) and found the $m$-mode decomposition by performing the relevant integrals numerically.


\subsection{Simulation details\label{subsec:simulation-details}}
In the following sections we employ the 4th-order puncture scheme unless otherwise stated. Therefore we will generally suppress the `order' index $[n]$ in the following sections, unless required for disambiguation.  

\subsubsection{Modal equations}

As discussed in Sec.~\ref{subsec:worldtube}, the individual $m$-modes of the field are governed by a set of equations given in Eq.~(\ref{eqs-worldtube}). To make use of Eq.~(\ref{eqs-worldtube}) we first require an explicit expression for the $2+1$D operator $\Box^m$ on the Schwarzschild spacetime, i.e.
\beq
\Box^m \Phi^m \equiv -f^{-1} \Phi^m_{,tt} + f \Phi^m_{,rr} +  2 r^{-2} (r-M) \Phi^m_{,r} + r^{-2} \left(\Phi^m_{,\theta\theta} + \cot \theta \Phi^m_{,\theta} \right) - m^2 r^{-2} \sin^{-2} \theta \Phi^m.
\eeq
Here $\Phi^m$ can represent either $\Phires^m$ or $\Phiret^m$, depending on whether we are inside or outside the worldtube. Let us note that stationary retarded solutions of Eq.~(\ref{eq:scalarfield}) fall off as $1/r$ towards spatial infinity, i.e.~$\Phiret^m \sim 1/r$ as $r \rightarrow \infty$. This motivates the introduction of new field variables,
\beq
\Psiret^m = r \Phiret^m , \quad \quad   \Psi^m_\mathcal{R} = r \Phires^m .
\eeq
The modes $\Psiret^m$ and $\Psires^m$ are governed by the set of equations
\beq
\left\{
\begin{array}{l l}
\Box^{m}_{\Psi} \Psires^m = -(fr/4) \Seff^m , & \text{ inside } \mathcal{T} ,\\
\Box^{m}_{\Psi} \Psiret^m = 0 , &  \text{ outside } \mathcal{T}, \\
\Psires^m = \Psiret^m - r \Phipunc^m , & \text{ across } \partial \mathcal{T} , \\
\end{array}
\right. 
\label{eqs-psi}
\eeq
where we recall that $\mathcal{T}$ and $\partial \mathcal{T}$ represent the interior and surface of a worldtube in the 2+1D domain, illustrated in Fig.~\ref{fig:worldtube}. Here,
\beq
\Box^m_{\Psi} \Psi^m \equiv \Psi^m_{, uv} - \frac{f}{4r^2} \left[ \Psi^m_{, \theta \theta} + \cot \theta \Psi^m_{,\theta} - \left(2M/r + m^2 \csc^2 \theta \right) \Psi^m \right] , 
 \label{BoxPsi1}
\eeq
where $u$ and $v$ are retarded and advanced Eddington-Finkelstein null coordinates, given by
\beq
u = t - r_\ast , \quad \quad v = t + r_\ast ,   \label{uvdef}
\eeq
with the tortoise coordinate
\beq
r_\ast = r + 2M \ln \left( \frac{r - 2M}{2M} \right).   \label{rstar-def}
\eeq

\subsubsection{2+1D grid and worldtube construction\label{subsec-grid}} 
\begin{figure}
 \begin{center}
  \includegraphics[width=7cm]{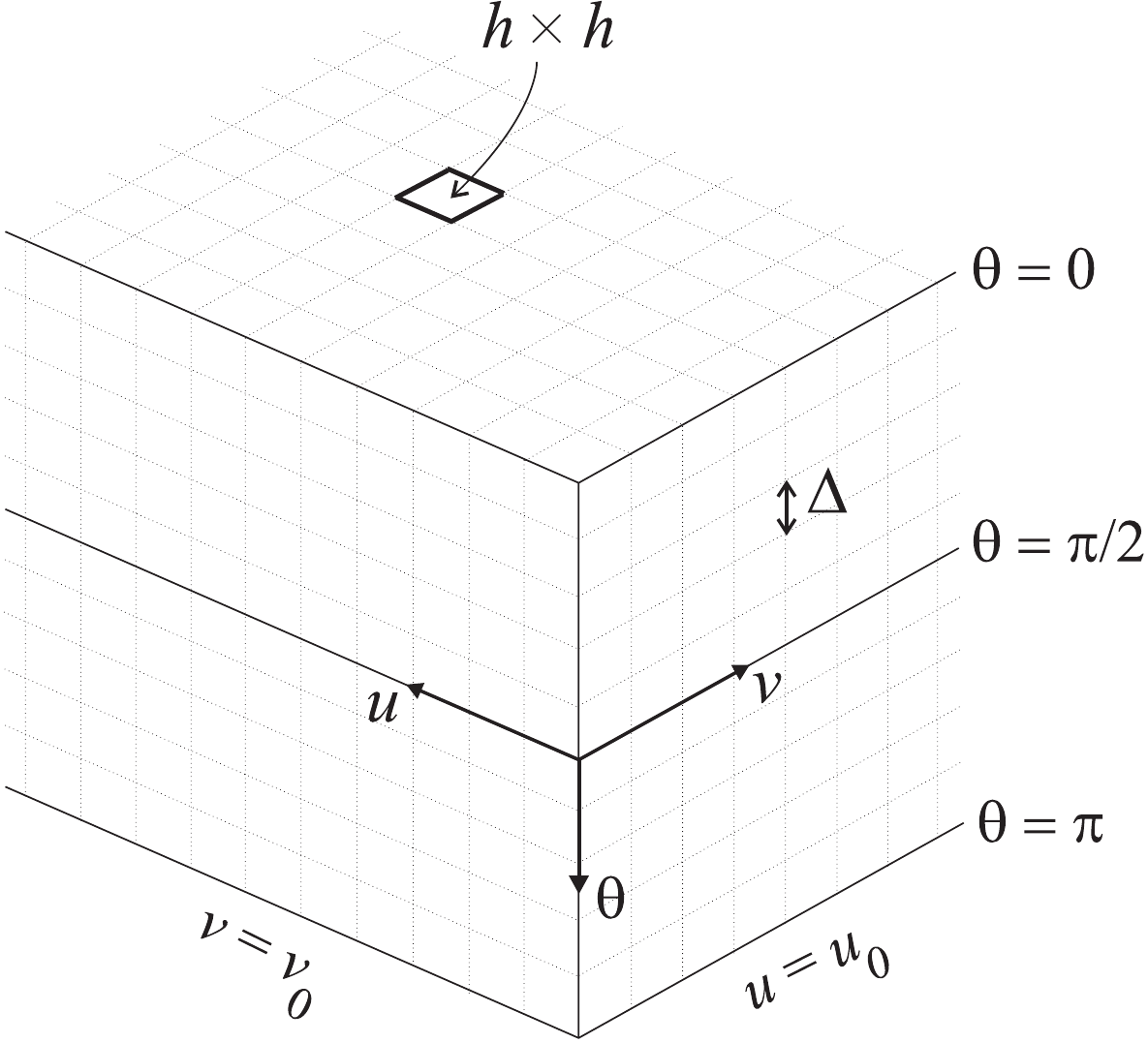}
  \includegraphics[width=7cm]{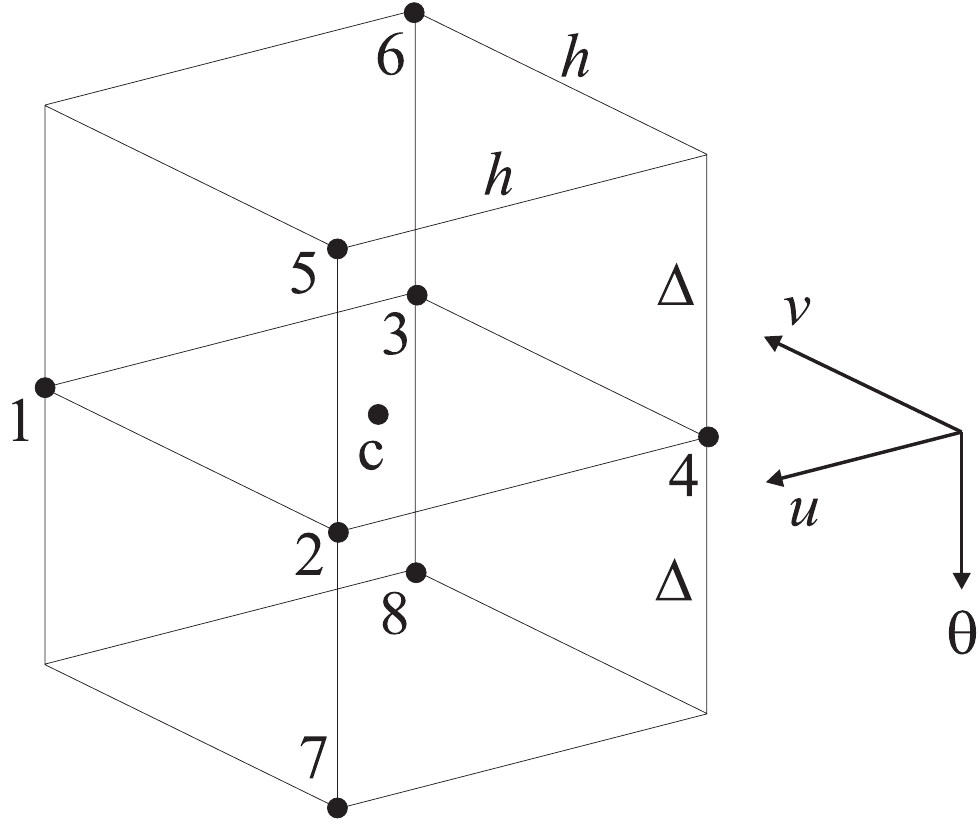}
 \end{center}
 \caption{2+1D finite difference scheme. The left plot shows the grid in $u$, $v$ and $\theta$. The right plot shows a single `cell' for the finite difference method. These plots are reproduced from \cite{Barack:Golbourn:2007}, Figs.~1 and 2.}
 \label{fig:grid}
\end{figure}

We construct a grid over coordinates $u$, $v$, $\theta$, with linear spacing $h$ in the $u$ and $v$ directions, and linear spacing $\Delta$ in the $\theta$ direction. See Fig.~\ref{fig:grid} for an illustration. The grid may be seen as a stack (in $\theta$) of causal diamonds (in $u$ and $v$). The central $r_\ast = \text{const}$ line of each diamond is taken at the orbital radius, $r_{\ast0} \equiv r_{\ast}(r_0)$ via (\ref{rstar-def}). The two initial surfaces, at $u=u_i \equiv -r_{\ast0}$ and $v=v_i=r_{\ast0}$ intersect at $t=0$, $r=r_0$. The final surfaces at $u = u_f$ and $v=v_f$ intersect at $t=\tmax$, $r=r_0$. The shape of the grid, and use of null coordinates $u$ and $v$, eliminates the need for boundary conditions except at the poles $\theta = 0, \pi$. 

Now let us introduce a worldtube of fixed coordinate widths $\{ \Gamma_{r_\ast}$, $\Gamma_\theta \}$, centred on the worldline at $r_{\ast} = r_{\ast0}$, $\theta = \pi/2$. Consider an arbitrary grid point with coordinates ($r_\ast$, $\theta$). If $|r_{\ast} - r_{\ast0}| \le \Gamma_{r_\ast} / 2$ and $|\theta - \pi/2| \le \Gamma_{\theta} / 2$, then the point lies within the worldtube; otherwise it lies outside. For convenience, we choose the worldtube widths $ \Gamma_{r_\ast}$ and $\Gamma_\theta $ to be integer multiples of the grid spacings $h$ and $\Delta$, respectively.

\subsubsection{Initial and boundary conditions}
For each $m$, we specify a `zero' initial condition on the initial surfaces $u = u_i$ and $v = v_i$, 
\beq \Psiret^m(u_i, v, \theta) = 0,  \quad \quad  \Psiret^m(u, v_i, \theta) = 0 .  \label{zero-ic} \eeq
The initial condition is not a solution of the sourced field equations. However, we expect that `junk' in the initial condition will radiate away, towards the horizon and infinity, and that after sufficient time the field near the particle will approach the correct (retarded) stationary solution.

At the poles, boundary conditions are required. As argued in \cite{Barack:Golbourn:2007}, the physical boundary conditions to be applied at the poles are
\beq
\partial_\theta \Psiret^{m=0}(\theta = 0, \pi) = 0,  \quad \quad \Psiret^{m \neq 0}(\theta = 0, \pi) = 0 . \label{polar-bc1} 
\eeq
To implement these conditions, we simply set $\Psiret^m = 0$ at the poles for $m\neq0$, and for $m=0$ we extrapolate to obtain
\begin{eqnarray}
\Psiret^{m=0}(\theta = 0) &=& \frac{1}{3} \left[ 4 \Psiret^{m=0}(\theta=\Delta) - \Psiret^{m=0}(\theta = 2 \Delta) \right] + \mathcal{O}\left( \Delta^4 \right),  \label{polar-bc2} \\
\Psiret^{m=0}(\theta = \pi) &=& \frac{1}{3} \left[  4 \Psiret^{m=0}(\theta = \pi - \Delta) - \Psiret^{m=0}(\theta = \pi - 2\Delta) \right] + \mathcal{O} \left( \Delta^4 \right) .   \label{polar-bc3}
\end{eqnarray}
Here the error term is $\mathcal{O}(\Delta^4)$, rather than $\mathcal{O}(\Delta^3)$, since
$\Psiret^m$ is an even function of $\theta$ at $\theta=0$ (and an even function of $\pi-\theta$ at $\theta = \pi$).

\subsubsection{Finite difference method\label{subsec:finite-diff-method}}
We employed the finite difference method which was described, applied and tested in \cite{Barack:Golbourn:2007}. Figure~\ref{fig:grid} shows a `cell' of grid points with centre $c$. Let us assume (for now) that all the grid points in the cell lie outside the worldtube so that $\Psim_1, \ldots ,\Psim_8$ represent values of the retarded field mode, $\Psiret^m$, at the positions shown. Finite-difference approximations for the values of the field and its derivatives at $c$ are obtained through the following expressions:
\begin{eqnarray}
\Psim_{c,uv} &=& \frac{\Psim_1 + \Psim_4 - \Psim_3 - \Psim_2}{h^2} + \mathcal{O} \left( h^2 \right) , \label{fdeq1} \\
\Psim_{c,\theta \theta} &=& \frac{\Psim_5 + \Psim_6 + \Psim_7 + \Psim_8 - 2\left(\Psim_2 + \Psim_3 \right)}{2 \Delta^2} + \mathcal{O} \left( \Delta^2 \right) ,\\
\Psim_{c,\theta} &=& \frac{\Psim_7 + \Psim_8 - \Psim_5 - \Psim_6}{4 \Delta} + \mathcal{O}\left( \Delta^2 \right), \\
\Psim_c &=& \frac{\Psim_2 + \Psim_3}{2} + \mathcal{O}\left(h^2\right) . \label{fdeq4}
\end{eqnarray}
Now, assuming that the values at points $2$ to $8$ have been obtained in previous steps, we may insert (\ref{fdeq1})--(\ref{fdeq4}) into Eq.~(\ref{eqs-psi}) and rearrange to find $\Psim_1$:
\begin{eqnarray}
\Psim_1 &=& \Psim_2 + \Psim_3 - \Psim_4 + \nonumber \\ &&  \frac{h^2 f}{8r^2} \left[ \left( \Psim_5 + \Psim_6 + \Psim_7 + \Psim_8 - 2\Psim_2 - 2\Psim_3 \right) / \Delta^2  + \cot \theta \left( \Psim_7 + \Psim_8 - \Psim_5 - \Psim_6 \right) / (2 \Delta) \right. \nonumber \\
&& \quad \quad \left. - \left( 2M/r + m^2 \csc^2 \theta \right) \left( \Psim_2 + \Psim_3 \right) \right] + \mathcal{O} \left( h^2 \Delta^2, h^4  \right).   \label{eq-finitediff}
\end{eqnarray}
Here, all $r$, $\theta$ dependent coefficients are evaluated at the centre of the cell.
Conversely, if all points in the cell lie \emph{inside} the worldtube, then $\Psim_1 , \ldots , \Psim_8$ represent values of the residual field mode, $\Psires^m$, and one may repeat the argument above to obtain
\beq
\Psim_1 = [\text{RHS of Eq.~(\ref{eq-finitediff})}] + h^2 \Zeff , \quad \quad \text{where} \quad \Zeff = - f r \Seff^m / 4. \label{eq-finitediff-source}
\eeq

In the case where the finite difference cell straddles the boundary of the tube (so that some points are `in' and some `out'), we make use of the puncture field $\Phipunc^m$ in the following way. If point 1 (Fig.~\ref{fig:grid}) is `out', then we first demote all `in' points in the cell to `out' points using $\Phi^m = \Phires^m + \Phipunc^m$ before applying (\ref{eq-finitediff}). If, conversely, point 1 is `in' then we promote all `out' points in the cell using $\Phires^m = \Phi^m - \Phipunc^m$ before applying (\ref{eq-finitediff-source}). This strategy is discussed in more detail in Sec.~VB of \cite{Barack:Golbourn:2007}.

For a fixed ratio $\Delta / h$, the finite difference equation (\ref{eq-finitediff}) has a local discretization error of $\mathcal{O}(h^4)$. In vacuum, therefore, we expect (and find) the scheme to be quadratically convergent [i.e.~to exhibit a global accumulated finite-differencing error which scales as $\mathcal{O}(h^2)$]. Unfortunately,  quadratic convergence of our simple scheme is by no means assured if a non-smooth source term is present, as in Eq.~(\ref{eq-finitediff-source}). Consider the special case of a grid cell whose centre $c$ lies exactly on the worldline. As discussed in Sec.~\ref{subsec:worldtube}, for puncture orders $n<4$ the effective source $\Seff^{[n<4]}$ is not continuous across the worldline, and $\Seff^{[n<4]m}$ cannot easily be evaluated for this cell. A strategy for dealing with this problem was discussed and implemented in \cite{Barack:Golbourn:2007}. The case of a cell centred on the worldline was handled separately, taking into account the singular structure of $\Seff^{[n]m}$. A similar method was employed here, in the cases $n=2$ and $3$. The procedure leads to a local error in the central cell that scales as $h^3 \ln h$, which leads to a term in the global accumulated finite-differencing error that scales with $h^2 \ln h$. Although this undesirable behaviour could perhaps be eliminated by using a more sophisticated finite-difference scheme in the vicinity of the worldline, we have not pursued such an approach here, principally because we now have a 4th-order puncture ($n=4$) available. In the 4th-order case, no difficulty is encountered for the cell on the worldline (since the source $\Seff^{n=4}$ is continuous across the worldline) and the global convergence rate is found to be quadratic.

\subsubsection{Numerical stability\label{subsec:stability}}
Finite difference methods can suffer from \emph{numerical instabilities}, where generic small-amplitude short-wavelength perturbations (originating for example from truncation errors) are amplified exponentially, eventually overwhelming the physical solution. Our finite difference method suffers from a numerical instability if the ratio of grid spacings $\Delta / h$ is set to be too small. We observed in our vacuum simulations that this instability arises first near the poles, and appears as a spurious oscillation in the $\theta$ direction with wavelength $2\Delta$ and an exponentially-growing amplitude. In Appendix \ref{appendix:von-neumann} we apply a von Neumann stability analysis (see, e.g., \cite{NumericalRecipes}) to our finite difference equation (\ref{eq-finitediff}) which suggests that a necessary condition for stability is 
\beq
\frac{\Delta}{h} \ge \frac{1}{2} \text{max} \left( r^{-1} f^{1/2} \right) \sqrt{ 1 + m^2 / 4 } \, ,  \label{stability-condition}
\eeq
where $\text{max} ( r^{-1} f^{1/2} ) \approx 0.19245 M^{-1}$. Equation (\ref{stability-condition}) becomes a highly restrictive condition when $m$ is large. For $m > 3$ this condition becomes stronger than the standard Courant condition,
\beq
\frac{\Delta}{h}  \ge  \text{max} \left( r^{-1} f^{1/2} \right),
\eeq
which is obtained by insisting that the numerical domain of dependence contains the physical, continuum domain of dependence at each point in the evolution. 

To mitigate the instability in large-$m$ modes we may move the grid boundary inwards from the poles. 
First we note that solutions have a simple asymptotic form near the poles, i.e.~ 
\begin{eqnarray}
\Psiret^m(\theta \ll 1) &=& A \theta^m + B \theta^{m+2} + \mathcal{O}(\theta^{m+4}) , \\
\Psiret^m(\pi - \theta \ll 1) &=& A (\pi - \theta)^m + B (\pi - \theta)^{m+2} + \mathcal{O}\left((\pi-\theta)^{m+4}\right) ,
\end{eqnarray}
where $A$ and $B$ are constants. In other words, the large-$m$ modes are very `flat' near the poles. If we move the boundary point inwards from $\theta = 0$ to $\theta = k \Delta$ (with $k$ being a small positive integer) then, repeating the analysis of Appendix \ref{appendix:von-neumann}, the stability condition becomes
\beq
\frac{\Delta}{h} \ge \frac{1}{2} \, \text{max} \left( r^{-1} f^{1/2} \right) \sqrt{ 1 + m^2 / [4 (k+1)^2] }  ,  \label{stability-modified}
\eeq
which is less restrictive than the original condition (\ref{stability-condition}). The boundary condition near $\theta=0$ changes to 
\begin{eqnarray}
k = 1  &:&   \Psiret^{m>0}(\theta = \Delta)    = \frac{1}{5} \left[ 2^{3-m} \Psiret^m(2 \Delta) - 3^{1-m} \Psiret^m(3 \Delta) \right] + \mathcal{O}( \Delta^{m+4} ) , \label{polar-bc-alt1}   \\
k = 2  &:&  \Psiret^{m>0}(\theta = 2 \Delta) = \frac{1}{7} \left[ 12 \times \left(2 / 3 \right)^m \Psiret^m(3 \Delta) - 5 \times 2^{-m} \Psiret^m(4 \Delta) \right]  + \mathcal{O}( \Delta^{m+4} ), \label{polar-bc-alt2}  
\end{eqnarray}
etc. The modes are symmetric under $\theta \rightarrow \pi - \theta$, so equivalent boundary conditions may be applied near the south pole at $\theta = \pi$. In our implementation, we fix the ratio $\Delta / h$ and, for a given mode $m$, we find the minimum value of $k$ required for stability using (\ref{stability-modified}).

\subsection{Simulations and data extraction\label{subsec:data-extraction}}
For a given $m$, we evolve the initial data (\ref{zero-ic}) according to the finite difference scheme (\ref{eq-finitediff})--(\ref{eq-finitediff-source}) with boundary conditions (\ref{polar-bc1})--(\ref{polar-bc3}) [or (\ref{polar-bc-alt1})--(\ref{polar-bc-alt2})] on the `diamond stack' 2+1D grid (Fig.~\ref{fig:grid}) of dimensions $u_f - u_i = v_f - v_i  = \tmax$, to obtain a numerical estimate for an $m$-mode residual field, $
\Psinum(t,r,\theta)$. Of course, the numerical solution obtained depends on the `physical' parameters, $r_0/M$ and $m$, and a set of `numerical' parameters, $\intl = \{ h, \Delta, \twrs, \twth, \tmax, \ldots \}$. We will refer to a simulation for a particular $m, r_0/M$ with a unique set of $\intl$ as a `run'.

As described in Sec.~\ref{subsec:mmode}, the SF and the R field (at a worldline point $z$) are computed from a sum over modal contributions $F_\mu^m$ and $\Phirestil^m(z)$ [see Eq.~(\ref{F-mmode}) and (\ref{Phi-mmode})]. Let us briefly describe how the modal contributions are obtained from our `runs'. First we assume that the total time $\tmax$ is sufficiently large that the residual field inside the worldtube has settled into a quasi-stationary state at late times (see Sec.~\ref{subsec:relaxation} for further consideration). Next, we read off the following quantities, 
\begin{eqnarray}
\PhiRtil^\m  \left( t_1 \right)  
&=&  r_0^{-1} \Psirestil^m \left( t_1, r_0, \pi/2 , \omega t_1 \right),   \label{mode-contribution-1} \\
\Ftil_r^\m  \left( t_1 \right) 
&=&  r_0^{-1} \left[ f_0^{-1} [ \partial_{r_\ast} \Psirestil^m ]  - r_0^{-1} \Psirestil^m  \right] (t_1, r_0, \pi/2, \omega t_1)  , \\
\Ftil_\varphi^\m \left( t_1\right) 
&=&  - 2 m r_0^{-1}  \text{Im} \left[ \Psires^m \left( t_1, r_0, \pi/2 \right) e^{i m \omega t_1} \right]  , \label{mode-contribution-3} 
\end{eqnarray}
where $\Psirestil^m = r_0 \Phirestil$ with $\Phirestil$ as defined in Eq.~(\ref{mmode-contribution-eq}). Here all quantities are evaluated at grid points on the worldline at a late time $t=t_1$, where $t_1 < \tmax$, and the derivative with respect to $r_\ast$ is found via central differencing on the grid. 
The SF and radiative field on the worldline are found by inserting (\ref{mode-contribution-1})--(\ref{mode-contribution-3}) into the mode sum reconstruction formulae (\ref{F-mmode}) and (\ref{Phi-mmode}), respectively. In practice, we compute $\Phi_R$, $\Fself_r$ and $\Fself_\varphi$ directly from mode sums; the remaining components are given by $\Fself_t = - \omega \Fself_\varphi$ and, by symmetry, $\Fself_\theta=0$.

\subsection{Sources of numerical error\label{subsec:error-sources}}
Of course, the values extracted from a particular run via Eqs.~(\ref{mode-contribution-1})--(\ref{mode-contribution-3}) depend in part on the set of numerical parameters $\intl$. Inevitably, the values contain numerical error, which we may define as the difference between a given numerical solution and the (unknown) exact solution. To compute accurate SF estimates we must first seek to understand the various sources of numerical error that arise in our implementation. By judiciously combining the results of multiple runs, we then attempt to quantify and minimize the error. 

Let us identify several key sources of numerical error. These will be more fully described in the next section. The evolution of a single mode is affected by the following:
\begin{itemize}
  \item \emph{Discretization error} (Sec.~\ref{subsec:discretization-error}), associated with use of a finite grid spacing $h, \Delta$, i.e. \beq \Psirestil^m(\{ h, \Delta \}) - \Psirestil^m( \{h \rightarrow 0, \Delta \rightarrow 0 \}) . \eeq
  \item \emph{Worldtube error} (Sec.~\ref{subsec:worldtube-error}). Changing the dimensions of the worldline $\{ \twr, \twth \}$ affects the amplitude of the discretization error, but should not affect its scaling with $h$.
  \item \emph{Source cancellation error} (Sec.~\ref{subsec:source-cancellation}), associated with roundoff error arising in the calculation of $\Seff$ close to the worldline, from the delicate mutual cancellations of large terms in the high-order puncture.
   \item \emph{Relaxation time error} (Sec.~\ref{subsec:relaxation}), associated with the time it takes for junk radiation to decay, and the solution to reach a steady state, i.e.\beq \Psirestil^m(  t_1 ) -  \Psirestil^m(  t_1 \rightarrow \infty ) . \eeq 
\end{itemize}
In computing mode sums, there arises further errors: 
\begin{itemize}
 \item \emph{$m$-mode summation error} (Sec.~\ref{subsec:mode-summation}). Only a finite number of modes may be calculated numerically. We impose a large-$m$ cutoff $\mmax$, and estimate the contribution from the remaining modes by fitting an appropriate model. 
 \item \emph{Mode cancellation error} (Sec.~\ref{subsec:mode-cancellation}). If the magnitude of individual modal contributions $F_\mu^m$ (or $\Phirestil^\m$) is large in comparison to magnitude of the total mode sum $\Fself_\mu$ (or $\Phi_R$), then the relative error in the mode sum may be much larger than the relative error in individual modes.
\end{itemize}


\section{Results and Analysis\label{sec:results}}
In this section we present a selection of results from our numerical simulations. We discuss the challenge of minimizing numerical errors by giving illustrative examples.

\subsection{Individual $m$-modes\label{subsec:results-modes}}

\subsubsection{Simulations and visualisation}
Let us consider a typical `run', i.e.~a simulation for a single $m$ and $r_0$ and a unique set of numerical parameters $\intl$. The results of a run can be visualised by examining particular slices through the $u$-$v$-$\theta$ grid. Three informative slicings are: (i) $t = \tmax / 2$, $\theta = \pi / 2$, i.e., across the central line of the $uv$ diamond in the equatorial plane, (ii) $t = \tmax / 2$, $r_\ast = r_{\ast0}$, i.e., from pole to pole, and (iii) $r_\ast = r_{\ast0}$, $\theta = \pi/2$, i.e., `along the worldline'. 

\begin{figure}
 \begin{center}
  \includegraphics[width=8.0cm]{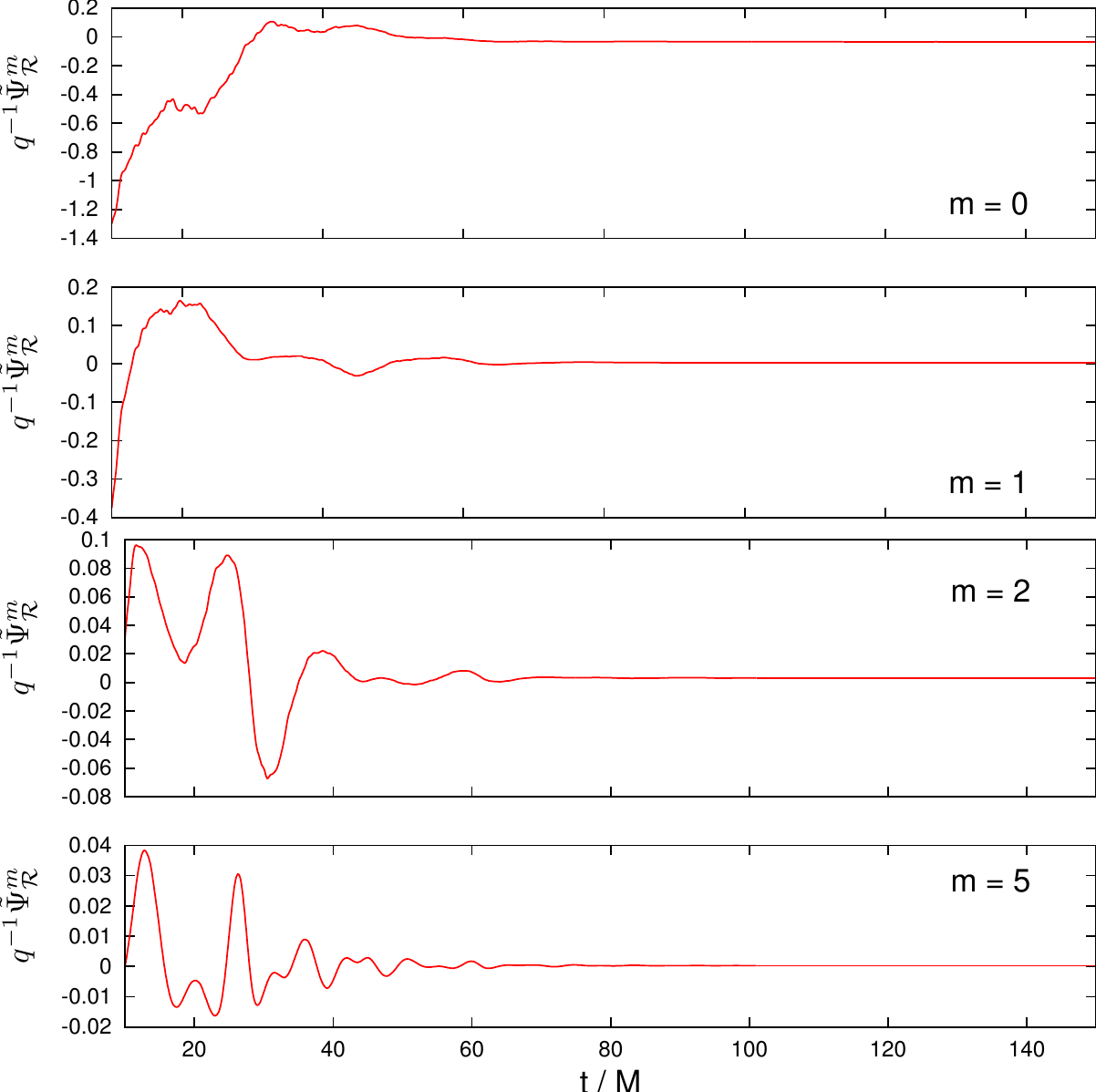}
 \end{center}
 \caption{Modes of the residual field, $\Psirestil^m$, as a function of time, evaluated along the worldline $r_\ast = r_{\ast0}$, $\theta = \pi/2$ [slice (iii) in text]. The initial burst of `junk radiation' (due to the imperfect initial condition)  radiates away, and the field approaches a steady-state in the vicinity of the worldline. }
 \label{fig:slice3}
 \end{figure}

Figure \ref{fig:slice1,2} shows $m$-mode contributions to the field modes along the constant-$t$ slices (i) and (ii), for the 2nd-order puncture scheme. The worldtube is visible as the `trough' in the centre of these plots. The residual field $\Psirestil^m$ (solid line) is continuous and differentiable across the worldline (at $\theta = \pi/2$, $r = r_0$), whereas the retarded field $\Psirettil^m$ (dotted line), found using $\Psiret^m = \Psires^m + r_0 \Phipunc^m$, diverges at the worldline. Figure~\ref{fig:slice3} shows $m$-mode contributions to the field modes `along the worldline' as a function of time, i.e., on slice (iii). At early times, the signal is dominated by `junk radiation' arising from our imperfect choice of initial condition. The effect of the junk radiation diminishes with time, and the field approaches a steady state. 

\begin{figure} 
 \begin{center}
  \includegraphics[width=8.1cm]{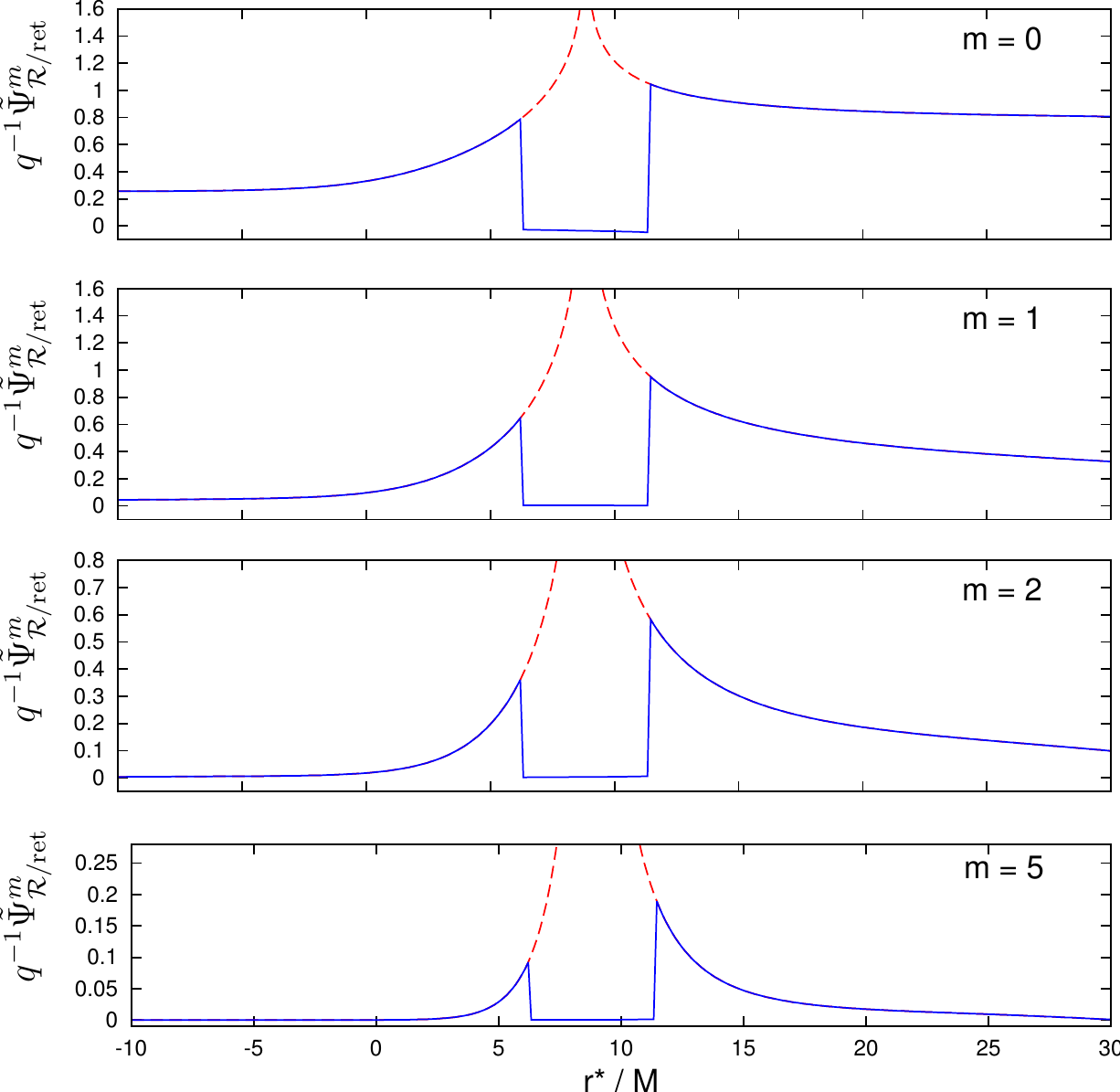}
  \includegraphics[width=8.1cm]{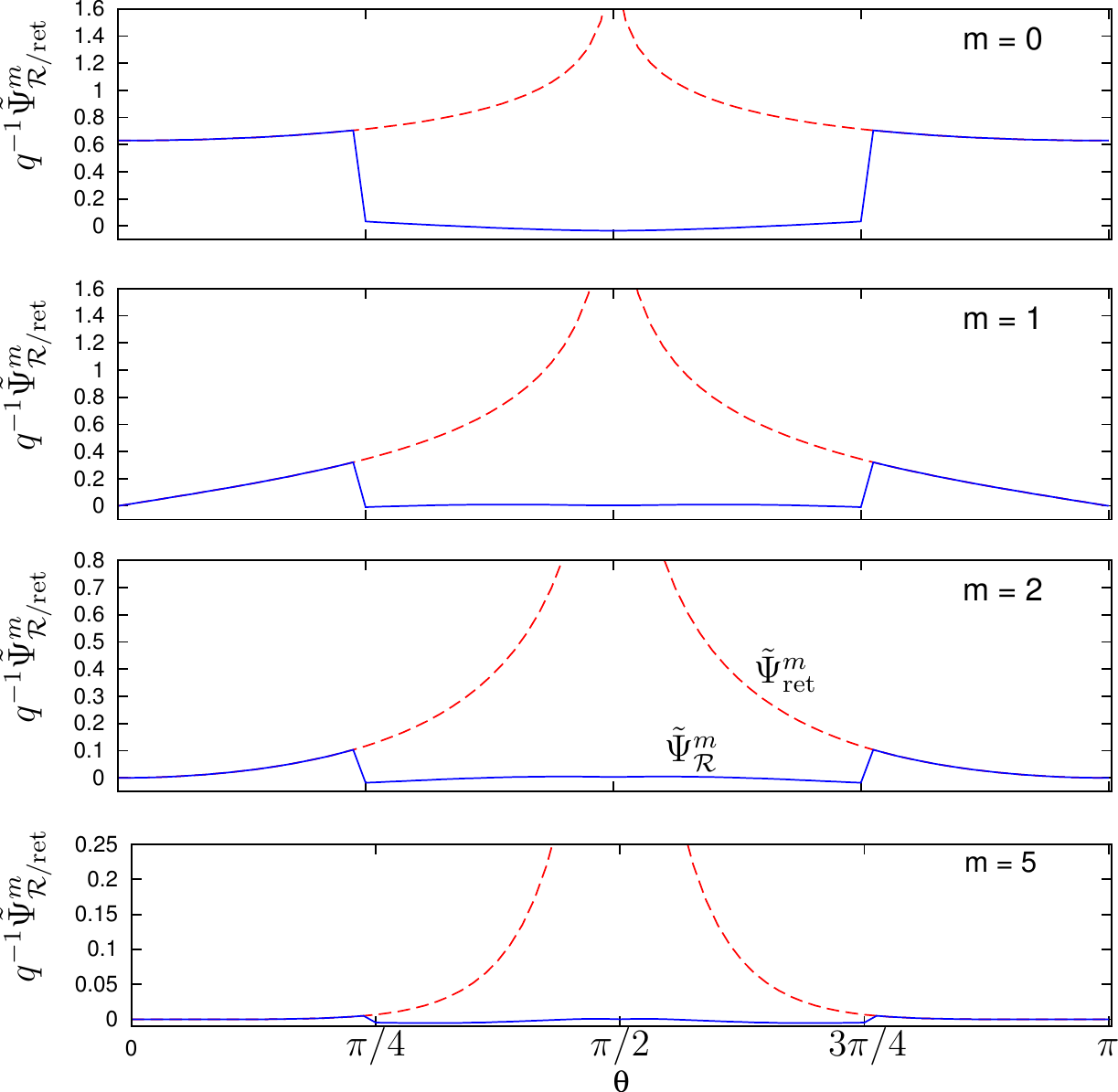} 
 \end{center}
 \caption{Field modes on constant time slices (at $t = \tmax / 2$) for a circular orbit at $r_0 = 7M$ ($r_{*0} \approx 8.8326 M$) with the 2nd-order puncture scheme. The left plots show field modes at fixed $\theta = \pi / 2$ and the right plots show field modes at fixed $r=r_0$, for a range of modes $m=0$, $1$, $2$, and $5$. Inside the worldtube (visible as the central `trough'), the dashed (red) line shows the full retarded field $\Psirettil^m$ and the solid (blue) line shows the residual field, $\Psirestil^m$. The numerical parameters are $\{h=M/8, \Delta = \pi / 40, \Gamma_\theta = \pi / 2, \Gamma_{r_\ast} = 5M\}$. }
 \label{fig:slice1,2}
 \end{figure}
 
\begin{figure}
 \begin{center}
  \includegraphics[width=8.0cm]{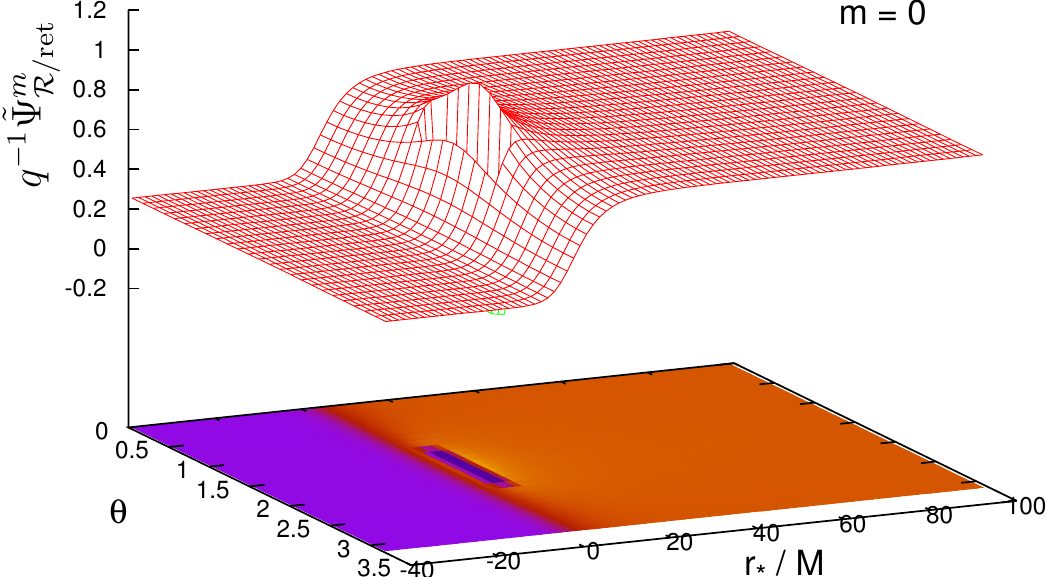}
  \includegraphics[width=8.0cm]{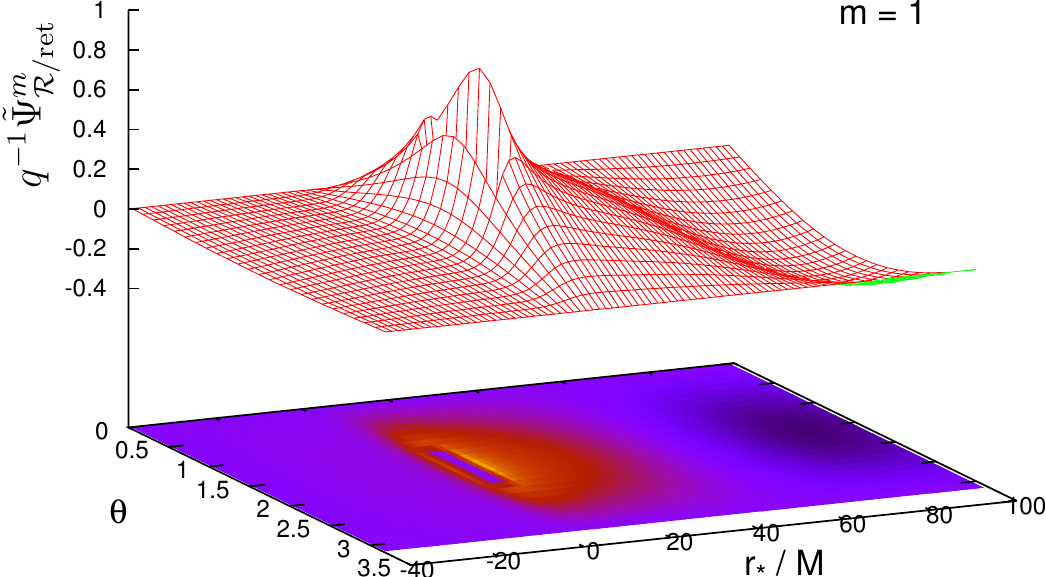} \\
  \includegraphics[width=8.0cm]{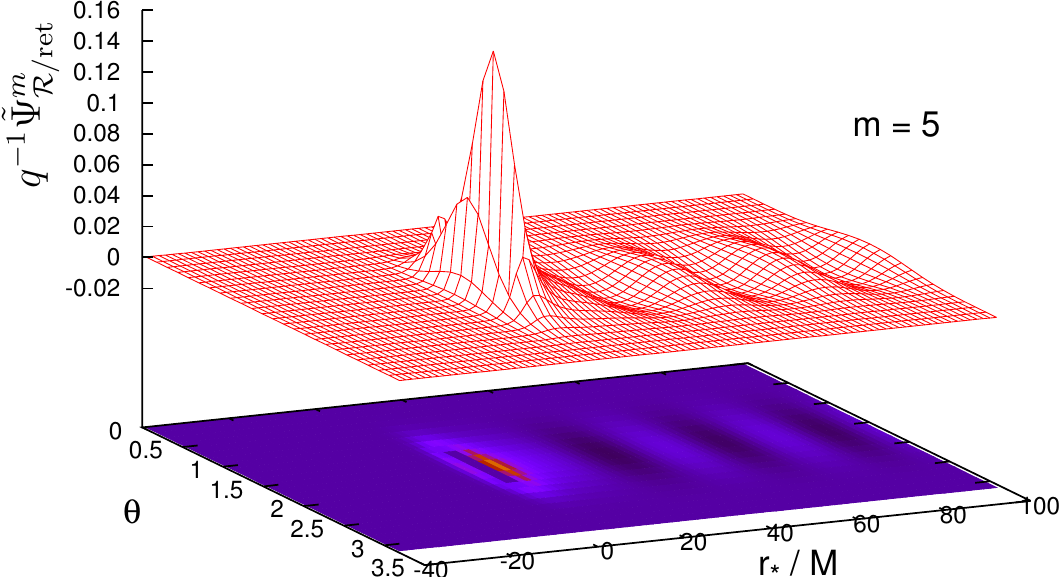}
  \includegraphics[width=8.0cm]{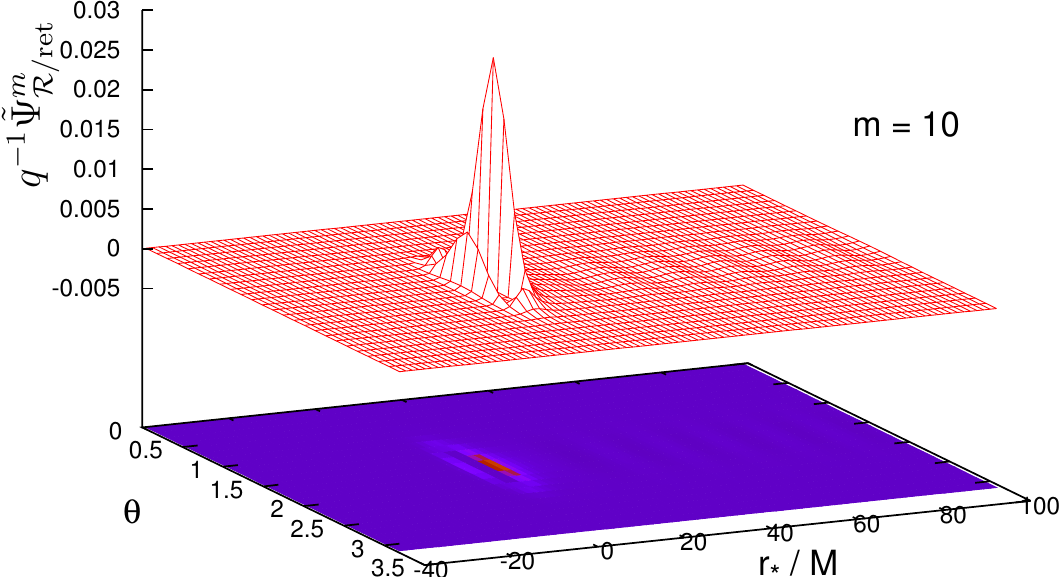}
 \end{center}
 \caption{Sample numerical results for the 4th-order puncture scheme at $r_0 = 7M$. Here we show $\Psirettil^m$ (outside the worldtube) and $\Psirestil^m$ (inside the worldtube) as a function of $r_\ast$ and $\theta$ at late time $t = 200M$ for $m=0$, $1$, $5$ and $10$. The worldtube is visible as a thin rectangle of fixed width $\twth$, $\twrs$ around the particle position at $\theta = \pi/2$, $r_\ast = r_\ast(r_0) \approx 8.83 M$.}
 \label{fig:slice3d}
 \end{figure}

Data on constant-$t$ surfaces can alternatively be visualised using 3D plots. Figure \ref{fig:slice3d} illustrates field modes as functions of $r_\ast$ and $\theta$, on constant-$t$ slices. Here, the worldtube is apparent as a thin interior rectangle. The central feature becomes sharper as $m$ is increased, with the solution becoming `flatter' at the poles (i.e.~near $\theta=0$ and $\pi$). The oscillations seen in Fig.~\ref{fig:slice3d} at large $r$ are outgoing waves emitted by the particle; they have a wavelength of $\sim 2 \pi / (m \omega)$.

\subsubsection{Discretization error, convergence tests and Richardson extrapolation\label{subsec:discretization-error}}

Figure~\ref{fig:slice3} illustrates how the system approaches a steady state once initial `junk' radiates away (we return to consider this point more carefully in Sec.~\ref{subsec:relaxation}). In the steady-state regime, we may extract estimates for the $m$-mode contributions on the worldline, Eqs.~(\ref{mode-contribution-1})--(\ref{mode-contribution-3}). The values obtained obviously depend upon the grid spacings, $h$ and $\Delta$. We set $h$ to be a simple fraction of $M$, i.e.~$h = M/\nr$ with $\nr$ an integer. Then, rather than varying $\{ h, \Delta \}$ separately, we fix the ratio $\Delta / h$. Here, we are limited by the stability condition (\ref{stability-modified}) which imposes an ($r_0$-independent) constraint upon $\Delta / h$. For convenience we choose $\Delta$, $h$ such that $\ar \equiv h \pi / (M \Delta)$ is a fixed integer (typically $\ar = 10$), and we determine $k$ (i.e.~the displacement in grid points of the numerical boundary from the poles) for each $m$ according to stability condition (\ref{stability-modified}). 

\begin{figure}
 \begin{center}
  \includegraphics[width=8.1cm]{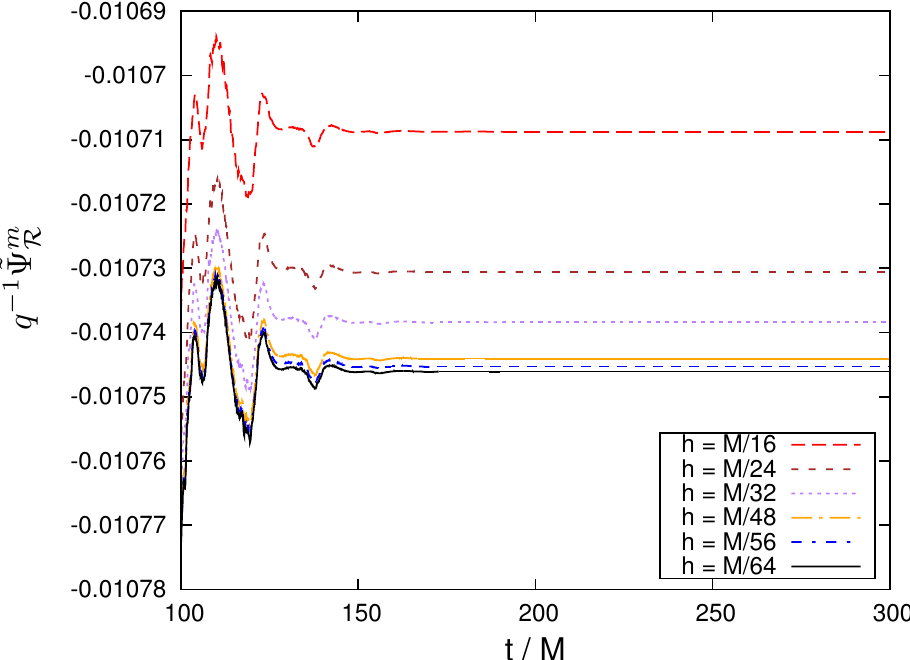}
  \includegraphics[width=8.1cm]{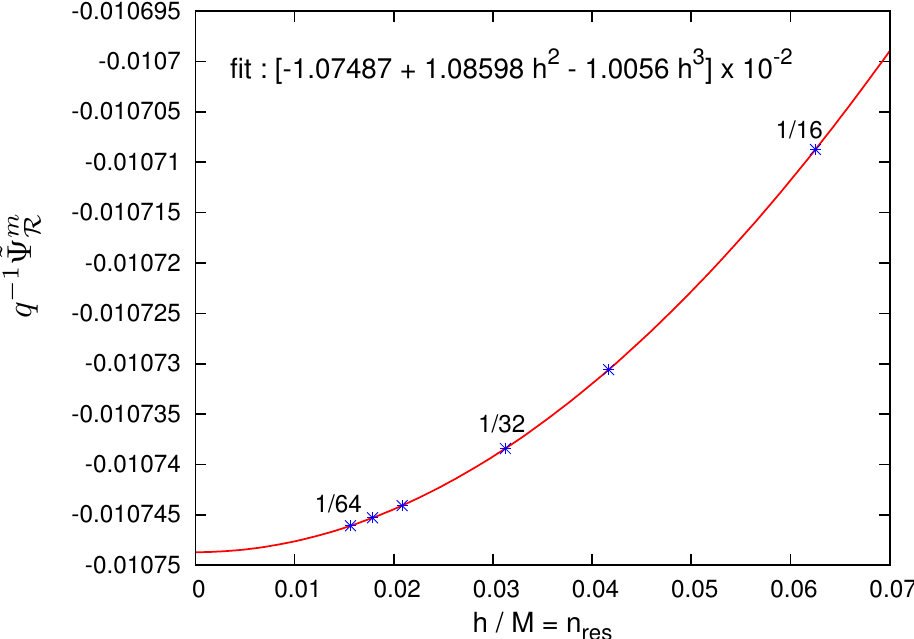}
 \end{center}
 \caption{Test of convergence. The left plot shows a mode of the residual field on the worldline, $\Psirestil^{m}$, as a function of time, for resolutions $\nr = 16, 24, 32, 48, 56$, and $64$ [here $h = M/\nr$, $\Delta = \pi/(10\nr)$, $m=2$, $r_0 = 6M$, with a 4th-order puncture]. The right plot shows the field extracted at late times ($t = 300M$) as a function of grid resolution. The line shows the best fit extrapolation model, $(-1.07487 + 1.08598 h^2 - 1.0056 h^3) \times 10^{-2}$. See also Fig.~\ref{fig:worldtube-error}.}
 \label{fig:extrapolation}
 \end{figure}

The left plot of Fig.~\ref{fig:extrapolation} illustrates the (4th-order) residual field as a function of time, for a range of resolutions $\nr = 16, 24, \ldots 64$. It suggests that the field converges towards a limiting curve as $\nr \rightarrow \infty$. We may test the convergence rate by taking ratios of the results of runs at different resolutions $\nr$. For example, consider the ratio
\beq
\ratio(h) = \frac{X(4h) - X(2h)}{X(2h) - X(h)} ,  \label{ratio-def}
\eeq
where $X \in \{\Psirestil^m, F_r^m, F_\varphi^m  \}$ and $X(kh)$ denotes the extracted result from a run with grid spacing $kh$. If the convergence rate is quadratic (i.e.~if the dominant term in the numerical error scales as $h^2$), then $\chi$ would approach the value of 4 as $h\rightarrow 0$; on the other hand, if the convergence is only linear then we expect $\chi \rightarrow 2$.  In Table \ref{table:convergence-ratios} we present sample values of $\ratio$ for both $\Psirestil^m$ and $F_r^m$ for the 2nd, 3rd and 4th-order puncture schemes. The data in Table \ref{table:convergence-ratios} shows convincingly that the 4th-order scheme is quadratically convergent (i.e. $\ratio \rightarrow 4$). The data for the 2nd and 3rd-order punctures is less conclusive. In \cite{Barack:Golbourn:2007} it was noted (for the 1st order puncture scheme) that the irregularity of $\Seff$ at the worldline disrupts the global quadratic convergence of our finite difference scheme. In Sec.~\ref{subsec:finite-diff-method} we described how the procedure for evaluating $\Seff^m$ in cells on the worldline is expected to introduce an additional term in the global discretization error which scales with $h^2 \ln h$. To test for the presence of this term, we construct the ratio
\beq
\ratiolog(h) = \frac{X(8h) - 5X(4h) + 4X(2h)}{X(4h) - 5X(2h) + 4X(h)} ,   \label{ratiolog-def}
\eeq
with $\ratiolog \rightarrow 4$ as $h \rightarrow 0$, if a $h^2 \ln h$ term is present. Table \ref{table:convergence-ratios} gives convincing numerical evidence in favour of the presence of an $h^2 \ln h$ term, at 2nd (field and SF) and 3rd order (SF only). 
 
\begin{table}

\begin{tabular}{l | r r | r r | r r}
\hline \hline
$\Psirestil^m$ & 2nd & & 3rd & & 4th & \\
$r_0 = 7M$  & $\ratio$ & $\ratiolog$ & $\ratio$ & $\ratiolog$ & $\ratio$  & $\bar{\ratio}$ \\
\hline
$m=5$ & \quad 3.30 & \quad 4.05  & \quad 3.92 & \quad 6.62 & \quad 3.85 & \quad 3.98 \\
$m=10$ & 3.24 & 4.16 & 3.95 & 8.87 & 3.97 & 4.00 \\
$m=15$ & 3.13 & 4.23  & 3.90 & 7.15  & 3.87 & 3.99 \\
\hline \hline
\end{tabular}

\begin{tabular}{l | r r | r r | r r}
\hline \hline
$F_r^m$ & 2nd  & & 3rd & & 4th & \\
$r_0 = 7M$  & $\ratio$ & $\ratiolog$ & $\ratio$ & $\ratiolog$ & $\ratio$  & $\bar{\ratio}$ \\
\hline
$m=5$ & \quad 3.39 & \quad 4.07 & \quad 4.25 & \quad 3.86 & \quad 3.96 & \quad 3.99 \\
$m=10$ & 3.08 & 4.11 & 4.72 & 3.87 & 3.92 & 3.99  \\
$m=15$ & 2.90 & 4.16 & 0.82 & 4.10 & 3.86 & 3.98 \\
\hline \hline
\end{tabular}
\caption{Sample convergence tests for $\Psirestil^m$ and $F_r^m$ for the 2nd, 3rd and 4th-order puncture schemes. The ratios $\ratio$ and $\ratiolog$ are defined in Eq.~(\ref{ratio-def}) and (\ref{ratiolog-def}). 
Here $\bar{\ratio}$ is an additional convergence ratio using points closer to the asymptotic regime, defined by $\bar{\ratio} = (135/52) [ F_r^m(\nr=48) - F_r^m(\nr=56)] / [ F_r^m(\nr=56) - F_r^m(\nr=64) ]$. For the 4th-order puncture, the ratios $\ratio$ and $\bar{\ratio}$ are $\sim 4$, implying that convergence is quadratic. Similar behaviour is apparent for the residual field $\Psirestil^m$  (but not the SF) at 3rd order. In all other cases shown ($\Psirestil$ and $F_r^m$ at 2nd order and $F_r^m$ at 3rd order) quadratic convergence is not clear. The data suggests that $\ratiolog \sim 4$ , which implies that global convergence is affected by the presence of an $\mathcal{O}(h^2 \ln h)$ term, due to the non-smoothness of the effective source on the worldline. Similar results are found for the angular component $F_\varphi^m$.  
}
\label{table:convergence-ratios}
\end{table}

In order to improve our estimates of the `physical' results, we used a fit model to extrapolate to $h \rightarrow 0$ (``Richardson's deferred approach to the limit'' \cite{NumericalRecipes}). As discussed above, the appropriate fit model depends on the order of the puncture. We use $X_{0} +  A h^2 + B h^2 \ln h + \mathcal{O}(h^3)$ for $\Psirestil^{[n=2]m}$ and $F_r^{[n=2,3]m}$, and $X_{0} +  A h^2 + \mathcal{O}(h^3)$ for all other cases. The fitting procedure is illustrated in the right panel of Fig.~\ref{fig:extrapolation}. The example shows that the 4th-order data for a field mode is well fitted by the simple model $\Psirestil^m (h) = \Psirestil^m (h=0) + Ah^2 + Bh^3$.  




\subsubsection{Worldtube error\label{subsec:worldtube-error}}
\begin{figure}
 \begin{center}
  \includegraphics[width=8.1cm]{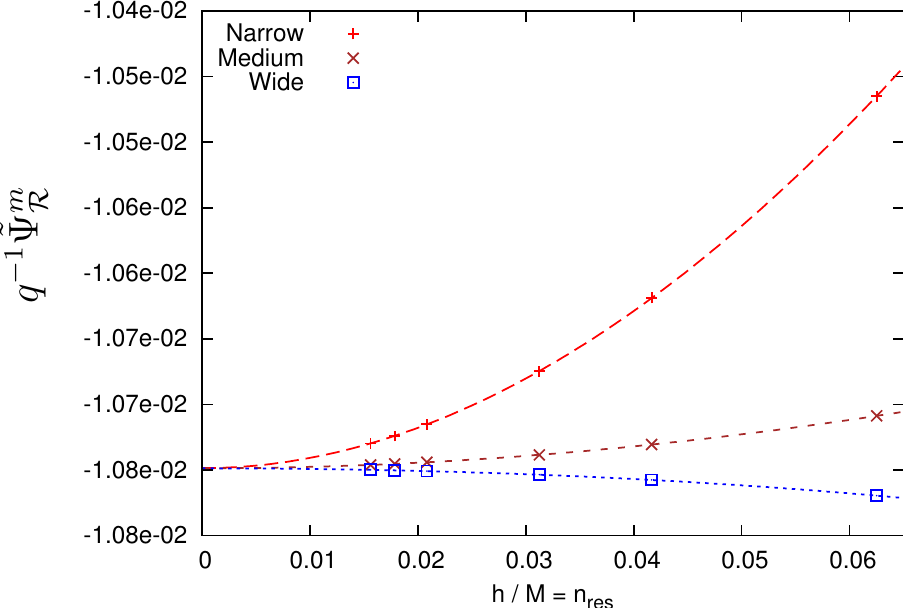}
  \includegraphics[width=8.1cm]{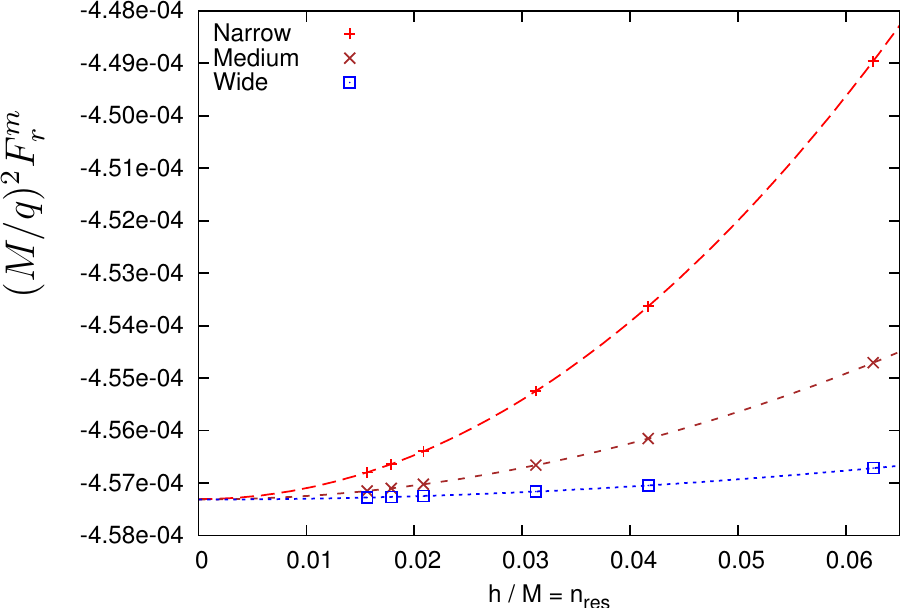}
 \end{center}
 \caption{Finite worldtube size effect and extrapolation to zero grid spacing. The plots show typical ($m=2$, $r_0=6M$, 4th-order puncture) modal contributions to the residual field (left) and radial SF (right) as a function of grid resolution ($h = M/\nr$, $\ar=10$), for worldtubes of three different widths, i.e.~(i) narrow: $\twrs = 1.25M$, $\twth = \pi/8$, (ii) medium: $\twrs = 2.5M$, $\twth = \pi/4$, and (iii) wide: $\twrs = 5M$, $\twth = \pi/2$. Numerical results from runs at six resolutions are shown ($h = M/\nr$ where $\nr = 24$, $32$, $48$, $56$, $64$ and $\ar = 10$) as data points.  
The best fits to the model $X_0 + A h^2 + Bh^3$ are shown as lines. The extrapolated values for the field (radial SF) vary only within $\sim 0.0004 \%$ ($\sim 0.0005\%$), which is considerably less than the relative error of $\sim 0.024 \%$ ($0.036\%$) obtained by comparing the highest resolution result with the extrapolated value. } 
 \label{fig:worldtube-error}
 \end{figure}

It is important to check that the dependence of the numerical results upon the dimensions of the worldtube (i.e., $\Gamma_{r\ast}$ and $\Gamma_\theta$) diminishes as $\nr \rightarrow 0$. 
Figure \ref{fig:worldtube-error} shows results from using three different worldtubes: narrow ($\twrs = 1.25M$, $\twth = \pi/8$), medium ($\twrs = 2.5M$, $\twth = \pi/4$) and wide ($\twrs = 5M$, $\twth = \pi/2$), as a function of grid resolution. The plots show that the magnitude of the discretization error (but not its scaling with $h$) depends on the worldtube dimensions. We would expect the magnitude of the discretization error to scale with the maximum absolute value of the numerical variable, which is typically the value of the field mode just outside the worldtube (see e.g. Fig.~\ref{fig:slice1,2}). Hence we expect that using a wider worldtube will generally decrease the magnitude of the grid resolution error, and this is what is observed in Fig.~\ref{fig:worldtube-error}. Note that arbitrarily large worldtubes are not practical however because (i) the computational expense of calculating $\Seff^m$ scales with the spatial cross-section of the tube, and (ii) $\Seff^m$ generally diverges far from the worldline, diminishing numerical accuracy.

\subsubsection{Source cancellation error\label{subsec:source-cancellation}}
The 4th-order effective source $\Seff^{[4]}$ is continuous across the worldline, and is zero on the worldline. It is calculated from the d'Alembertian of the puncture field, which is divergent at the worldline. The calculation of $\Seff$ near the worldline involves the delicate cancellation of large terms. This calculation is susceptible to numerical round-off error.

We found that, unmitigated, source cancellation error has greatest relative impact on high-resolution runs (which have a greater density of grid points in the vicinity of the worldline), and the error disrupts the smooth convergence to infinite resolution exhibited in Fig.~\ref{fig:extrapolation} and Fig.~\ref{fig:worldtube-error}. In consequence, unmitigated source cancellation error affects the validity of the extrapolation described in Sec.~\ref{subsec:discretization-error}. 
To deal with the problem, we used a symbolic algebra package ({\sc Maple}) to obtain an approximation for the source close to the worldline. First we introduced a scaling parameter $\lambda$ (see Sec.~\ref{subsec:puncture-order}) via $\delta x^{\mu} \equiv \lambda \delta \bar{x}^\mu$, and expanded the full expression for $\Seff$ in powers of $\lambda$ at $\lambda = 0$. We verified that (for the 4th-order puncture) the coefficients of the divergent terms (at orders $\lambda^{-3}, \lambda^{-2}, \lambda^{-1}$), as well as the constant term $\lambda^0$, are identically zero,
\beq
\Seff(\lambda \delta \bar{x}^\mu )  =  \lambda s_1( \delta \bar{x}^\mu ) + \mathcal{O}(\lambda^2) , \label{Seff-expansion}
\eeq
where $s_1$ is a $C^{-1}$ (i.e.~discontinuous but bounded) function of rescaled coordinate differences $\delta \bar{x}^\mu$.  Equation (\ref{Seff-expansion}) is not sensitive to large numerical round-off errors in the vicinity of the worldline. Using Eq.~(\ref{Seff-expansion}) very close to the worldline, and the full expression further away, significantly reduces the effect of source cancellation error. 

\subsubsection{Relaxation time error\label{subsec:relaxation}}

We wish to estimate the steady-state values for the mode-sum contributions. Obviously it is not possible to run the simulation for an infinite amount of time, and long runs are computationally expensive. With our 2+1D grid (Fig.~\ref{fig:grid}), doubling the physical simulation time ($\tmax \rightarrow 2 \tmax$) quadrupoles the run-time (i.e.~the CPU time). 
We explored a range of methods to obtain accurate estimates of steady-state values from simulations with finite 
$\tmax$, which we briefly describe below. We begin with an estimate of the approximate magnitude of the errors. 

\paragraph{Magnitude of error.}
Table \ref{table:relaxation-error} provides data on the approximate magnitude of the relaxation error in the lowest modes $m =0$, $1$, $2$, for orbits at $r_0 = 6M$ and $r_0 = 20M$. Let us define $\DelPhi^m = \Phirestil^m(t=300M) - \Phirestil^m(250M)$, i.e.~the difference between modal contributions `read off' at $t=300M$ and at $t=250M$, and define $\DelF^m$ in a similar way. We make the following simple observations: (i) $|\DelPhi^m|$ and $|\DelF^m|$ decrease in magnitude as $m$ increases, i.e.~the dominant error is in the $m=0$ mode. This is expected, since the $m=0$ mode contains the monopole which relaxes most slowly (see below); (ii) for the $m=0$ mode, the relative error $\DelPhi^{m=0} / \Phirestil^{m=0}$ is larger than the relative error $\DelF^{m=0} / F_r^{m=0}$; (iii) the absolute errors $\DelPhi^m$ and $\DelF^m$ increase somewhat in magnitude in going from $r_0 = 6M$ to $r_0 = 20M$; consequently, the relative errors $\DelPhi^{m} / \Phirestil^{m}$ and $\DelF^{m} / F_r^{m}$ are significantly worse at $r_0 = 20M$ than at $r_0 = 6M$ because the total field and SF diminish rapidly as $r_0$ increases. 

\begin{table}
\begin{tabular}{l | r r r}
\hline \hline
$r_0 = 6M$  & $m=0$ & $m=1$ & $m = 2$ \\
\hline
$\DelPhi^m$ &  $-1.6 \times 10^{-5}$  &  $9.6 \times 10^{-9}$  &  $6.3 \times 10^{-12}$    \\
$\DelPhi^m / \Phirestil^m$ &  $2.9 \times 10^{-3} $ & $1.8 \times  10^{-6} $ & $1.2 \times 10^{-9} $ \\
\hline
$\DelF^m$ & $1.5 \times 10^{-8}$ & $1.9\times 10^{-9}$ & $-8.0 \times 10^{-12}$ \\
$\DelF^m / F_r^m$ & $9.1 \times 10^{-5} $  & $1.1 \times 10^{-5}$ & $-4.7 \times 10^{-8}$  \\
\hline
\hline
$r_0 = 20M$ & & & \\
\hline
$\DelPhi^m$ & $-2.4 \times 10^{-5}$ & $-4.4 \times 10^{-7}$ & $-3.0 \times 10^{-9}$ \\
$\DelPhi^m / \Phirestil^m$ & $1.5 \times 10^{-2} $ & $-2.2 \times 10^{-4}$ & $6.2 \times 10^{-6}$ \\
\hline
$\DelF^m$  & $-3.9 \times 10^{-8}$ & $-2.4 \times 10^{-8}$ & $-1.6 \times 10^{-10}$ \\
$\DelF^m / F_r^m$ & $-1.4  \times 10^{-3}$ & $7.6  \times 10^{-4}$ & $-8.6  \times 10^{-5}$ \\
\hline
\hline
\end{tabular}
\caption{Sample data for relaxation error due to dissipation of junk radiation in low modes ($m = 0$, $1$, $2$) for $r_0 = 6M$ and $r_0= 20M$. Here we give numerical data for $\DelPhi^m = \Phirestil^m(t=300M) - \Phirestil^m(250M)$ and $\DelF^m = F_r^m( t=300M) - \Phirestil^m(250M)$, i.e~the difference between field and SF values extracted at $t=300M$ and $t=250M$, after extrapolation to $h \rightarrow 0$ has been performed. The data shows that (i) the magnitude of the error decreases with $m$, as expected from considering power-law decay; (ii) for $m=0$, the relative error in the field mode, $\DelPhi^{m=0} / \Phirestil^{m=0}$, is larger than the relative error in the radial SF mode, $\DelF^{m=0} / F_r^{m=0}$; (iii) for $m=0$, the absolute error increases somewhat with radius $r_0$; hence the relative error increases rapidly with $r_0$ (see text). 
}
\label{table:relaxation-error}
\end{table}

\begin{figure}
 \begin{center}
   \includegraphics[width=8.0cm]{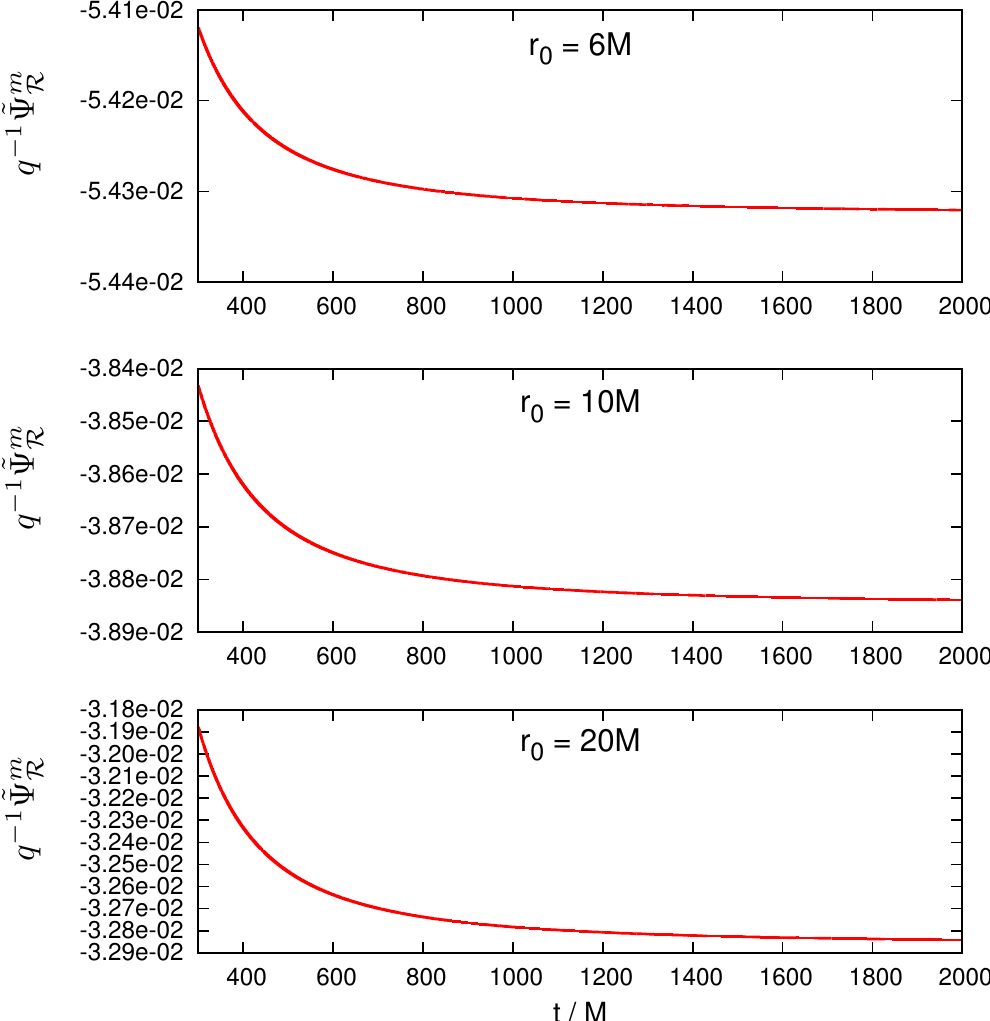}
  \includegraphics[width=8.0cm]{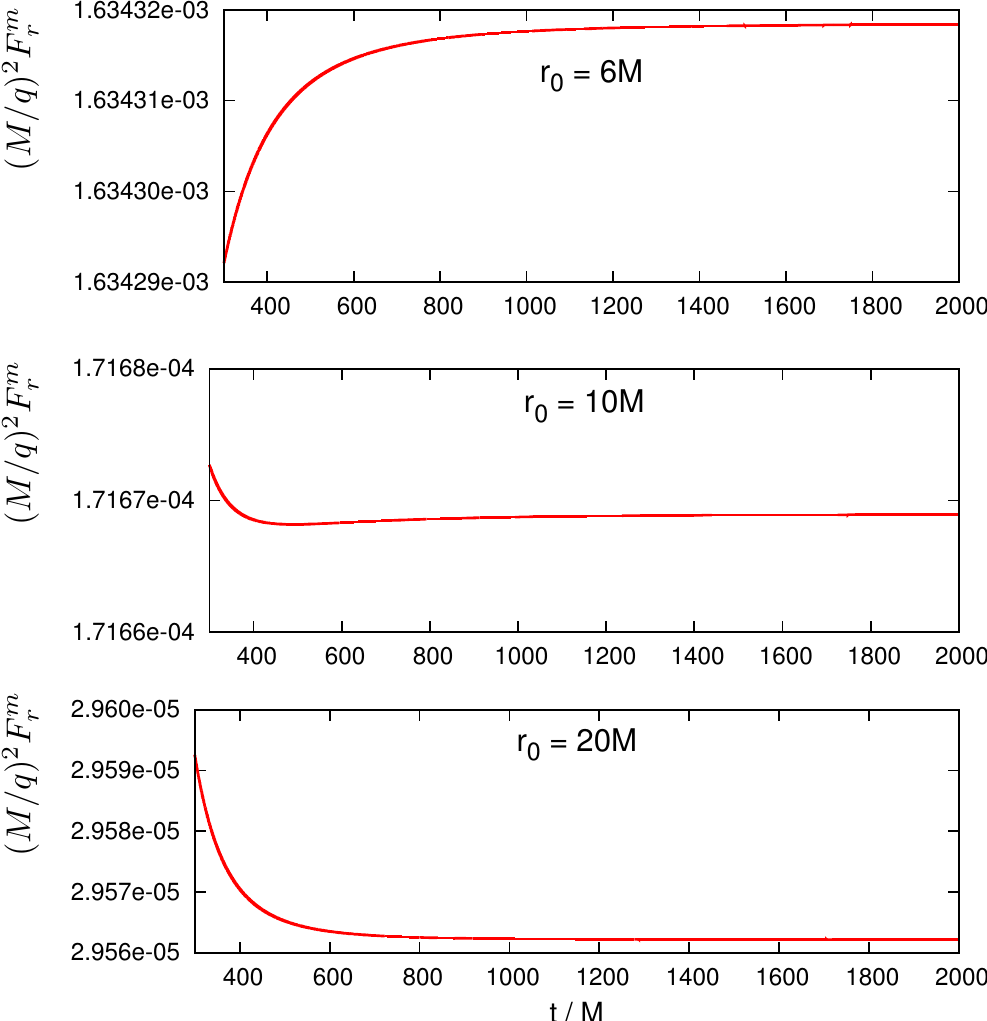}
 \end{center}
 \caption{Relaxation towards equilibrium of the $m=0$ modes for a range of orbital radii ($r_0 = 6M$, $10M$ and $20M$). The left plots show the $m=0$ mode of the field, $\Psirestil^{m=0}$, evaluated on the worldline, and the right plots show the same mode of the radial SF, $\Ftil^{m=0}_r$, as a function of time $t$, for a low-resolution long-time run ($\nr = 16$, $\ar = 10$, $\tmax=2000M$). In the appropriate late-time regime, the data are well-fitted by simple power-law relaxation models (see Fig.~\ref{fig:powerlaw}). 
 }
 \label{fig:m0relaxation}
\end{figure}

\begin{figure}
 \begin{center}
  \includegraphics[width=8.0cm]{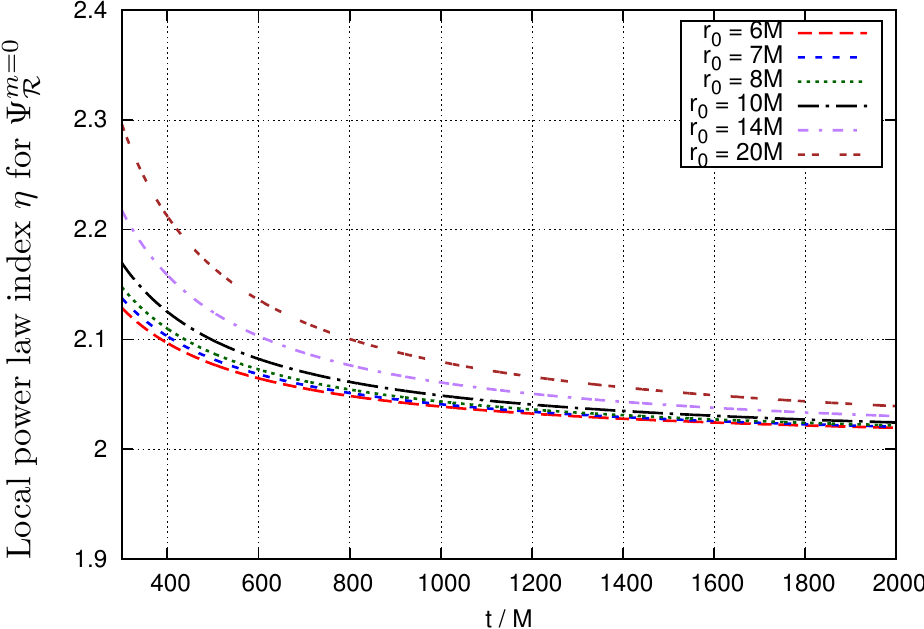}
  \includegraphics[width=8.0cm]{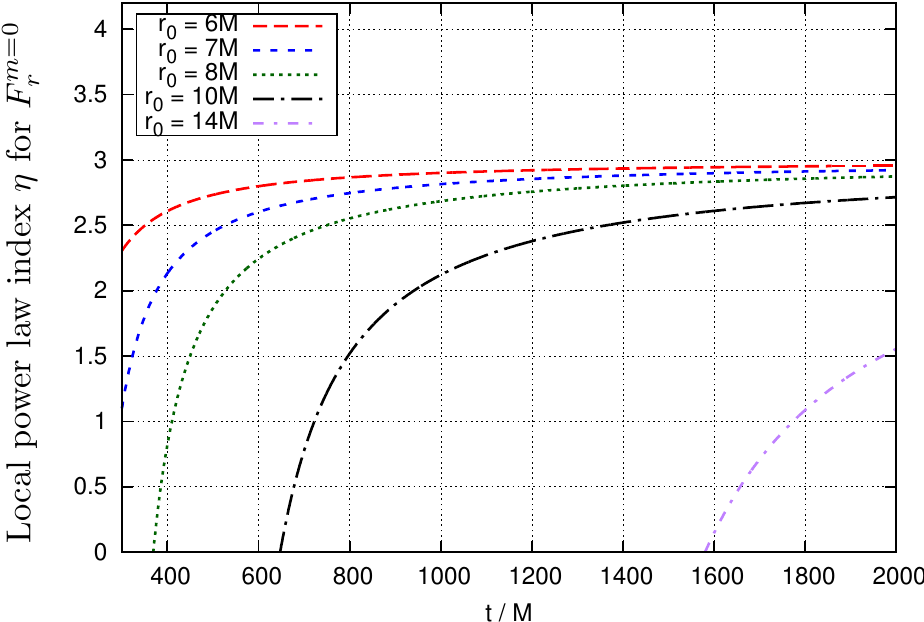}
 \end{center}
 \caption{Power law relaxation of the $m=0$ mode. The left-hand plot shows the local power-law index $\eta$ of the relaxation of the $m=0$ mode of the field, determined from $\eta(t) = - t \ddot{\Psi}^{m=0}_\mathcal{R} / \dot{\Psi}^{m=0}_\mathcal{R} - 1$ (where overdot denotes differentiation with respect to $t$, and derivatives are evaluated numerically). The right-hand plot shows $\eta(t)$ for the radial SF mode $F_r^{m=0}$. In the case of the field (left), the index tends towards $\eta = 2$, as expected for a non-compact $l=0$ perturbation. In the case of the radial SF (right), the index tends towards $\eta = 3$, although for large $r_0$ it takes a long time to reach this asymptotic regime. The higher index in the right plot (i.e. $\eta= 3$) is due to the fact that the slowest-decaying part of the monopole is spatially constant \cite{Barack:1999}.}
 \label{fig:powerlaw}
\end{figure}

\paragraph{Power-law relaxation.}
The data in Table \ref{table:relaxation-error} suggests that the $m=0$ mode relaxes to equilibrium most slowly, as expected. Figure~\ref{fig:m0relaxation} shows the relaxation of the $m=0$ mode of the residual field (left) and the radial SF (right) as a function of time, for a range of radii. The field exhibits a power-law relaxation, i.e.,
\beq
\Psirestil^{m=0}(t) = \Psirestil^{m=0}(t\rightarrow \infty)  + A t^{-\eta} + \mathcal{O}(t^{-\eta-1}).  \label{powerlaw-m0}
\eeq
It is expected from theory \cite{Price:1972} that the appropriate index for the monopole component of the $m=0$ mode (i.e.~for the $l=0$ multipole) is $\eta = 3$ if the initial junk radiation is localized in space and $\eta = 2$ otherwise. In this case, the latter index is applicable because the steady-state solution for $\Psiret^{m=0}$ tends to a non-zero value in the limit $r \rightarrow \infty$. Figure~\ref{fig:powerlaw} shows the numerically-determined local power-law index, defined by $\eta(t) = - t \ddot{\Psi}^{m=0}_\mathcal{R} / \dot{\Psi}^{m=0}_\mathcal{R} - 1$, plotted as a function of time. For all orbital radii, the local index asymptotes to $2$ in the late-time regime, as expected. 
Making use of this observation, we may minimise relaxation error in the $m=0$ mode of the field by fitting the numerical data to a power-law model (\ref{powerlaw-m0}), with $\eta=2$, to extract the steady-state value. Whilst this procedure is straightforward for the $m=0$ mode, it is more difficult for higher modes ($m > 0$) which also exhibit damped oscillations of frequency $m \omega$. However, it suffices for our purpose to fit only the $m=0$ mode, since this is by far the dominant source of relaxation error.

The relaxation of the $m=0$ mode of the radial SF also exhibits power-law decay, but in this case, the appropriate index is $\eta = 3$ as shown in the right-hand plot of Fig.~\ref{fig:powerlaw}. It turns out that the slowest-decaying $t^{-2}$ part of the monopole ($l=0$) perturbation in $\Phiret^{m=0}$ does not depend on radius [see Ref.~\cite{Barack:1999}, in particular Eq.~(89)] and as a result the relaxation of $F_r^{m=0}$ is one power of $1/t$ faster than naively expected. The right-hand plot of Fig.~\ref{fig:powerlaw} shows that the onset of the late-time regime, where power-law relaxation is manifest, increases with orbital radius. Unless one can evolve for very long time, fitting a simple power law to the radial SF generally does not give good results. Longer runs are computationally expensive, since the runtime and memory usage scales as $\tmax^2$. In practice, the computational burden associated with high-resolution ($\nr \gtrsim 64$, $\ar = 10$), long-time ($\tmax \gtrsim 1000M$) runs may be prohibitive. This leads us on to consider an alternative strategy.


\paragraph{Multigrid refinement.\label{subsec:multigrid}}
A complementary solution to fitting a power-law relaxation model is to use a kind of `mesh refinement'. The key idea here is to run the bulk of the simulation at low resolution (which is computationally cheap) and then improve the resolution in the late-time regime. In other words, we use the results of low-resolution runs to improve the initial data used in the final high-resolution run, which in turn reduces the amplitude of the final relaxation error. 
 
\begin{figure}
 \begin{center}
  \includegraphics[width=6.0cm]{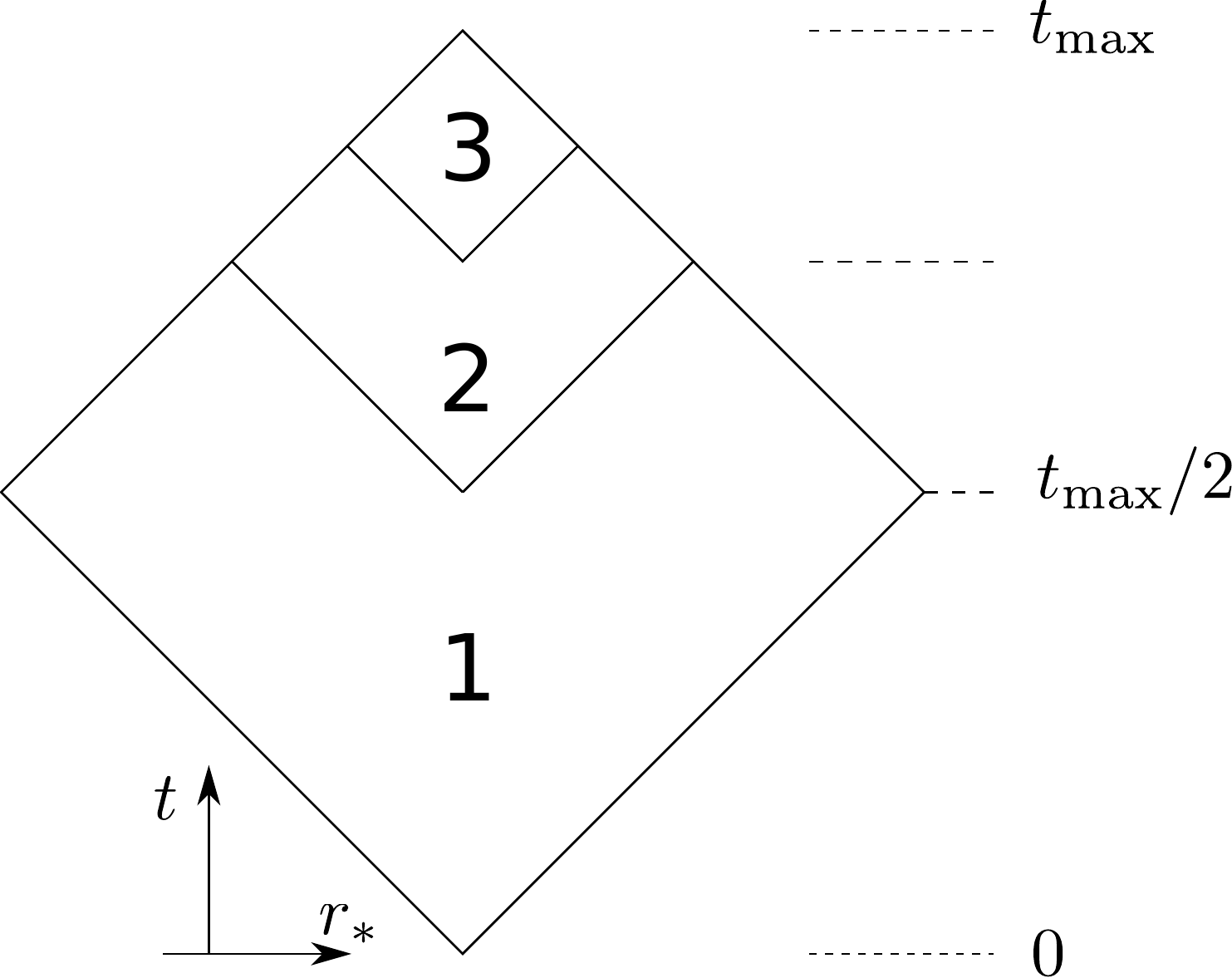}
 \end{center}
 \caption{Three-stage multigrid refinement method. A `crude' run (1) provides initial data for an `intermediate' run (2) which in turn provides initial data for a `fine' run (3).  Here we show a single slice of the grid at $\theta = \text{const}$.}
  \label{fig:grid-refinement-diagram}
\end{figure}

Figure \ref{fig:grid-refinement-diagram} gives an illustration of a three-stage process of grid refinement. Here the `fine' grid (3) has twice the resolution (in both radial and angular directions) of the `intermediate' grid (2), which in turn has twice the resolution of the `crude' grid (1).  A `crude' run (1) provides a rough estimate for the field everywhere in the largest grid. Initial data for the `intermediate' run (2) is then obtained by interpolating the values of the field read off along the initial boundary of grid 2. In a similar way, the intermediate run (2) then provides data for the fine run (3). The `fine' grid takes approximately twice as long to run as the intermediate grid, and four times as long as the crudest grid. It is much faster to run the multigrid scheme than to run the single large grid (1) at the highest resolution. The speed-up factor for a three-level grid is approximately $8^2 / (1+2+4) \approx 9$, and it is $8^3 / (1+2+4+8) \approx 34$ for a four-level grid.

\begin{figure}
 \begin{center}
  \includegraphics[width=8.0cm]{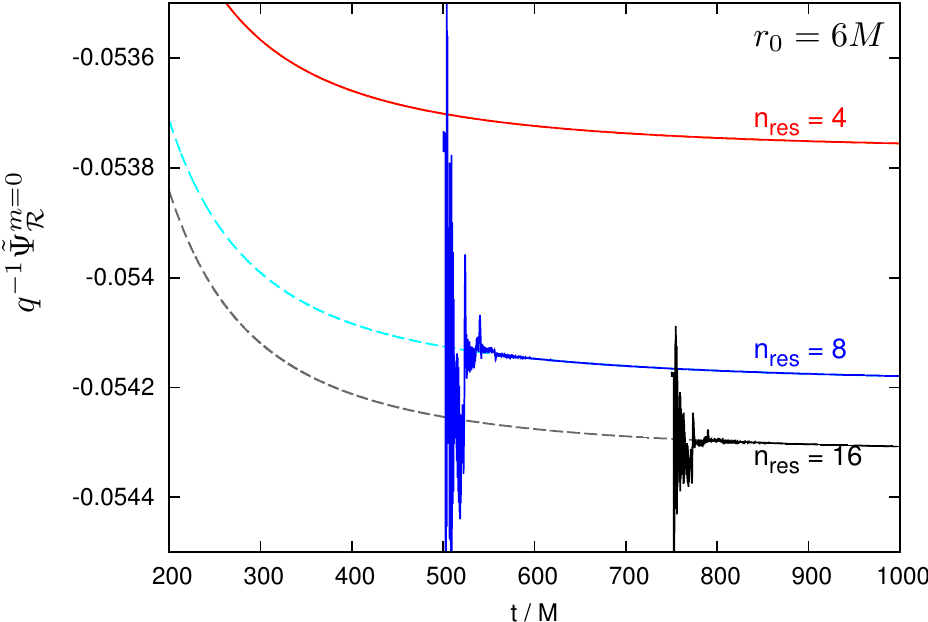}
  \includegraphics[width=8.0cm]{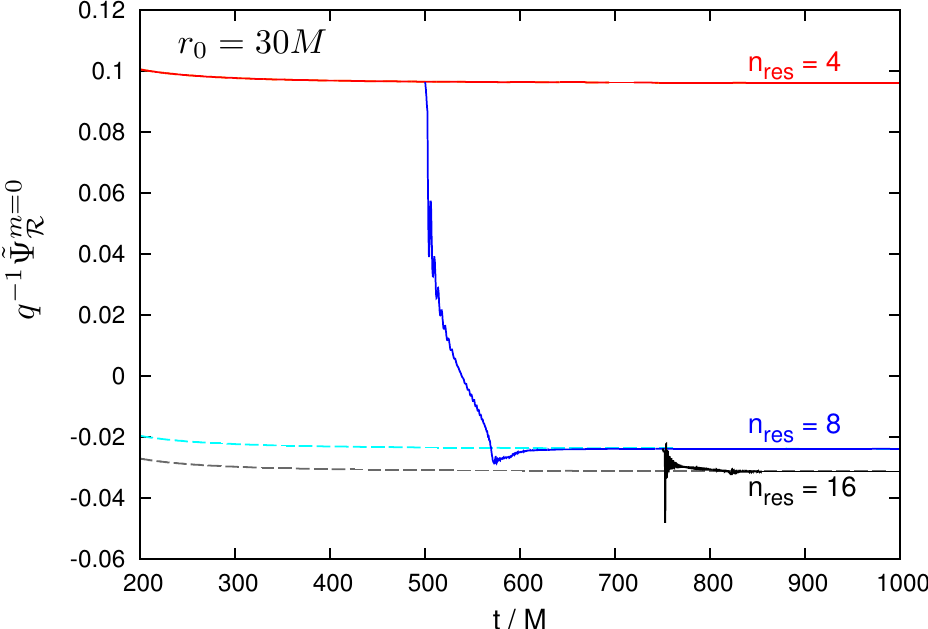}
  \includegraphics[width=8.0cm]{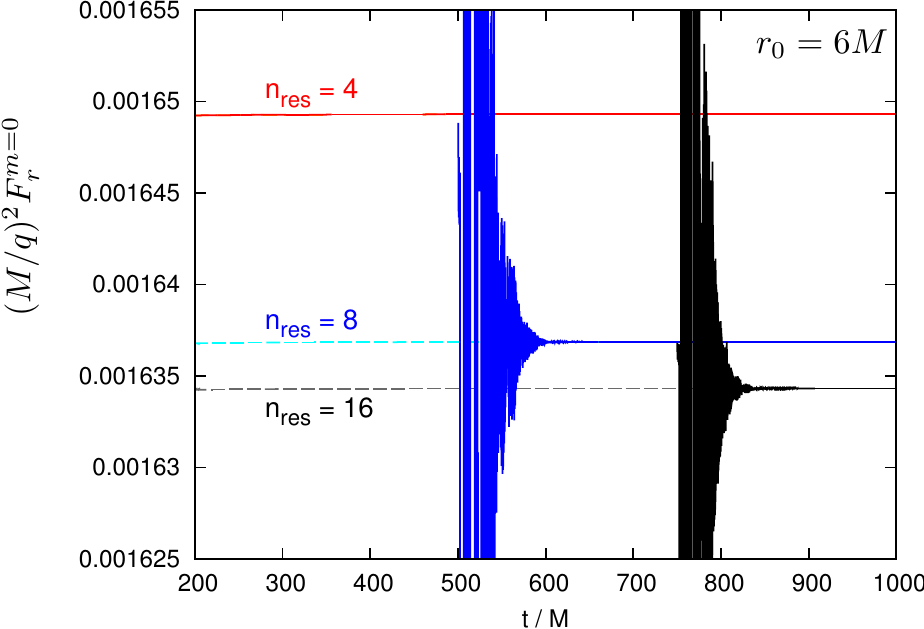}
  \includegraphics[width=8.0cm]{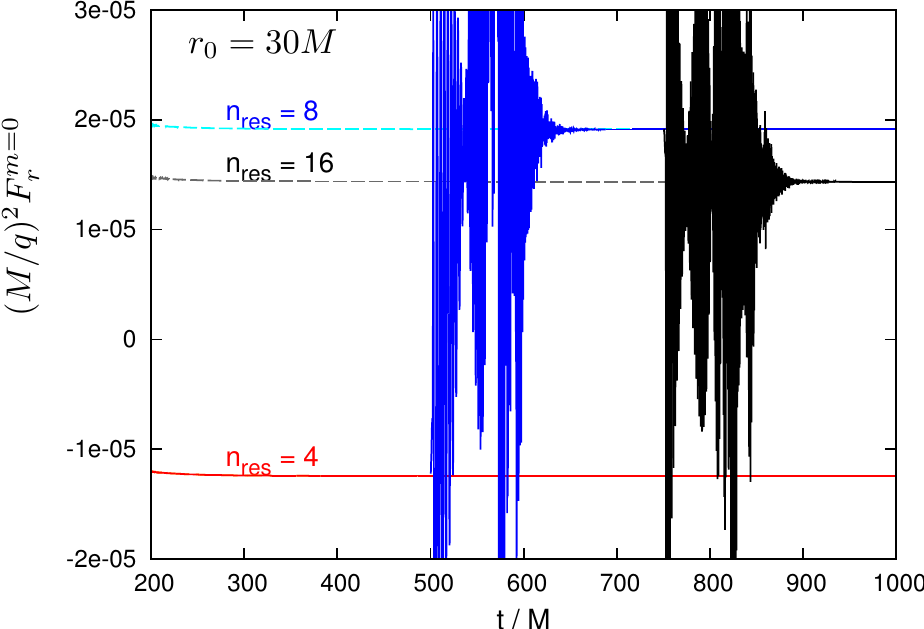}
 \end{center}
 \caption{Sample results comparing multigrid and unigrid evolutions of the $m=0$ mode. The left-hand plots show the evolution of the field mode $\Psirestil^{m=0}$ [upper] and radial SF mode $\Ftil_r^{m=0}$ [lower] at $r_0 = 6M$. The right-hand plots show similar data at $r_0 = 30M$. The dashed lines show unigrid evolutions at three different resolutions, $\nr = 4$, $8$ and $16$. The solid lines show the results of the multigrid scheme. Refinements in resolution occur at $t=500M$ ($\nr=4$ to $\nr=8$) and $t=750M$ ($\nr = 8$ to $\nr=16$). After a period of transition dominated by junk radiation, we find that the `refined' data asymptotes to the unigrid data. Note that the multigrid evolution is approximately $9$ times faster than the unigrid evolution at equivalent resolution.}
 \label{fig:multigrid-results}
\end{figure}

Figure \ref{fig:multigrid-results} shows typical results from a multigrid implementation, for orbits of radii $r_0 = 6M$ and $r_0 = 30M$. Here, the results of multigrid refinement are compared against unigrid results (which, as argued above, take much longer to run). Junk radiation is visible in the `intermediate' resolution which starts at $t = 500M$ and in the `fine' resolution which starts at $t=750M$. After the high-frequency junk has dissipated, at late times, the multigrid and unigrid results are found to be in close agreement. Importantly, the difference between multigrid and unigrid results at the same resolution is observed to be much less than the difference between unigrid runs at different resolutions. In other words, these plots demonstrate that the left-over error associated with refinement (e.g.~from the interpolation procedure) is much smaller than the discretization (`grid resolution') error, and that grid refinement is a useful technique which greatly diminishes the computational burden. The only constraint upon the scheme is that the finest grid must be large enough that the high-frequency junk arising from interpolation has time to dissipate. 

The simple method of multigrid refinement outlined here is crude in comparison with the more systematic adaptive mesh refinement (AMR) algorithm implemented by Thornburg \cite{Thornburg:2009} in 1+1D. Our hope is that an AMR scheme could  be applied to 2+1D simulations in the future \cite{Thornburg:inprog}.

\subsection{Mode sums\label{subsec:modesums}}
Now let us turn attention to the mode sums given in Eqs.~(\ref{Phi-mmode})--(\ref{F-mmode}) and their numerical calculation. 

\subsubsection{Large-$m$ asymptotics and convergence}
Let us consider the behaviour of the modal contributions $\Psirestil^\m$, $\Ftil^\m_r$ and $\Ftil^\m_\varphi$ in the large-$m$ limit. As we argued in Sec.~\ref{subsec:mmode}, their limiting behaviour depends on the order of the puncture scheme. In this section, we give numerical evidence in support of the following conclusions:
(i) $\Psirestil^\m \sim \mathcal{O}(m^{-2})$ for 2nd-order punctures, and $ \Psirestil^\m \sim \mathcal{O}(m^{-4})$ for 3rd and 4th-order punctures; (ii) $\Ftil^\m_r \sim \mathcal{O}(m^{-2})$ for 2nd and 3rd-order punctures, and $ \Ftil^\m_r \sim \mathcal{O}(m^{-4})$ for 4th-order punctures; and (iii) $\Ftil^\m_t , \Ftil^\m_\varphi \propto \exp(- \beta m )$, where $\beta$ is an $m$-independent positive constant which depends on orbital radius $r_0$. 
A heuristic explanation for these behaviours was given in Sec.~\ref{subsec:mmode}. 

\begin{figure}
\begin{center}
\includegraphics[width=10cm]{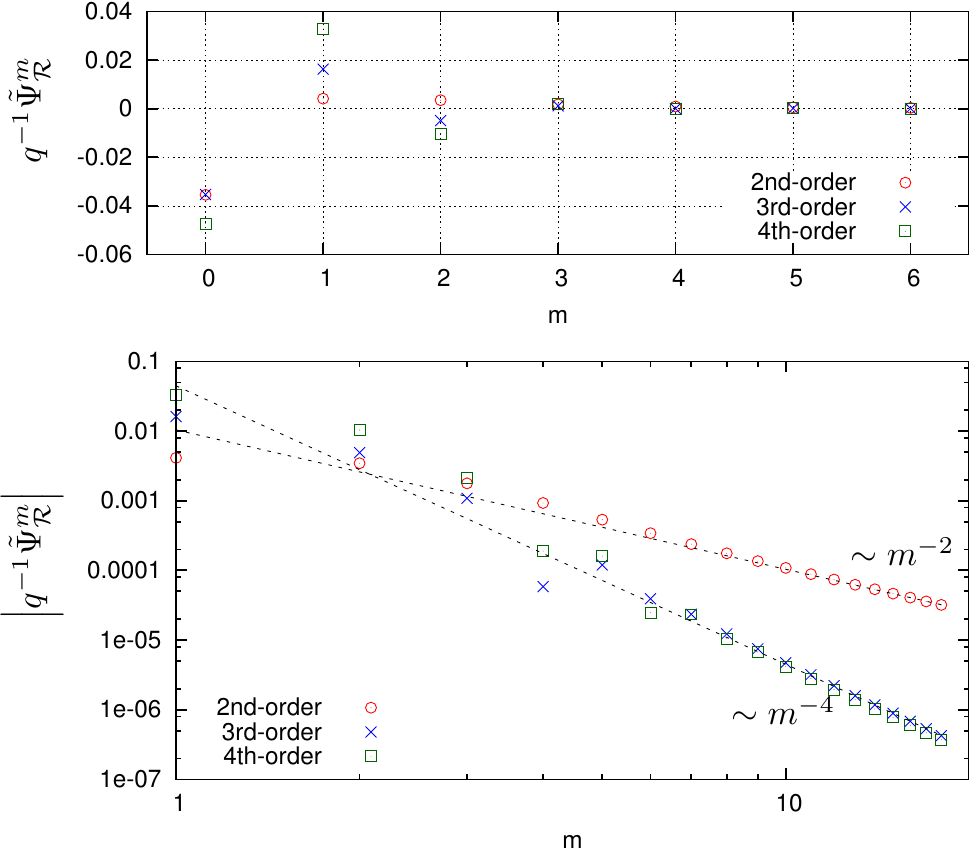}
\end{center}
\caption{Modes of the residual field, $\Psirestil^\m$, at $r_0 = 7M$, for puncture orders $n=2$, $3$ and $4$. The upper plot shows that the small-$m$ modes may take either sign. The lower plot shows power-law fall-off $\Psirestil^m \sim \mathcal{O}(m^{-\zeta})$ at large $m$, with exponent $\zeta = 2$ for the 2nd-order puncture, and $\zeta = 4$ for the 3rd and 4th-order punctures. The dotted lines are reference lines $\propto m^{-2}$ and $\propto m^{-4}$.}
\label{fig:modes-field}
\end{figure}

\begin{figure}
 \begin{center}
    \includegraphics[width=10cm]{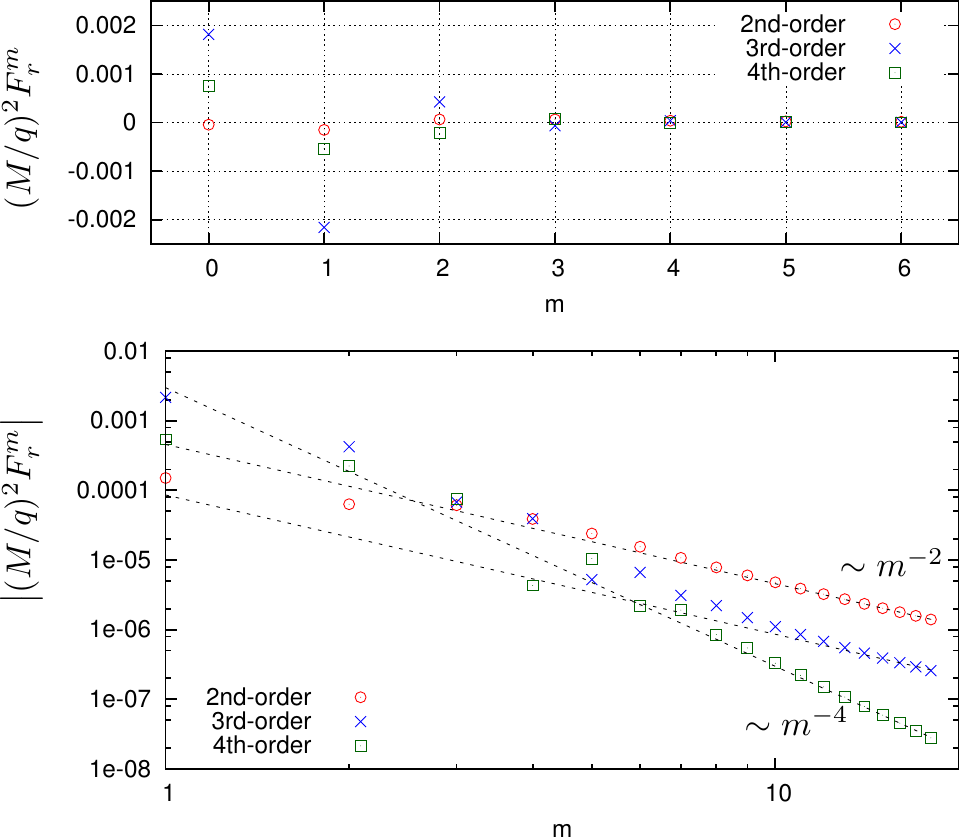}
 \end{center}
 \caption{Modes of the radial (conservative) SF component, $\Ftil_r^m$, at $r_0 = 7M$, for puncture orders $n=2$, $3$ and $4$. The upper plot shows that small-$m$ modes may take either sign. The lower plot shows power-law fall-off $\Ftil_r^m \sim \mathcal{O}(m^{-\zeta})$ at large $m$, with exponent $\zeta = 2$ for the 2nd and 3rd-order punctures, and $\zeta = 4$ for the 4th-order puncture. The dotted lines are reference lines $\propto m^{-2}$ and $\propto m^{-4}$.}
  \label{fig:modes-Fr}
 \end{figure}

Let us examine the magnitude of the modal contributions for a circular orbit of radius $r_0 = 7M$, for implementations of the 2nd, 3rd and 4th-order puncture schemes. Figure \ref{fig:modes-field} shows the modal contributions to the residual field, $\Psirestil^m$. The upper plot shows that the modal contributions can change sign; in particular, the $m=0$ and $m=1$ modes have opposite signs. The lower, log-log plot suggests the scaling $\Psirestil^m \sim \mathcal{O}(m^{-\zeta})$ in the large-$m$ limit, with $\zeta = 2$ for the 2nd-order puncture and $\zeta = 4$ for the 3rd and 4th-order punctures, as anticipated in Sec.~\ref{subsec:mmode} (see Table \ref{table-modesum-convergence}). Figure \ref{fig:modes-Fr} shows modal contributions $F_r^m$, for the conservative component of the SF. Here again we see in the upper plot that low-$m$ modal contributions can take either sign. We also see strong evidence for power-law convergence, i.e.~$F_r^m \sim \mathcal{O}(m^{-\zeta})$, with $\zeta = 2$ for 2nd and 3rd-order punctures, and $\zeta = 4$ for the 4th-order puncture. This, again, is consistent with the predictions of Sec.~\ref{subsec:mmode}.

\begin{figure}
 \begin{center}
    \includegraphics[width=10cm]{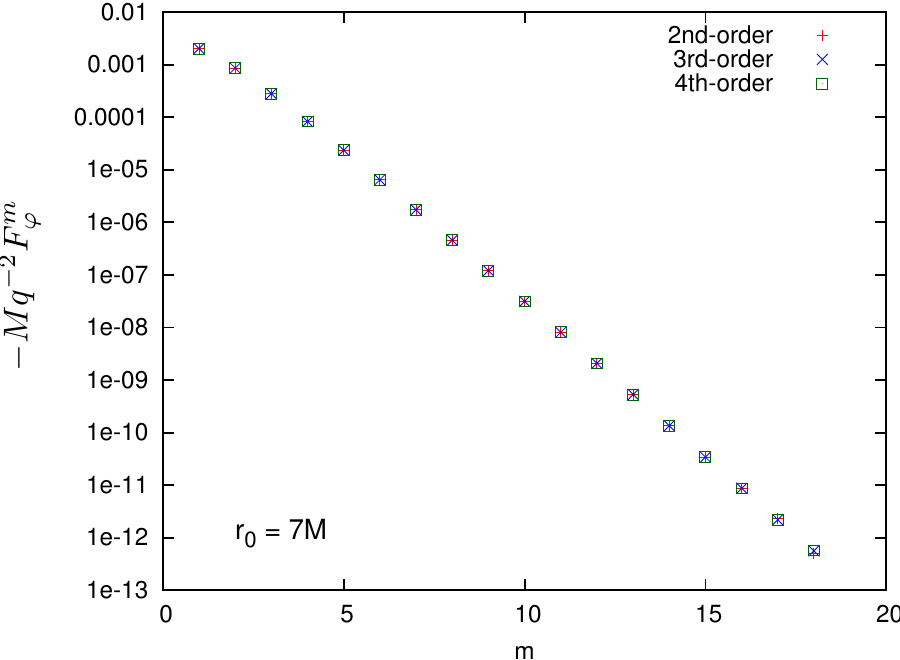}
 \end{center}
 \caption{Modes of the angular (dissipative) SF component, $F_\varphi^m$, at $r_0 = 7M$ for puncture orders $n=2$, $3$ and $4$. The plot shows the modal contributions $-F_\varphi^m$ on a semi-log scale. It illustrates that (i) the modes $F_\varphi^m$ do not depend on the order of the puncture to within numerical error (note that the points in the plot are superimposed upon one another), and (ii) the modal contributions diminish exponentially fast with $m$.}
 \label{fig:modes-Fphi}
 \end{figure}

Figure \ref{fig:modes-Fphi} displays the modal contributions to $\Fself_\varphi$, the dissipative component of the SF. The plot shows that the modal values are independent of the order of the puncture (up to numerical error), as foreseen in Sec.~\ref{subsec:exponential-convergence}. Furthermore, the modes exhibit a clear exponential convergence, $\Ftil^\m_\varphi \propto \exp(- \beta m )$ (with $\beta > 0$) at large $m$.
 
\begin{figure} 
  \begin{center}
    \includegraphics[width=10cm]{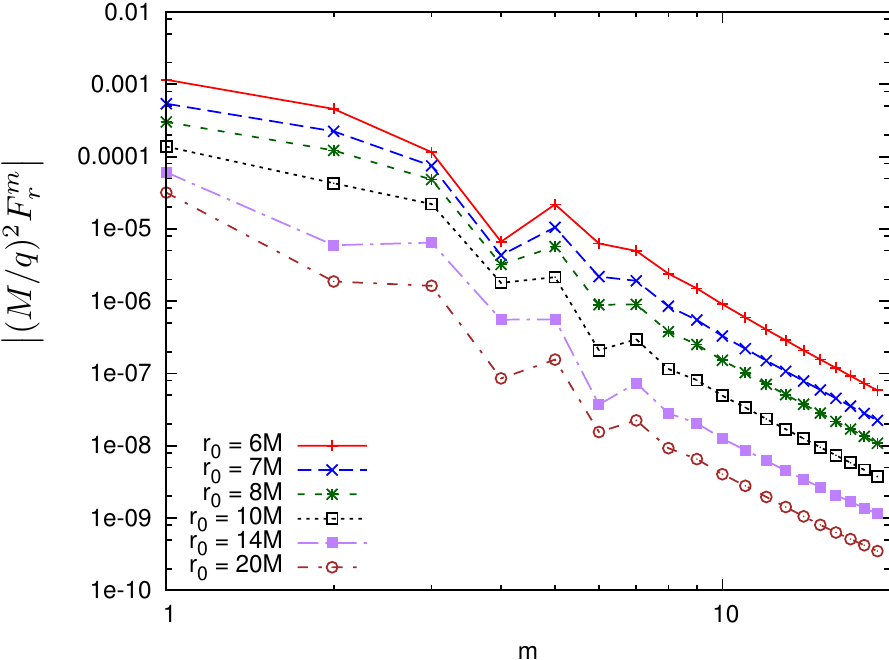}
  \end{center}
 \caption{Modal contributions to the radial component $\Fself_r$ for a range of orbital radii, with a 4th-order puncture.  Plotted here on a log-log scale is the absolute magnitude of $\Ftil^\m_r$ as a function of $m$. The modal contributions diminish in magnitude as $r_0$ increases (see also Fig.~\ref{fig:m0}), and it appears that all modes are scaled by approximately the same factor. For large $m$, the amplitude of the modal contributions falls off as $\sim m^{-4}$. The asymptotic regime begins around $m \sim 10$ for all radii.}
 \label{fig:modes-Fr-radius}
 \end{figure}

Finally let us examine the dependence of the modal contributions upon the orbital radius $r_0$, focussing on the 4th-order puncture. Figure \ref{fig:modes-Fr-radius} shows the magnitude of various modal contributions to the radial SF for a range of radii. The magnitudes of the modal contributions diminish with increasing $r_0$, though the relative contributions of different modes do not change substantially. The plot makes it clear that the `asymptotic regime', in which the modal contributions follow an inverse power law in $m$, begins at around $m \sim 10$ for all radii.
 
\begin{figure}
  \begin{center}
    \includegraphics[width=10cm]{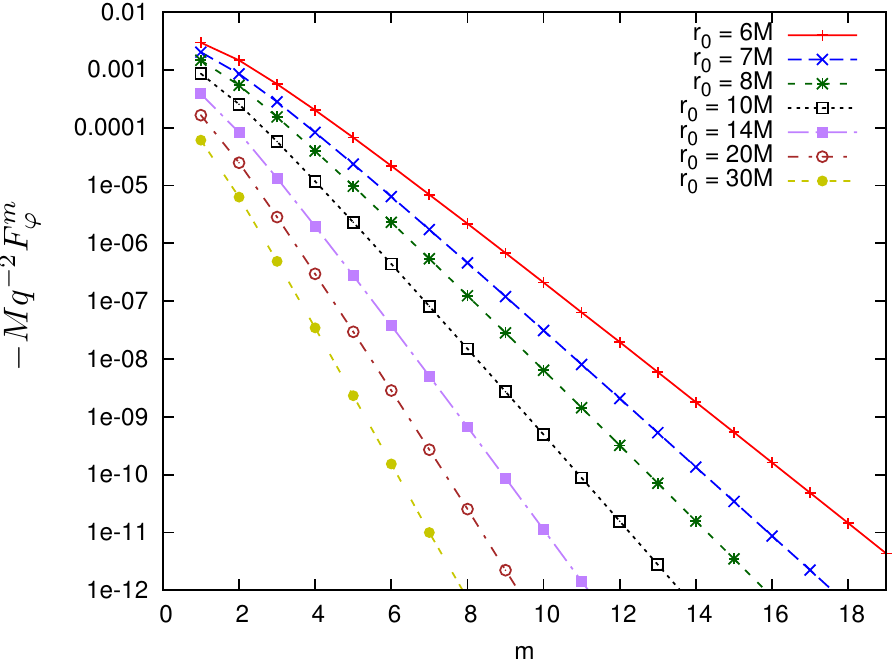}
  \end{center}
  \caption{Modal contributions to the angular component $\Fself_\varphi$ for a range of orbital radii. Plotted here on a semi-log scale is $-\Ftil^\m_\varphi$ (as all modes are negative) as a function of $m$. The modal contributions $\Ftil^\m_\varphi$ fall off exponentially with $m$ and the decay rate increases with $r_0$. }
  \label{fig:modes-Fphi-radius}
 \end{figure}

Figure \ref{fig:modes-Fphi-radius} shows the magnitude of modal contributions to the angular component $\Fself_\varphi$, for a range of radii. Again, the magnitudes of the modes diminish as $r_0$ is increased. Exponential decay of the modes with $m$ is seen for all radii, with the decay rate $\beta$ increasing with $r_0$.
 
\subsubsection{Mode summation and large-m fitting\label{subsec:mode-summation}}
For the field ($\Phi_R$) and conservative SF ($\Fself_r$) it is important to account for the modes in the large-$m$ `tail', whereas for the dissipative SF ($\Fself_\varphi$) this is not necessary (as the large-$m$ modes converge exponentially fast in the latter case).  For $X^m \in \{ \Phirestil^\m, \Ftil^\m_r \}$, we fit the simple power-law asymptotic model
\beq
X^m = m^{-\zeta} \left( A + B / m + C / m^2 + \ldots  \right) ,   \label{mmode-fit}
\eeq
where $A,B,C, \ldots$ are constant coefficients and $\zeta$ depends on the puncture order, as detailed at the start of Sec.~\ref{subsec:modesums}. We are free to choose the number of terms $N$ in the fit (\ref{mmode-fit}), and the part of the $m$-mode spectrum that we use for the fitting, i.e. $\mmin \le m \le \mmax$, provided that $N \le \mmax - \mmin + 1$. Typical values in our analysis are $N=3$, $\mmin = 12$, and $\mmax = 19$. We split the sum into two parts,
\beq
\sum_{m=0}^{\infty} X^m = \sum_{m=0}^{\mmax} X^m  +  \sum_{m=\mmax+1}^{\infty} X^m .
\eeq
The first sum is found by adding the numerically-determined modal contributions. The second sum is found by analytically summing the fit formula, Eq.~(\ref{mmode-fit}). The values of the fit parameters $A, B, C, \ldots$ depend somewhat on the set $\{ N, \mmin, \mmax \}$. By varying this set we may estimate the `mode summation error'. Sample values of this error are quoted in the Tables presented in Sec.~\ref{subsec:modesumresults}.

\subsubsection{Mode cancellation error\label{subsec:mode-cancellation}}
\begin{figure}
  \begin{center}
    \includegraphics[width=8cm]{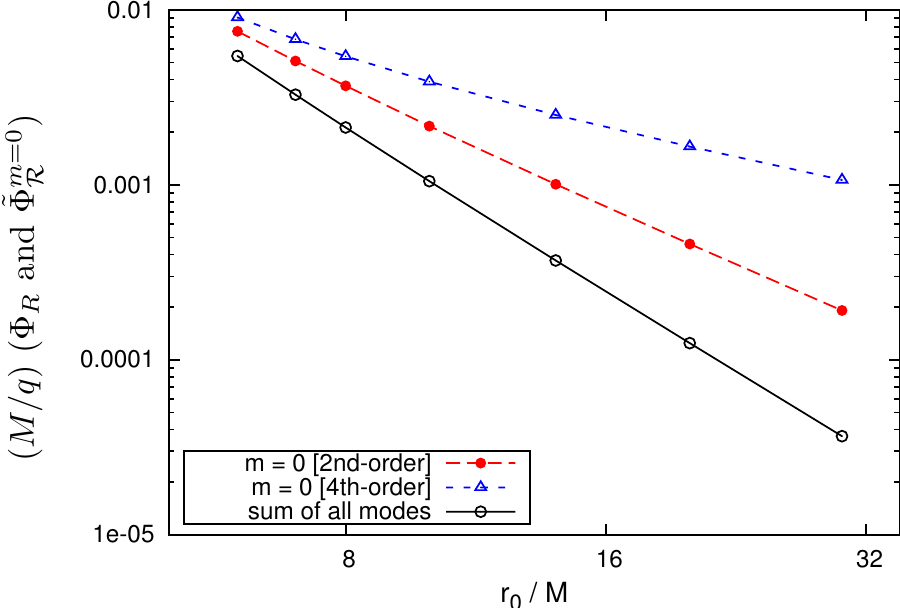}
    \includegraphics[width=8cm]{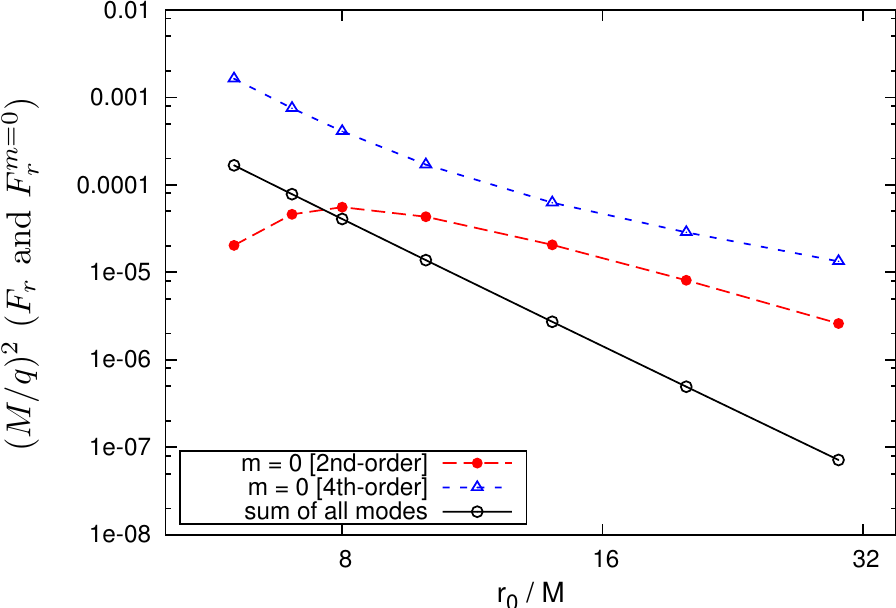}
  \end{center}
 \caption{Comparing the $m=0$ mode $\Phirestil^{m=0}$ (left plot) and $F_r^{m=0}$ (right plot) with the totals $\Phi_R$ and $\Fself_r$, across a range of orbital radii $r_0$. The dashed (red) line shows the magnitude of the monopole component for our 2nd-order puncture, and the dotted (blue) line shows the same for our 4th-order puncture. The solid (black) line shows the magnitude of the total radiative field (left) and radial SF (right), which scale as $r_0^{-3}$ and $r_0^{-5}$ (respectively) in the limit $r_0 \rightarrow \infty$. The ratios $\Phirestil^{m=0} /  \Phi_R$ and $\Ftil_r^{m=0}/\Fself_r$ increase as $r_0$ increases, and hence the accuracy of the mode sum degrades at large $r_0$ due to `mode cancellation error' (see text for discussion).}
  \label{fig:m0}
 \end{figure}

Figures \ref{fig:modes-field} and \ref{fig:modes-Fr} show that the modal contributions at small $m$ may take either sign, and that the magnitude of individual modes can be substantially larger than the total sum. This is illustrated in Fig.~\ref{fig:m0}, which compares the magnitude of the $m=0$ modal contribution (for 2nd and 4th-order punctures) to the magnitude of the total mode sum. For both the field and radial SF, the $m=0$ mode is substantially larger in absolute value than the total. The total field (radial SF) diminishes as $r_0^{-3}$ ($r_0^{-5}$) in the large-$m$ limit (see \cite{Diaz-Rivera:2004}), whereas the $m=0$ mode (and other modal contributions) are found to diminish far less rapidly. Hence the ratios $\Phirestil^{m=0} /  \Phi_R$ and $\Ftil_r^{m=0}/\Fself_r$ increase with $r_0$. At $r_0 = 30M$, the former ratio is $\sim 29$ and the latter ratio is $\sim 186$.

The phenomenon of cancellation between modes to leave a small remainder has a detrimental effect on the accuracy that can be achieved when computing the mode sum. At large $r_0$ we are reliant on delicate cancellations, with the total mode sum being orders of magnitude smaller than the $m=0$ contribution; this results in a relative error in the total mode sum that is much larger than the relative error in the individual modes. Unfortunately, this sets a practical limit upon the range of radii for which an accurate SF can be calculated using our version of the 4th-order puncture (Sec.~\ref{subsec:4thordpunc}). It may be possible to find an alternate version of the 4th-order puncture which alleviates this problem somewhat; this issue is certainly worth further investigation. Note, however, that we do not anticipate mode cancellation to be a significant problem in the case of gravitational SF calculations, because in the gravitational case the radial SF diminishes as $r_0^{-2}$ at large $r_0$ (as opposed to $r_0^{-5}$ in the scalar case).

 \subsection{Computational resource}
The computational workload in the $m$-mode regularization scheme is `embarrassingly parallel' in the sense that each run (i.e.~each 2+1D evolution for given $m, r_0, \intl$) may be assigned to a separate thread, and little or no communication is required between threads. The mode sums are computed by post-processing the results from multiple runs.
  
To run multiple threads in parallel, we made use of the {\sc Iridis} 3 HPC resource. To obtain a SF estimate at a given radius, we typically compute $20$ modes ($m=0,\ldots, 19$), at four different resolutions (e.g., $\nr=32$, $48$, $56$, $64$ with $\ar = 10$). Thus we require approximately $80$ nodes for each $r_0$. For fixed grid dimensions, the runtime scales as $\nr^3$; hence the $\nr=64$ run takes eight times longer than the $\nr=32$ run. With $\tmax=300M$ and $\nr=64$ a single run takes approximately $12$ hours. 

Additional resource is devoted to the $m=0$ and $m=1$ modes, to mitigate the problem of relaxation error (Sec.~\ref{subsec:relaxation}). Typically we ran these modes up to $t=1000M$ with a maximum resolution $\nr=64$ using the multigrid refinement scheme of Sec.~\ref{subsec:multigrid}. This illustrates a key flexibility of the $m$-mode scheme: the slow-decaying part of the initial junk is dominated by the lowest modes which can be handled separately.

 \subsection{Mode sum results\label{subsec:modesumresults}}

In this section we present sample numerical results for the SF and radiative field, obtained using our (debut)  implementation of the $m$-mode regularization scheme. Let us begin by considering results for the radial SF obtained with the 2nd, 3rd and 4th-order puncture schemes. Table \ref{table:Fr-orders} shows numerical data for the radial SF $\Fself_r$ for an orbital radius of $r_0 = 7M$, and compares with the frequency-domain $l$-mode results of \cite{Diaz-Rivera:2004}. As expected, the 4th-order scheme is most accurate and has the narrowest error bar, and the 2nd-order is least accurate and has the largest error bar. In the case of the 2nd-order puncture, the error  is dominated by the `tail-fitting error', i.e.~the error in summing the large-$m$ tail after fitting to an appropriate asymptotic model (see Sec.~\ref{subsec:mode-summation}). The tail-fitting error is large at 2nd-order for two reasons: (i) the tail decays slowly with $m$, as $\mathcal{O}(m^{-2})$, and (ii) the contribution from the modes in the high-$m$ tail represents a sizable proportion of the total (see Fig.~\ref{fig:modes-Fr}). To illustrate the latter point, in Table \ref{table:Fr-orders} we give the proportion of the total contained in the modes $m > 15$. We see that, although modes of the 3rd-order puncture also decay slowly, as $\mathcal{O}(m^{-2})$, the tail at 3rd order has a smaller magnitude than the tail at 2nd order ($\sim 31\%$ vs.~$\sim 5.4\%$), and hence the tail-fitting error in the 3rd-order result is commensurately smaller. In the case of the 4th-order puncture, the modes in the tail are rapidly-decaying [$\mathcal{O}(m^{-4})$], and the magnitude of the tail is small (see Fig.~\ref{fig:modes-Fr}); hence tail-fitting error is much reduced and, in fact, no longer the dominant source of error.
 
 \begin{table}
  \begin{tabular}{l | l r r r}
  \hline
  \hline
  Punc. order & \quad \ $(M/q)^2 \Fself_r$ & \quad Tail contribution & \quad Large-$m$ behaviour \\
  \hline
  2nd & \quad $7.86(7) \phantom{99} \e5$ & 31\% & $\quad \mathcal{O}(m^{-2})$ \\
  3rd & \quad $7.86(1) \phantom{99} \e5$ & 5.4\% & $\quad \mathcal{O}(m^{-2})$ \\
  4th & \quad $7.8507(3)\e5$ & 0.2\% & $\quad \mathcal{O}(m^{-4})$ \\
 \hline
    f-domain &  \quad $7.850679\,\e5$ \quad \\ 
  \hline
  \hline
  \end{tabular}
  \caption{Numerical results for the radial SF at $r_0=7M$. This table compares the results from implementations of 2nd, 3rd and 4th-order puncture schemes against the frequency domain results (final row) of Diaz-Rivera \etal~(see Table I in \cite{Diaz-Rivera:2004}). The digit in paratheses indicates the estimated error in the final digit quoted; for example, $7.86(7) \e5$ implies $(7.86 \pm 0.07)\e5$. The third column (`Tail contribution') lists the proportion of the total radial SF which comes from the sum of the modes $m > 15$, i.e.~the ratio $\sum_{m=16}^\infty \Ftil_r^m / \sum_{m=0}^\infty \Ftil_r^m$. The final column indicates the asymptotic behaviour of modes $F_r^m$ at large $m$ (see, e.g., Fig.~\ref{fig:modes-Fr}).}
   \label{table:Fr-orders}
\end{table}

In Table \ref{table:PhiR} we present numerical results for the radiative field $\Phi_R$ obtained via the 4th-order puncture scheme, for a range of orbital radii. The results shown in the second column were obtained by post-processing the results of multiple runs. We used unigrid runs up to $\tmax=300M$ for modes $m=2,\ldots,19$ at a range of resolutions $\nr = 32$, $48$, $56$, and $64$ with $\ar = 10$. We used the various resolutions to extrapolate to zero grid spacing (see Sec.~\ref{subsec:discretization-error}). To mitigate the relaxation error (Sec.~\ref{subsec:relaxation}), we ran the modes $m=0$ and $m=1$ up to $\tmax=1000M$ on a three-level multigrid (Sec.~\ref{subsec:multigrid}) with maximum resolution $\nr = 64$, and for $m=0$ fitted the late-time data ($t=900M$ to $1000M$) with the appropriate power-law relaxation model. To sum the large-$m$ tail (Sec.~\ref{subsec:mode-summation}), i.e.~the modes $m>19$, we fitted the modes $m = 12, \ldots, 19$ with a three-term model $Am^{-4} + Bm^{-5} + C^{m-6}$. Estimates of the residual errors that remain after performing these steps are given in the final three columns. We find that, although the residual errors are broadly similar in magnitude, the residual relaxation error remains the largest. 

In Tables \ref{table:Fr} and \ref{table:Fphi} we present numerical results for the conservative and dissipative components (respectively) of the SF, for a range of radii. The radial component of the SF, shown in Table \ref{table:Fr}, was computed in a similar manner to $\Phi_R$, that is, by post-processing the results of multiple runs to minimize discretization error (\ref{subsec:discretization-error}), relaxation error (\ref{subsec:relaxation}) and tail-fitting error (\ref{subsec:mode-summation}). The angular component of the SF, shown in Table \ref{table:Fphi}, was simpler to compute because in this case it was not necessary to model the large-$m$ tail, since the modal contributions $F_\varphi^m$ decay exponentially-fast (see Fig.~\ref{fig:modes-Fphi}). In addition, relaxation error is less significant for $\Fself_\varphi$ because the slowly-relaxing $m=0$ mode of $\Fself_\varphi$ is zero. We remind that the temporal component of the SF, $\Fself_t$, may be found directly from $\Fself_\varphi$ using (\ref{Ft-eq}). 

Tables \ref{table:PhiR}--\ref{table:Fphi} demonstrate that the $m$-mode method can yield highly-accurate SF estimates for circular orbits in the strong field (i.e. $6M \le r_0 \lesssim 10M$). For example, at $r_0 = 6M$, the conservative part of SF (i.e.~$\Fself_r$) is in error by one part in $\sim 5 \times 10^4$, and the dissipative part of SF (i.e.~$\Fself_\varphi$, $\Fself_t$) by one part in $\sim 5 \times 10^8$. This level of accuracy would not have been possible without careful modelling and mitigation of the sources of error described in the previous sections. Unsurprisingly, the results for $\Fself_\varphi$, whose large-$m$ modes converge exponentially fast, are substantially more accurate than the results for $\Fself_r$, whose large-$m$ modes exhibit power-law decay, $\mathcal{O}(m^{-4})$. 

It is clear from Tables \ref{table:PhiR}--\ref{table:Fphi} that the results of the $m$-mode method degrade in accuracy  as $r_0$ increases. For example, the relative error in our result for $\Fself_r$ at $r_0 = 30M$ is approximately $4500$ times greater than the relative error at $r_0 = 6M$. A loss of relative accuracy with radius can be explained from two points of view. Firstly, the magnitude of the singular part of the field (and thus the magnitude of the retarded field outside the worldtube, which is related to the magnitude of the discretization error) depends only weakly upon the orbital radius. In other words, the absolute error in our simulations depends only weakly upon $r_0$, and therefore, since the radial SF falls off very rapidly (as $r_0^{-5}$), the {\it relative} error in our results increases rapidly with $r_0$.  An alternative point of view is to see the relative loss of accuracy as a consequence of `mode cancellation error', which was described in Sec.~\ref{subsec:mode-cancellation}. Fig.~\ref{fig:m0} shows that the ratio of the magnitude of a typical mode to the total mode sum increases rapidly with $r_0$; hence small relative errors in individual modes may become large relative errors in the total. 

The loss of accuracy at large $r_0$ demonstrated here is not a particular concern to the prospect of accurate time-domain SF calculations, for two reasons. Firstly, in the more interesting case of the gravitational SF, the radial component of the SF falls off only as $r_0^{-2}$, rather than $r_0^{-5}$, and so we expect the loss of accuracy to be far less pronounced in that case. Secondly, we note that complementary approaches, such as Post-Newtonian methods, are well-suited to modelling orbits in the weak-field (large-$r_0$) regime, whereas the primary goal of the SF approach is to accurately describe EMRI physics in the strong-field  (small-$r_0$) regime.


  
    \begin{table}
  \begin{tabular}{l | l l |  l l l}
  \hline \hline
  &  \multicolumn{2}{c |}{$(M/q) \Phi_R$} & \multicolumn{3}{c}{Relative error estimates}   \\
  $r_0 / M$ &  $m$-mode, t-domain &  $l$-mode, f-domain \cite{Diaz-Rivera:2004}  & Relaxation & Discret. & Tail fit  \\
  \hline
  $6$  & $-5.45482(5)\e3$ \quad \quad & $-5.45480\e3$ & 8.4\e6 &  2.1\e6 & 4.2\e6\\
  $7$ & $-3.27533(5)\e3$ & $-3.27534\e3$ & 1.4\e5 & 3.4\e6 & 2.5\e6 \\
  $8$ & $-2.12750(4)\e3$ & $-2.12751\e3$ & 1.8\e5 & 5.3\e6 & 1.8\e6 \\
  $10$ & $-1.04976(3)\e3$ & $-1.04979\e3$ & 2.0\e5 & 1.1\e5 & 6.9\e7 \\
  $14$ & $-3.7014(4)\phantom{9}\e4$ & $-3.70065\e4$ & 6.5\e5 & 7.1\e5 & 3.2\e6 \\
  $20$ & $-1.244(1)\phantom{99}\e4$ & $-1.24673\e4$ & 8.7\e4 & 1.0\e4 & 3.3\e5 \\
  $30$ & $-3.59(3)\phantom{999}\e5$ & $-3.66171\e5$ & 6.9\e3 & 4.1\e4 & 7.6\e4 \\
  \hline \hline
  \end{tabular}
  
  \caption{Numerical results for the field $\Phi_R$ for a range of orbital radii. The second column gives the numerical result from our implementation of the time-domain $m$-mode method using the 4th-order puncture, a maximum resolution $\nr = 64$, $\ar=10$, unigrid runs with $\tmax = 300M$ for modes $m=0 , \ldots , 19$, and multigrid runs for modes $m=0$ and $m=1$ with $\tmax = 1000M$. 
  The third column quotes the (highly-accurate) frequency-domain $l$-mode method results of \cite{Diaz-Rivera:2004} for comparison. The remaining columns give estimates of sources of numerical error. The relaxation error (`Relaxation') was estimated from alternative extrapolations to infinite time for the $m=0$ mode, using $\nr=64$ multigrid data up to $t = 1000M$. The discretization error (`Discret.') was estimated by summing in quadrature the modal discretization errors found by comparing alternative extrapolations to infinite resolution based on data at $\nr=64,56,48,32$. The large $m$ tail-fitting error (`Tail fit') is an estimate of error in summing modes in the large-$m$ tail above $m > 19$ by using a fitting model $A m^{-4} + B m^{-5} + C m^{-6}$. These estimates suggest that relaxation is the dominant source of error in $\Phi_R$, for all radii. The error bar on the final result (parenthetical figures in the second column) was found by combining these error estimates in quadrature.}
  \label{table:PhiR}
 \end{table}

   
    \begin{table}
   \begin{tabular}{l | l l | l l l}
   \hline \hline
       & \multicolumn{2}{c |}{$(M/q)^2 \Fself_r$} & \multicolumn{3}{c}{Relative error estimates}  \\
    $r_0 / M$ & $m$-mode, t-domain & $l$-mode, f-domain \cite{Diaz-Rivera:2004} & Relaxation & Discret. &  Tail fit \\
    \hline
    $6$ & \quad $1.67731(4)\e4$ & \quad $1.67728\e4$ & 6.3\e6 & 1.2\e5 &  1.8\e5 \\
    $7$ & \quad $7.8507(3) \phantom{9} \e5$ & \quad $7.85068\e5$ & 5.8\e6 & 2.2\e5 & 2.5\e5 \\
    $8$ & \quad $4.0826(6) \phantom{9} \e5$ & \quad $4.08250\e5$ & 8.0\e6 & 3.8\e5 & 1.4\e4 \\
    $10$ & \quad $1.3785(7) \phantom{9} \e5$ & \quad $1.37845\e5$ & 1.2\e5 & 9.3\e5 &  5.2\e4 \\
    $14$ & \quad $2.721(3) \phantom{99} \e6$ & \quad $2.72008\e6$ & 2.5\e4 & 2.3\e4 &  9.6\e4 \\
    $20$ & \quad $4.96(3) \phantom{999} \e7$ & \quad $4.93790\e7$ & 5.5\e3 & 2.7\e3 &  2.9\e3 \\
    $30$ & \quad $8.5(8) \; \, \phantom{999}  \e8$  & \quad $7.17192\e8$ & 2.0\e2 &7.9\e2 &  6.5\e2 \\
    \hline \hline
   \end{tabular}
   
  \caption{Numerical results for the radial SF $\Fself_r$ for a range of orbital radii. 
  The second column gives the numerical result from our implementation of the time-domain $m$-mode method. The structure of the table is similar to that of Table \ref{table:PhiR}, and the same numerical parameters have been used. 
  }
  \label{table:Fr}
 \end{table}

 \begin{table}
   \begin{tabular}{l | r r }
  \hline
  \hline
    & \multicolumn{2}{c}{$(M/q^{2}) \Fself_\varphi$}  \\
  $r_0 / M$ & $m$-mode, t-domain  &  $l$-mode, f-domain \cite{Diaz-Rivera:2004} \\
  \hline 
  $6$   & $-5.30423170(3)\e3$ & \quad \quad $-5.30423170\e3$   \\ 
  $7$   & $-3.2731229(1) \phantom{9} \e3$ & \quad \quad $-3.27312280\e3$   \\
  $8$   & $-2.2111614(1) \phantom{9} \e3$ & \quad \quad $-2.21116134\e3$   \\ 
  $10$ & $-1.1859260(2) \phantom{9} \e3$ & \quad \quad $-1.18592599\e3$   \\ 
  $14$ & $-4.838491(2) \phantom{99} \e4$ & \quad \quad $-4.83849328\e4$   \\ 
  $20$ & $-1.92445(2)\phantom{999} \e4$ & \quad \quad $-1.92444253\e4$   \\ 
  $30$ & $-6.8629(3) \phantom{9999} \e5$ & \quad \quad $-6.86315934\e5$   \\   
  \hline \hline
  \end{tabular}
  \caption{Numerical results for the angular component of SF, $\Fself_\varphi$, for a range of orbital radii. This data were obtained with the same numerical parameters as in Tables \ref{table:PhiR} and \ref{table:Fr}. The third column shows a comparison with the results of \cite{Diaz-Rivera:2004}. The error here is predominantly discretization error, arising from extrapolation to zero grid resolution.
 }
   \label{table:Fphi}
 \end{table}


\section{Discussion and Conclusions\label{sec:conclusions}}

In the foregoing sections, we have presented details of the first implementation of the $m$-mode regularization scheme for SF calculations. Our objective has been to establish the scheme, first proposed in \cite{Barack:Golbourn:2007, Barack:Golbourn:Sago:2007}, as a practical alternative to (i) $l$-mode regularization in cases where separation of variables seems infeasible, e.g., gravitational SF on Kerr in the Lorenz gauge, and (ii) $3+1$D evolution schemes (see, e.g., \cite{Vega:Diener:Tichy:Detweiler:2009}). We have demonstrated here that (at least in the simple case of circular orbits on Schwarzschild) the results from $l$-mode and $m$-mode regularization are fully consistent, to within the limits of numerical accuracy. This work provides reassuring evidence that the $m$-mode scheme is well-founded and has been correctly implemented. 

To attain accurate results (such as those presented in Sec.~\ref{sec:results}), we were compelled to improve our understanding of a number of issues, including (i) the effect of puncture order upon the rate of convergence of the $m$-mode sum (Sec.~\ref{subsec:mmode} and \ref{subsec:modesums}), (ii) the stability and convergence of the finite difference scheme (Sec.~\ref{subsec:stability}), (iii) the influence of various sources of error upon numerical accuracy (Sec.~\ref{subsec:error-sources} and \ref{subsec:results-modes}), and (iv) the use of a simple mesh refinement algorithm to improve computational efficiency (Sec.~\ref{subsec:multigrid}). A crucial step forward was taken in moving from a 2nd-order puncture scheme (described in \cite{Barack:Golbourn:Sago:2007}) to a 4th-order puncture scheme. We believe that the 4th-order scheme will be the foundation for a range of $m$-mode implementations.

An obvious way to achieve greater accuracy is to increase the grid resolution. However, given that the scaling of runtime with resolution is problematic (runtime $\propto \nr^3$), more advanced methods may yet be needed. One possibility is to use a higher-order (e.g., 4th-order) finite-differencing scheme. A subtlety here is that global 4th-order convergence may be difficult to achieve, given the limited differentiability of the effective source on the worldline. Case-by-case treatment of finite difference molecules near the worldline will be necessary, but somewhat arduous. Another possibility is to employ a systematic adaptive mesh refinement scheme \emph{a la} Thornburg \cite{Thornburg:2009}. This is a promising route for the future, but one which will require a much more sophisticated code architecture. 

It is possible that the use of higher-order punctures may become feasible in future. Higher-order punctures would improve the power-law convergence of the $m$-mode tail of the conservative component of the SF, and generate an effective source which is smoother and flatter near the worldline. However, a naive implementation would face at least two difficulties: (i) computing the d'Alembertian of the puncture algebraically leads to an extremely cumbersome expression at high orders, and (ii) computing a finite source from a divergent puncture may rely on delicate cancellations between terms, a problem which (without care) will get worse at higher order (but see Ref.~\cite{Wardell:2010} for possible resolutions of these problems). For practical reasons, we believe the 4th-order puncture may represent a `sweet spot' since it is the lowest-order puncture to exhibit $m^{-4}$ convergence for the SF modes, which renders the tail-fitting error sub-dominant. To obtain improved convergence for the SF modes, one must go up to \emph{sixth} order.

An intriguing possibility is that calculations of `infinite order' (i.e., quasi-exact) punctures may become possible by, e.g., obtaining $U$ and $V$ in the Hadamard form (\ref{eq:Hadamard}) by integrating transport equations along the family of geodesics that join a field point to the worldline \cite{BarryWardell}. In this case the effective source would be identically zero, and all components of the SF would exhibit exponential convergence with $m$. See Ref.~\cite{Wardell:2010} for further discussion of this idea.

In this work, we have solved the wave equation in the region exterior to the horizon, using a tortoise coordinate and a $uv$ grid. This approach has the benefit of simplicity, in that boundary conditions are not required at the horizon or at spatial infinity. However, the $uv$ method has several drawbacks: (i) doubling the simulation time quadruples the grid area and thus the run-time, (ii) much runtime is `wasted' in the near-horizon regime of small-$r_\ast$, (iii) the reflection of `junk radiation' from large $r$ leads to rather slow power law relaxation. An alternative spacetime slicing, such as the asymptotically-null hyperboloidal slicing proposed in \cite{Zenginoglu:Tiglio:2009}, may bring advantages. The benefits of such a slicing for SF calculations were recently highlighted \cite{PeterDiener}. This example of `technology transfer' highlights the fact that many of the techniques routinely employed by numerical relativists could also benefit time-domain SF simulations. 
 
A companion paper (in progress) will describe the first implementation of the $m$-mode scheme for scalar-field SF on circular orbits of the Kerr spacetime. As the $m$-mode scheme is axisymmetric by construction, the method outlined here actually requires little modification. In the forthcoming work, we will focus only on the additional issues that arise, such as the problem of finding a stable finite-differencing method. Fortunately, we may test the accuracy of our implementation by comparing against recent $l$-mode results \cite{Warburton:Barack:2010}.

In progressing the $m$-mode scheme, we remain mindful of three open challenges for the future. The first challenge is to compute the gravitational SF in the Lorenz gauge. We hope to break new ground by examining the gravitational SF for circular orbits on Kerr within the $m$-mode scheme. Such a calculation has not been possible with the $l$-mode scheme, due to the apparent inseparability of the field equations in the Lorenz gauge. Of course, it may be possible to work in an alternative gauge; recent progress on a radiation gauge calculation is described in \cite{Keidl:Shah:Friedman:2010}. The second challenge is to adapt the scheme to treat highly eccentric or unbound orbits, which are beyond the scope of frequency-domain approaches. The third challenge is to achieve accurate, self-consistent long-term orbital evolutions of EMRIs using gravitational SF calculations. We refer the reader to Refs.~\cite{Pound:Poisson:2008, Pound:2010, Hinderer:Flanagan:2008} for first steps in this direction.  
 


\acknowledgments

SD acknowledges support from EPSRC through Grant No.~EP/G049092/1. LB acknowledges support from STFC through Grant No.~PP/E001025/1. We are grateful for the use of the {\sc Iridis} 3 cluster at the University of Southampton. We are indebted to Barry Wardell for many patient discussions, for providing explicit expressions for the 3rd and 4th-order punctures, and for checking a draft of this manuscript. We are grateful to Jonathan Thornburg for many stimulating discussions.

\appendix

\section{Toy model for the residual field: Details\label{appendix:toymodel}}
Here we show that the function $H^{[n]}$ introduced in Sec.~\ref{subsec:mmode} Eq.~(\ref{Fdef}), has an $m$-mode contribution given by Eq.~(\ref{toymodel-mmodes}). 

Although it is non-smooth at $\varphi=0$, $H^{[n]}$ is differentiable an infinite number of times in a piecewise sense. Hence we may express its derivatives in terms of distributions. In particular, near $\varphi=0$, its $(n+2)$th derivative has the expansion
\begin{equation}
H^{[n] (n+2)}= 2 h_n \delta''(\varphi)+ 2 h_{n+1} \delta'(\varphi)+ 2 h_{n+2}\delta(\varphi) 
+{\rm sign}(\varphi)(\cdots) .
\end{equation}
Here $\delta(\cdot)$ is the Dirac delta distribution, a prime denotes differentiation with respect to $\varphi$, and $(\cdots)$ represents a regular Taylor expansion in $\varphi$ about $\varphi=0$. We have made use of the distributional identities $\varphi / |\varphi| = 2 \Theta(\varphi) - 1$, $\Theta^\prime(\varphi) = \delta(\varphi)$, $\varphi^k \delta^{(k)}(\varphi) = k! (-1)^k \delta(\varphi)$ and $\varphi^k \delta(\varphi) = 0$ for integer $k > 0$. Note that the globally defined function 
\begin{equation}
\delta H^{[n](n+2)} \equiv H^{[n](n+2)}-2 \left[h_n \delta''(\varphi) + h_{n+1}\delta'(\varphi) + h_{n+2}\delta(\varphi)\right]
\end{equation}
is {\em bounded} in magnitude everywhere on $-\pi\leq\varphi\leq\pi$.

Now consider the $m$ mode coefficient of $H^{[n]}$, given (for $m\ne 0$) by
\begin{eqnarray}\label{split}
H^{[n] \m}&\equiv&\frac{1}{2\pi}\int_{-\pi}^{\pi} H^{[n]}e^{-im\varphi}d\varphi
=\frac{1}{2\pi(im)^{n+2}}\int_{-\pi}^{\pi}H^{[n](n+2)} e^{-im\varphi}d\varphi
\nonumber\\
&=&
\frac{1}{2\pi(im)^{n+2}}\int_{-\pi}^{\pi} \delta H^{[n](n+2)} e^{-im\varphi}d\varphi
\nonumber\\
&&+\frac{1}{\pi(im)^{n+2}}\int_{-\pi}^{\pi}
\left[h_n \delta''(\varphi)+h_{n+1}\delta'(\varphi)+h_{n+2}\delta(\varphi)\right]
e^{-im\varphi}d\varphi,
\end{eqnarray}
where in the second equality we integrated by parts $n+2$ times (note that we used the condition of smoothness across $\varphi = -\pi$, $\pi$ to eliminate boundary terms). Consider the first integral in the final expression: Since $|\delta H^{[n](n+2)} e^{-im\varphi}|$ is bounded on $-\pi\leq\varphi\leq\pi$, the magnitude of this integral can be bounded by $Cm^{-n-2}$ with some positive ($m$-independent) constant $C$. The second integral in the final expression is readily evaluated in explicit form. Altogether we get (for $m\ne 0$)
\begin{eqnarray}
\pi H^{[n] \m}&=&\frac{h_n}{(im)^n}+\frac{h_{n+1}}{(im)^{n+1}}
+\frac{h_{n+2}}{(im)^{n+2}} + \mathcal{O}(m^{-n-2})
\nonumber\\
&=&
\frac{h_n}{(im)^n}+\frac{h_{n+1}}{(im)^{n+1}} + \mathcal{O}(m^{-n-2}),  \label{Fkmmode}
\end{eqnarray}
where in the first line $\mathcal{O}(m^{-n-2})$ represents the contribution from the $\delta H^{[n](n+2)}$ integral in Eq.~(\ref{split}), and in the second line we have absorbed the $\propto h_{n+2}$ term within $\mathcal{O}(m^{-n-2})$.
Inserting (\ref{Fkmmode}) into the modal contribution formula (\ref{mmode-contribution-eq}) leads directly to Eq.~(\ref{toymodel-mmodes}).

\section{$m$-mode decomposition in the 2nd-order scheme\label{sec:appendix:second-order}}
The individual modes of the puncture field are obtained via integrals,
\beq
\Phipunc^{m} = \frac{1}{2\pi} \int_{-\pi}^{\pi} \Phipunc e^{- i m \varphi} d \varphi .
\eeq 
With $\Phipunc$ as defined in (\ref{punc-2nd-order}), and after the replacement in Eq.~(\ref{phi-replacement-2nd}), these integrals can be performed analytically, in a similar manner to \cite{Barack:Golbourn:2007}. We find
\beq
\Phipunc^m = \frac{q e^{-i m \omega t_p}}{2 \pi B^{1/2}} \gamma \left[ p_K^m(\p) \ellK(\gamma) + p_E^m(\p) \ellE(\gamma)  \right]
\eeq
where
\beq
\p^2 \equiv A / (4 B) ,    \quad \quad \gamma = \left( 1 + \p^2 \right)^{-1/2},   \label{rhotilde-def}
\eeq
and the quantities $A$ and $B$ are simply
\beq
A = \Prr \dr^2 + \Pth \dth^2 + \Qrr \dr^3 + \Qth \dr \dth^2 , \quad \quad B = \Pph + \Qph \dr,
\eeq
with coefficients $P_{ij}$ and $Q_{ij}$ defined in Eq.~(\ref{Pco-def}) and (\ref{Qco-def}). Here $\ellK(\cdot)$ and $\ellE(\cdot)$ are complete elliptic integrals of the first and second kinds, respectively, defined by
\beq
\ellK(k) = \int_0^{\pi/2} \left(1 -  k^2 \sin^2 x \right)^{-1/2} dx, \quad 
\ellE(k) = \int_0^{\pi/2} \left(1 -  k^2 \sin^2 x \right)^{1/2} dx .
\eeq
The polynomials $p_K^m(\p)$ and $p_E^m(\p)$ were given explicitly for $m=0, \ldots ,5$ in the tables of  Appendix A of \cite{Barack:Golbourn:2007} (note that our $\p$ plays the role of $\tilde{\rho}$ in Ref.~\cite{Barack:Golbourn:2007}). Polynomials for $m > 5$ are straightforward to calculate using a symbolic algebra package. 

The $m$ modes of the effective source are obtained via the integral
\beq
\Seff^{m} = \frac{1}{2\pi} \int_{-\pi}^{\pi} \Seff e^{- i m \varphi} d \varphi ,
\eeq
where $\Seff = S - \Box \Phipunc$. These integrals may be expressed analytically in the form
\beq
\Seff^m(r,\theta) = \frac{q}{2\pi} e^{-im \omega t_p} \left(S_1 I_1^m + S_2 I_2^m + S_3 I_3^m + S_4 I_4^m + S_5 I_5^m \right) ,  \label{Seff-m-2nd}
\eeq
where
\begin{eqnarray}
S_1 &=& (r-M) r^{-2} X(r,\theta) + f(r) \left( \Prr + 3 \Qrr \dr \right) + r^{-2} \left(1 + \dth \cot \theta \right)\left( \Pth + \Qth \dr \right) , \label{eqS1}  \\
S_2 &=& \left( r^{-2} \sin^{-2} \theta - \omega^2 / f(r) \right) B - 2 (r-M) r^{-2} \Qph ,  \\
S_3 &=& -3 f(r) \left( X(r,\theta) / 2 - \Qph \right)^2 - 3 r^{-2} \dth^2 \left( \Pth + \Qth \dr  \right)^2 , \\
S_4 &=& -3 B^2 \left( r^{-2} \sin^{-2} \theta - \omega^2 / f(r) \right) + 3 f(r) \Qph^2 ,  \\
S_5 &=& -3 f(r) \left( X(r,\theta) / 2 + \Qph \right)^2 - 3 r^{-2} \dth^2 \left(\Pth + \Qth \dr \right)^2 ,
\end{eqnarray}
and
\beq
X(r, \theta) = 2 \Prr \dr + 3 \Qrr \dr^2 + \Qth \dth^2 + 2 \Qph .
\eeq
In Eq.~(\ref{Seff-m-2nd}) the quantities $I_n^m$ are
\begin{eqnarray}
I_{1}^m &\equiv&  \int_{-\pi}^{\pi} \epsilon_P^{-3} \, e^{-i m \delta \varphi} d(\delta \varphi)   
 =  \frac{\gamma}{B^{3/2}} \left[ p_{1K}^m(\p) \ellK(\gamma) + \p^{-2} p_{1E}^m(\p) \ellE(\gamma) \right]   ,\label{integralI1}  \\
I_{2}^m &\equiv&  \int_{-\pi}^{\pi} \epsilon_P^{-3} \, \cos \delta \varphi \, e^{-i m \delta \varphi} d(\delta \varphi)  
 =  \frac{\gamma}{B^{3/2}} \left[ p_{2K}^m(\p) \ellK(\gamma) + \p^{-2} p_{2E}^m(\p) \ellE(\gamma) \right]  ,  \\
I_{3}^m &\equiv&  \int_{-\pi}^{\pi} \epsilon_P^{-5} \cos^2 \left( \frac{\delta \varphi}{2} \right) \, e^{-i m \delta \varphi} d(\delta \varphi)  
 =  \frac{\gamma}{\p^2 B^{5/2}} \left[ p_{3K}^m(\p) \ellK(\gamma) + \p^{-2} p_{3E}^m(\p) \ellE(\gamma) \right]  , \label{integralI3}  \\
I_{4}^m &\equiv& \int_{-\pi}^{\pi} \epsilon_P^{-5} \sin^2 \left( \delta \varphi \right) \, e^{-i m \delta \varphi} d(\delta \varphi) 
 =  \frac{\gamma}{B^{5/2}} \left[ p_{4K}^m(\p) \ellK(\gamma) + \p^{-2} p_{4E}^m(\p) \ellE(\gamma) \right] ,  \\
I_{5}^m &\equiv& \int_{-\pi}^{\pi} \epsilon_P^{-5} \sin^2 \left( \frac{\delta \varphi}{2} \right) \, e^{-i m \delta \varphi} d(\delta \varphi) 
=  \frac{\gamma^3}{B^{5/2}} \left[ p_{5K}^m(\p) \ellK(\gamma) + \p^{-2} p_{5E}^m(\p) \ellE(\gamma) \right]  .\label{integralI5}
\end{eqnarray}
Note that $I_1^m$, $I_2^m$ and $I_4^m$ are the same integrals as defined in Eqs.~(46), (49) of \cite{Barack:Golbourn:2007} but with $\Pph$ replaced by $B$ and with $\tilde{\rho}^2$ replaced by $A / (4B)$. Note also that we have introduced new definitions for $I_3$ and $I_5$. The polynomials $p_{1K}^m$, $p_{1E}^m$, $p_{2K}^m$, $p_{2E}^m$, $p_{4K}^m$ and $p_{4E}^m$ were given for $m=0,\ldots,5$ in Appendix A of \cite{Barack:Golbourn:2007}. The polynomials $p_{3K}^m$, $p_{3E}^m$, $p_{5K}^m$, and $p_{5E}^m$, again for $m=0, \ldots, 5$, are given in Table \ref{table:p-polynomials} here. Polynomials for $m > 5$ were calculated using a symbolic algebra package. 


\begin{table}
\begin{tabular}{l l}
\hline
\hline
$m$ & $p_{3K}^m(\rho)$ \\
$0$ & $ -\frac{1}{24}$ \\
$1$ & $\frac{1}{24}\left(4 \p^2 - 1\right)$  \\
$2$ & $\frac{1}{24}\left( 64 \p^4 + 40\p^2 - 1 \right) $ \\
$3$ & $\frac{1}{24}\left(512\p^6 + 640\p^4 + 180\p^2 - 1\right)$ \\
$4$ & $\frac{1}{120}\left( 16384\p^8 + 29696\p^6 + 16512\p^4 + 2720\p^2 - 5 \right)$ \\
$5$ & $\frac{1}{840}\left( 655360\p^{10} + 1540096\p^8 + 1263104 \p^6 + 418688\p^4 + 45500\p^2 - 35\right)$  \\
\hline
\vspace{0.2cm} \\
\hline
$m$ & $p_{3E}^m(\rho)$ \\
$0$ & $ \frac{1}{24} \left( \p^2 + 2 \right)$ \\
$1$ & $ -\frac{1}{24} \left( 4 \p^4 + \p^2 - 2 \right)$ \\
$2$ & $ -\frac{1}{24} \left( 64\p^6 + 72\p^4 + 7\p^2 - 2 \right)$ \\
$3$ & $ -\frac{1}{24} \left( 512\p^8+896\p^6+404\p^4+17\p^2-2 \right)$ \\
$4$ & $ -\frac{1}{120} \left( 16384\p^{10} + 37888\p^8 + 28288\p^6 + 6944\p^4 + 155\p^2 - 10\right)$ \\
$5$ & $ -\frac{1}{840} \left( 655360\p^{12} + 1867776 \p^{10} + 1910272\p^8 + 822912\p^6 + 126876\p^4 + 1715\p^2 - 70 \right)$ \\
\hline
\vspace{0.2cm} \\
\hline
$m$ & $p_{5K}^m(\rho)$ \\
$0$ & $ \frac{1}{24}$ \\
$1$ & $ -\frac{1}{24} \left(4\p^2 + 5 \right)$ \\
$2$ & $ -\frac{1}{24} \left(64\p^4 + 88\p^2 + 23  \right)$ \\
$3$ & $ -\frac{1}{24} \left(512\p^6 + 896\p^4 + 436\p^2 + 53  \right)$ \\
$4$ & $ -\frac{1}{120} \left(16384\p^8 + 35840\p^6 + 25728\p^4 + 6752\p^2 + 475 \right)$ \\
$5$ & $ -\frac{1}{840} \left(655360\p^{10} + 1736704 \p^8 + 1656320\p^6 + 683648\p^4 + 113852\p^2 + 5215 \right)$ \\
\hline
\vspace{0.2cm} \\
\hline
$m$ & $p_{5E}^m(\rho)$ \\
$0$ & $ \frac{1}{24} \left( -\p^2 + 1\right)$ \\
$1$ & $ \frac{1}{24} \left( 4\p^4 + 7\p^2 + 1 \right)$ \\
$2$ & $ \frac{1}{24} \left( 64\p^6 + 120\p^4 + 55\p^2 + 1\right)$ \\
$3$ & $ \frac{1}{24} \left( 512 \p^8 + 1152\p^6 + 788\p^4 + 151\p^2 + 1 \right)$ \\
$4$ & $ \frac{1}{120} \left( 16384\p^{10} + 44032\p^8 + 40576\p^6 + 14432\p^4 + 1499\p^2 + 5\right)$ \\
$5$ & $ \frac{1}{840} \left( 655360\p^{12} + 2064384\p^{10} + 2401792\p^8 + 1247616\p^6 + 272412\p^4 + 17669\p^2 + 35\right)$ \\
\hline
\hline
\end{tabular}
\caption{The polynomials $p_{nK}^m$ and $p_{nE}^m$ appearing in Eqs.~(\ref{integralI3}) and (\ref{integralI5}), for $m=0, \ldots, 5$.}
\label{table:p-polynomials}
\end{table}

\section{Stability of the finite difference method\label{appendix:von-neumann}}
As discussed in Sec.~\ref{subsec:stability}, in vacuum simulations we observed a numerical instability arising first near the poles, with a short wavelength $2\Delta$ in the $\theta$ direction and an exponentially-growing amplitude. The origin of the instability can be better understood by applying a von Neumann stability analysis (see, e.g., \cite{NumericalRecipes}) to the finite difference equations (\ref{eq-finitediff}) in vacuum (i.e., for $\Seff^m = 0$). Let us consider some numerical `noise' on a timeslice $t=t_c$ with a short angular wavelength of $2\pi / \kappa$ (with $\kappa \gg 1$) and an amplitude $\epsilon_\kappa$.  If it turns out that the finite difference method amplifies this noise exponentially, then we expect the method to be unstable. Let us begin with an ansatz
\beq
\Psi^m(t_i,r_j,\theta_k) = \epsilon_\kappa \xi^{(t_i - t_c) / (h/2)} \exp \left( i \kappa \theta_k \right) ,\label{stability-ansatz}
\eeq 
where $t_i$, $r_j$, and $\theta_k$ are the values of coordinates at points 1--8 in the grid cell shown in Fig.~\ref{fig:grid}. Note that here we have ignored variation in the $r$ direction to focus only upon the angular instability. Next we insert (\ref{stability-ansatz}) into the finite difference equation (\ref{eq-finitediff}) to obtain a quadratic equation for the (complex) amplification factor $\xi$,
\beq
\xi^2 + \Upsilon \xi + 1 = 0,  \label{eq:quadratic}
\eeq
where 
\beq
\Upsilon = - 2 + \frac{f}{4 r^2} \frac{h^2}{\Delta^2} \left[ 2 (1 - \cos (\kappa \Delta) ) - i \Delta \cot \theta_k \sin(\kappa \Delta) + \Delta^2 \left( \frac{2M}{r} + \frac{m^2}{\sin^2 \theta_k}  \right) \right] .   \label{stab-eq}
\eeq

The general case is difficult to analyse, so let us focus on the relevant case of the short-wavelength angular mode with $\kappa = \pi / \Delta$, in the grid cell closest to the North pole with $\theta_k = \Delta$. This is precisely the mode which was observed to be the source of the instability in our numerical implementation. In this case, $\Upsilon$ is the real quantity
\beq
\Upsilon = - 2 + \frac{f}{4 r^2} \frac{h^2}{\Delta^2} \left( m^2 + 4 \right)  +  \mathcal{O}\left[ \Delta^2 \times \left(h^2/\Delta^2\right) \right]. \label{eq:nearpole}
\eeq
If $-2 \le \Upsilon \le 2$ then the roots $\xi$ of (\ref{eq:quadratic}) are a complex-conjugate pair with $|\xi| = 1$ and so we expect the method to be stable. Conversely, if $|\Upsilon| > 2$ then at least one root of (\ref{eq:quadratic}) has a magnitude larger then unity, and we expect the method to be unstable. The requirement $\left| \Upsilon \right| \le 2$ in (\ref{eq:nearpole}) leads immediately to the stability condition (\ref{stability-condition}).

\end{document}